\documentclass{aa}  
\usepackage{natbib,wasysym}
\usepackage{longtable}
\bibpunct{(}{)}{;}{a}{}{,} 
\usepackage{graphicx}
\usepackage{txfonts}
\newcommand{\zphot}{\ensuremath{z_{\rm phot}} }
\newcommand{\zspec}{\ensuremath{z_{\rm spec}} } 
\newcommand{\mfpf}{\ensuremath{\rm m_{4.5\mu m}} }
\begin{document}
\title{GMASS ultradeep spectroscopy of galaxies at $z \sim 2$
  \footnote{Based on observations of the Very Large Telescope Large
    Programme 173.A-0687 carried out at the European Southern
    Observatory, Paranal, Chile.}}

   \subtitle{VII. Sample selection and spectroscopy}

   \author{J. Kurk\inst{1,2,3} \and
          A. Cimatti\inst{4,3} \and
          E. Daddi\inst{5} \and
          M. Mignoli\inst{6} \and
          L. Pozzetti\inst{6} \and
          M. Dickinson\inst{7} \and
          M. Bolzonella\inst{6} \and
          G. Zamorani\inst{6} \and
          P. Cassata\inst{8} \and
          G. Rodighiero\inst{8} \and
          A. Franceschini\inst{8} \and
          A. Renzini\inst{9} \and
          P. Rosati\inst{10} \and
          C. Halliday\inst{3} \and
          S. Berta\inst{1}
          }

   \offprints{J. Kurk}

   \institute{Max-Planck-Institut f\"ur Extraterrestrial Physik,
              Gie\ss enbachstrasse, D-85748, Garching bei M\"unchen\\
              \email{kurk@mpe.mpg.de}
         \and
              Max-Planck-Institut f\"ur Astronomy, K\"onigstuhl 17, 
              D-69117, Heidelberg\\
         \and
             INAF-Osservatorio Astrofisico di Arcetri, 
               Largo E. Fermi 5, I-50125, Firenze\\
         \and
             Universit\`a di di Bologna,Dipartimento di Astronomia, 
               Via Ranzani 1, I-40127, Bologna\\
         \and
            CEA, Laboratoire AIM, Irfu/SAp, F-91191, Gif-sur-Yvette\\
         \and
             INAF-Osservatorio Astronomico di Bologna,
               Via Ranzani 1, I-40127, Bologna\\
         \and
             NOAO-Tucson, 950 North Cherry Avenue, Tucson, AZ 85719, USA\\
         \and
             Department of Astronomy, University of Massachusetts,
             710 North Pleasant Street, Amherst, MA 01003, USA\\ 
         \and
             INAF-Osservatorio Astronomico di Padova,
               Vicolo dell'Osservatorio 5, I-35122, Padova\\
         \and
             European Southern Observatory, 
               Karl-Schwarzschild-Strasse 2, D-85748, Garching bei M\"unchen\\
             }

   \date{Received ; accepted }

  \abstract
  {Ultra-deep imaging of small parts of the sky has revealed many
    populations of distant galaxies, providing insight into the early
    stages of galaxy evolution.  Spectroscopic follow-up has mostly
    targeted galaxies with strong emission lines at $z>2$ or
    concentrated on galaxies at $z<1$.
  }
  {The populations of both quiescent and actively star-forming
    galaxies at $1<z<2$ are still under-represented in our general
    census of galaxies throughout the history of the Universe.  In the
    light of galaxy formation models, however, the evolution of
    galaxies at these redshifts is of pivotal importance and merits
    further investigation.  In addition, photometry provides only
    limited clues about the nature and evolutionary status of these
    galaxies.
    We therefore designed a spectroscopic observing campaign of a
    sample of both massive, quiescent and star-forming galaxies at
    $z>1.4$.}
  {To determine redshifts and physical properties, such as
    metallicity, dust content, dynamical masses, and star formation
    history, we performed ultra-deep spectroscopy with the
    red-sensitive optical spectrograph FORS2 at the Very Large
    Telescope.  We first constructed a sample of objects, within the
    CDFS/GOODS area, detected at 4.5\,$\mu$m, to be sensitive to
    stellar mass rather than star formation intensity.  The
    spectroscopic targets were selected with a photometric redshift
    constraint ($z > 1.4$) and magnitude constraints ($B_{\rm AB} <
    26$, $I_{\rm AB} < 26.5$), which should ensure that these are
    faint, distant, and fairly massive galaxies.}
  {We present the sample selection, survey design, observations, data
    reduction, and spectroscopic redshifts.  Up to 30 hours of
    spectroscopy of 174 spectroscopic targets and 70 additional
      objects enabled us to determine 210 redshifts, of which 145 are
    at $z > 1.4$.  The redshift distribution is clearly inhomogeneous
    with several pronounced redshift peaks.  From the redshifts and
    photometry, we deduce that the \emph{BzK} selection criteria are
    efficient (82\%) and suffer low contamination (11\%).  Several
    papers based on the GMASS survey show its value for studies of
    galaxy formation and evolution.  We publicly release the redshifts
    and reduced spectra. In combination with existing and on-going
    additional observations in CDFS/GOODS, this data set provides a
    legacy for future studies of distant galaxies.}
   {}


   \maketitle
%

\section{Introduction}\label{sec:introduction}

Multi--wavelength surveys have provided stringent constraints on the
evolution of galaxies up to $z \sim 1$. In this framework, massive
galaxies (${\cal M} > 10^{10.5}$ M$_{\odot}$) play a special role
because they host most of the stellar mass at $z\sim0$, hence are very
suitable tracers of the cosmic history of galaxy mass assembly and
provide a benchmark for the comparison of observations with the
predictions of galaxy formation models.

While the cosmic star formation density strongly decreases from
$z\sim1$ to $z\sim0$ \citep[see][ and references therein]{hop06}, the
evolution of the galaxy stellar mass function in the same redshift
range differs markedly as a consequence of the different evolutionary
trends that galaxies have depending on their mass. In particular,
near-infrared (NIR) surveys, which are more sensitive to changes in
stellar mass up to $z\sim1-2$ than optical surveys, indicate that the
number density of massive galaxies shows only a moderate increase from
$z\sim1$ to $z\sim0$, thus suggesting that the majority of massive
galaxies were already in place at $z\sim0.7-1$, whereas lower mass
galaxies display a much faster increase in their number density from
$z\sim1$ to $z\sim0$ \citep[see e.g.,][]{fon04, gla04, dro05, cap05,
  cap06, bun06}.

These results had previously been inferred from the evolution of the
NIR luminosity function and density
\citep[e.g.,][]{poz03,feu03}, and are in broad agreement with the
\emph{downsizing} scenario proposed more than ten years ago by
\citet{cow96}, where star formation activity was stronger, earlier, and
faster for massive galaxies while low mass systems continued their
activity to later cosmic times. The downsizing is consistent with
several results obtained at low and high redshifts, such as the
mass--dependent star formation histories of early-type galaxies
\citep{tho05}, the evolution of the fundamental plane
\citep[e.g.,][]{tre05,vdw05,ser05}, the evolution of the optical
luminosity function of early-type galaxies to $z\sim1$
\citep{cim06,sca07}, the evolution of the cosmic star formation density
and specific star formation \citep{gab06,feu05,jun05}, and the
evolution of the colour--magnitude relation \citep{tan04}.
However, the results of studies aimed at constraining the star
formation rates (SFRs) and dust content of $z\sim2$ galaxies show that
dust attenuation is a strong function of galaxy stellar mass with more
massive galaxies being more obscured than lower mass objects, and
therefore that specific star formation rates (SSFRs) are constant over
about 1 dex in stellar mass up to the highest stellar masses probed
\citep[$\sim$10$^{11}$M$_\odot$,][]{pan09}.  In addition,
\citet{kar11} find that since $z = 1.5$, there is no direct evidence
that galaxies of higher mass experienced a more rapid waning of their
SSFR than lower mass star-forming systems and that since $z \sim 3$
the majority of all new stars were always formed in galaxies of M$_*$
= $10^{10.6\pm0.4}$ M$_\odot$. They conclude that the data rule out
any strong downsizing in the SSFR.  In contrast, \citet{rod10} find,
using \emph{Herschel}/PACS far-infrared photometry, that the most
massive galaxies have the lowest SSFR at any redshift.

In this framework, a key role is played by the substantial population
of distant early-type galaxies that have been spectroscopically
identified at $1<z<2$ \citep{cim04,mcc04,dad05a,sar05,doh05}.
These galaxies are very red ($R-K_s>5$, $I-H>3$ in the Vega
photometric system), display the spectral features of passively evolving
old stars with ages of 1--4 Gyr, have large stellar masses with ${\cal
  M}>10^{11} $M$_{\odot}$, E/S0 morphologies, and are strongly
clustered, with a comoving $r_0 \sim 10$ Mpc at $z\sim1$ similar to
that of present-day luminous early-type galaxies \citep[e.g.,][ see
also \citealt{kon06}]{mcc01,dad02}.

The properties of these distant early-type galaxies seemed to imply
that their precursors were characterised by (1) a strong ($>100$
M$_{\odot}$ yr$^{-1}$) and short-lived ($\tau \sim$0.1-0.3 Gyr)
starburst (where SFR $\propto \exp\,(t/ \tau)$), (2) an onset of star
formation occurring at high redshift ($z_{f}>1.5-3$), (3) a
passive--like evolution after the major starburst, and (4) the strong
clustering expected in the Lambda cold dark matter ($\Lambda$CDM)
models for the populations located in massive dark matter halos and
strongly biased environments.  However, recent studies suggest that
stars in these galaxies were formed instead by a quasi-steady SFH,
increasing with time and extending over timescales of order a few
billion years \citep[e.g., ][]{dad07a,gen08,ren09}.  
\emph{Herschel} observations indeed show that starbursts contribute only
$\sim$10\% to the total SFR density at $z\sim2$ \citep{rod11}.

All the results discussed above imply that the critical epoch for the
formation of the massive galaxies is the redshift range of
$1.5<z<3$. To properly investigate galaxy evolution in this cosmic
epoch, we started a new project called GMASS ({\it ``Galaxy Mass
  Assembly ultra-deep Spectroscopic Survey''}) based on an ESO Large
Programme (PI A.  Cimatti). The main scientific aims of GMASS can be
summarised as follows: (1) to identify and study old, passive, massive
early-type galaxies at the highest possible redshifts; (2) to search
for and study the progenitors of massive galaxies at $z>1.5$; (3) to
investigate the physical and evolutionary processes that lead to the
assembly of massive galaxies; and (4) to trace the evolution of the
stellar mass function up to $z\sim3$. In addition, the GMASS
observations allow us to study the properties of a large sample of
$z>1.4$ star-forming galaxies, including outflows, dust extinction,
and stellar metallicity.

Photometric redshifts are insufficient to fully address the above
questions because they provide limited clues on the physical and
evolutionary statuses of the observed galaxies.  Spectroscopy is
therefore essential to derive reliable and accurate spectroscopic
redshifts, perform detailed spectral and photometric SED fitting (with
known spectroscopic redshift), and characterise the nature and
diversity of galaxies in the $1.5<z<3$ redshift range.  However, the
spectroscopic approach is very challenging because a typical
$M^{\ast}$ galaxy in the local universe would be faint in the NIR,
with $K\approx21$ if observed at $z\sim2$ (in the absence of strong
star formation, as in the case of early-type galaxies), and with very
faint optical magnitudes (e.g. $R>26$, $I>25$). To attempt to overcome
these problems, we decided to push the \emph{European Southern
  Observatory} (ESO) 8.2m \emph{Very Large Telescope} (VLT) beyond the
conventional limits by performing {\it ultra-deep} multi-slit
spectroscopy in the optical with the second \emph{FOcal Reducer and
  low dispersion Spectrograph} \citep[FORS2, ][]{app98}.  The choice
of optical spectroscopy is driven by the absence of efficient NIR
multi--object spectrographs at 8--10m class telescopes. The choice of
ultra-deep spectroscopy (i.e., integrations up to 30 hours) is driven,
on the one hand, by the need to derive secure spectroscopic redshifts
for the faintest galaxies, and on the other hand by the desire to
obtain high quality and high signal--to--noise spectra for the
brighter galaxies to have the possibility of detailed and possibly,
spatially resolved spectral studies.  The GMASS project can also be
seen as an experiment to assess the spectroscopic limits of the
current generation of 8--10m class telescopes and place constraints
on the requirements of the future Extremely Large Telescopes (ELTs).

In this paper, we present the GMASS project, the definition of the
sample, the multi--band photometry, the estimates of photometric
redshifts, the details of the strategy of the spectroscopic
observations and data reduction, the redshift determination method and
results, and notes about some particular objects.  In several other
papers, more results based on the GMASS observations were reported.
\citet{cim08} described the discovery of superdense passive galaxies
at $1.4<z<2.0$ using a stack of 13 GMASS spectra.  Fits of different
stellar populations to this spectrum indicated that the bulk of the
stars in these passively evolving galaxies must have formed at $2 < z
< 3$. The galaxy radii are smaller by a factor 2$\sim$3 than those
observed in early types with the same stellar mass in the local
Universe, implying that the stellar mass surface density of passive
galaxies at $<$$z$$>$$\sim$1.6 is five to ten times higher. Such
superdense early type galaxies are extremely rare or even completely
absent in the local Universe.  \citet{cap09} confirmed that these
early-type galaxies are intrinsically massive by measuring stellar
velocity dispersions in two individual spectra at $z \approx 1.4$ and
a stacked spectrum of seven galaxies at $1.6<z<2.0$. \citet{hal08}
measured the iron-abundance, stellar metallicity of star-forming
galaxies at redshift $z\sim2$ in a spectrum created by combining 75
galaxy spectra from the GMASS survey.  The stellar metallicity is 0.25
dex lower than the oxygen-abundance gas-phase metallicity for $z\sim2$
galaxies of similar stellar mass.  \citet{hal08} concluded that that
this is due to the establishment of a light-element overabundance in
galaxies as they are being formed at redshift $z\sim2$.  \citet{cas08}
studied the evolution of the rest-frame colour distribution of
galaxies with redshift, in particular in the critical interval
$1.4<z<3$.  They used the GMASS spectroscopy and photometry to show
that the distribution of galaxies in the ($U$$-$$B$) colour vs.\
stellar mass plane is bimodal up to at least redshift $z=2$.
\citet{nol09} measured the shape of the ultraviolet (UV) extinction
curve in a sample of 78 galaxies from the GMASS survey at $1<z<2.5$
and concluded that diversification of the small-size dust component
has already started in the most evolved star-forming systems in this
redshift range.  In \citet{kur09}, we described the properties of a
structure of galaxies at $z=1.6$, which form a strong peak in the
redshift distribution within the GMASS field and an overdensity in
redshift space by a factor of six.  The deep GMASS spectroscopy also
include red, quiescent galaxies and, combined with 10 redshifts from
public surveys, provide redshifts for 42 galaxies within this
structure, from which we measured a velocity dispersion of 450
km\,s$^{-1}$.  This dispersion, together with the low (undetected)
X-ray emission, classify the structure as a group, rather than a rich
cluster, despite the presence of a red sequence of evolved galaxies,
which may have formed their stars in a short burst at $z=3$.
\citet{gia11} presented the first (tentative) evidence, based on
spectra from GMASS and other surveys, of accretion of cold, chemically
young gas onto galaxies in this structure at $z=1.6$, possibly feeding
their star formation activity.  Finally, \citep{tal12} presented
evidence for outflowing gas of galaxies at $z\sim2$, with typical
velocities of the order of $\sim$100 km\,s$^{-1}$, as measured in a
stack of 74 GMASS spectra of star forming galaxies.  Furthermore, they
found a correlation between dust-corrected SFR and stellar mass, with
a slope that agrees with other measurements at $z\sim2$.

In addition, \citet{dad07a} used GMASS and other surveys' redshifts,
to test the agreement between different tracers of star formation
rates, finding a tight and roughly linear correlation between stellar
mass and SFR for 24\,$\mu$m-detected galaxies.  However, 20\%--30\% of
the massive galaxies in the sample, show a mid-infrared (MIR) excess
that is likely due to the presence of obscured active nuclei
\citep{dad07b}, as suggested by their stacked X-ray spectrum. These
MIR excess galaxies are part of the long sought after population of
distant heavily obscured AGNs predicted by synthesis models of the
X-ray background.  We note that GMASS galaxies are also part of the
sample of high-redshift galaxies observed by the Spectroscopic Imaging
survey in the NIR with SINFONI \citep[SINS,][]{for09,cre09}.

We adopt $H_0=70$ km s$^{-1}$ Mpc$^{-1}$, $\Omega_{\rm m}= 0.3$, and
$\Omega_{\Lambda}=0.7$ and give magnitudes in the AB photometric
system \citep[AB $\equiv - 2.5 \log f_\nu$ - 48.60, where $f_\nu$ is
in erg s$^{-1}$ cm$^{-2}$ Hz$^{-1}$, ][]{oke74}, unless otherwise
stated.


\section{Sample definition}

\subsection{Project set--up}

An important ingredient of the GMASS project, apart from the
above--mentioned ultra-deep spectroscopy, is MIR imaging by the
\emph{Infrared Array Camera} \citep[IRAC, ][]{faz04} at the
\emph{Spitzer} Space Telescope \citep{wer04}.  Our MIR photometry
combined with existing ground and space-based UV to NIR photometry
allowed us to perform a pre--selection of targets based on reliable
photometric redshifts and derive more reliable estimates of the
stellar mass than those based on spectral energy distribution (SED)
fitting of objects that lack MIR photometry.  Using this
multi--wavelength data, we constructed a catalogue of 1277 objects,
called the \emph{GMASS catalogue}. After the spectroscopy was
performed, we added 28 objects for which we could determine
redshift. These were not among the 1277 objects but included as
fillers or serendipitously. The final GMASS catalogue therefore
contains 1305 objects. Obviously, it was impracticable to obtain
spectra for all of these objects.  The requested and allocated amount
of observing time for spectroscopy was 145 hours, which were
distributed over six masks including 221 unique objects, 176 of which
were present in the GMASS catalogue and 141 of which were
pre--selected for spectroscopy (the \emph{GMASS spectroscopic
  sample}).  Three of the masks were observed by employing a grism
sensitive to blue wavelengths (starting at $\sim$ 3300\,\AA) and three
others employing a grism sensitive instead to red wavelengths (ranging
from $\sim$ 0.6 to about 1\,$\mu$m).  These are called the \emph{blue}
and \emph{red} masks, respectively.  We note that in some of the
studies presented in Sec.\ \ref{sec:introduction} the complete GMASS
catalogue was used, not only those for which we have carried out
spectroscopy.

In the following subsections, we describe how the GMASS catalogue was
constructed, how photometric redshifts were determined for the objects
in the catalogue, and how the GMASS spectroscopic sample was defined.

\subsection{The GMASS field}
In terms of multi-wavelength coverage, the \emph{Chandra Deep Field
  South} \citep[CDFS,][]{gia01} is one of the most intensively studied
fields.  This field has the following properties: (1) a very low
Galactic neutral-hydrogen column, comparable to that of the Lockman
Hole; (2) no stars brighter than $m_v$ = 14; and (3) is well--suited
to observations with 8\,m class telescopes from the southern
hemisphere, such as the VLT \citep{gia01}.  The field was targeted by
a \emph{Spitzer} Legacy Programme to carry out the deepest
observations with that facility from 3.6 to 24 microns (Dickinson et
al., in preparation), the deepest existing \emph{Herschel}/PACS data
\citep{elb11,lut11}, the deepest \emph{Chandra} 4Ms imaging
\citep{xue11}, \emph{XMM} observations \citep{com11},
\emph{APEX}/LABOCA submm imaging \citep{wei09}, and \emph{AzTEC/ASTE}
mm imaging \citep{sco10}.

\begin{figure}
  \centering
  \includegraphics[width=\columnwidth,clip=]{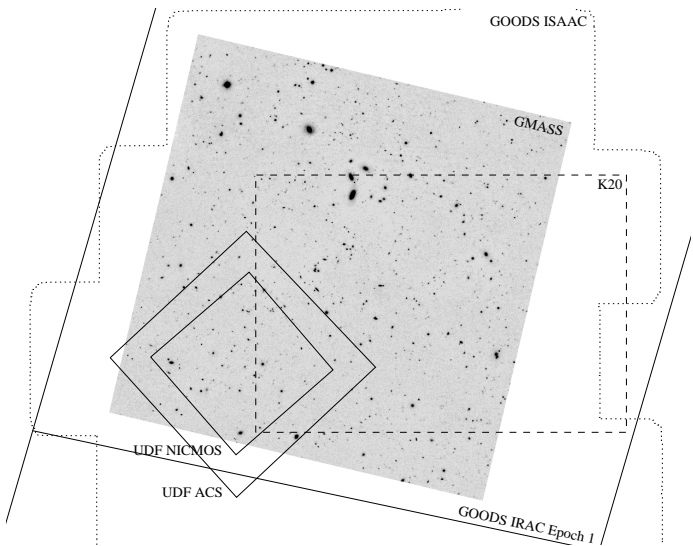}
  \caption{Location of the GMASS field (greyscale, $K_s$ band)
    compared to other fields (K20, dashed) and instrument imaging (UDF
    NICMOS and ACS, diamonds; GOODS ISAAC, dotted; GOODS IRAC, large
    rectangular, with Epoch 1 indicated) coverages.  The GMASS field
    was chosen to be inside the IRAC Epoch 1 and ISAAC imaging,
    covering as much as possible of the UDF and K20 field.  North is
    up, east to the left.}
  \label{fig:gmass_field}
\end{figure}

The GMASS sample was constrained to objects detected within a square
field of $6\farcm8\times6\farcm8$, centred at R.A.\ =
3$^h$32$^m$31$^s$3 and DEC = -27$^\circ$46\arcmin07 (J2000) and with
position angle -13.2$^\circ$ (north to east, see
Fig.~\ref{fig:gmass_field}).  The field geometry is equal to that of
the 46.2 square arcmin field of view of the FORS2 instrument and
contains enough spectroscopic targets to fill the six masks designed
for the GMASS spectroscopic survey.  It was chosen to be completely
within the area covered by the IRAC observations of CDFS, but at the
same time cover as much of the Hubble \emph{Ultra Deep Field} (UDF)
and K20 field \citep{cim02b} as possible.

\subsection{IRAC observations and photometry}

As the main contributors to the light of massive galaxies are, even at
high redshift, old stars that emit most of their light at wavelengths
above 4000\,\AA, it is important to analyse this red light when estimating
the mass of a galaxy.  This is illustrated by the properties of the
galaxies found by the successful Lyman-break technique, which
identifies high redshift galaxies based on their strong
emission in the rest-frame UV and therefore selects almost exclusively
young, low-mass, strongly star-forming galaxies \citep{ste03}.  The
red, more massive, and (relatively) less active distant galaxies are
more difficult to find, but progress also has been made here, for
example using the $BzK$ selection technique \citep{dad04b}.  However,
to select distant galaxies mainly on the basis of mass, radiation
redward of one micron in the rest-frame needs to be detected, as
variations in the mass-to-light ratio with stellar population age are
smaller at longer wavelengths, where longer-lived, cooler stars
contribute a larger fraction of the integrated luminosity.  This
became possible with the launch of the Spitzer Space Telescope, which
is equipped with a sensitive MIR camera (IRAC).

IRAC is a four-channel camera that provided (at the time of cryogenic
operation) simultaneous 5\farcm2$\times$5\farcm2 images at 3.6, 4.5,
5.8, and 8.0 microns \citep{faz04}.  The spatial resolution of the
IRAC images is limited primarily by the telescope itself, i.e.\ by its
aperture of 85~cm, resulting in a point spread function (PSF) full
width at half maximum (FWHM) of $\sim$1.6\arcsec\ at 4.5 microns.

The IRAC CDFS observations were obtained as part of the Great
Observatories Origins Deep Survey (GOODS) \emph{Spitzer} campaign and
targeted at R.A.\ = 3$^h$32$^m$30$^s$37 and DEC =
-27$^\circ$48\arcmin16\farcs8 (J2000) with a mean position angle of
-14 degrees.  The exposure time per channel is approximately 23 hours.
The data was reduced by the (\emph{Spitzer}) GOODS team and have
magnitude limits at signal-to-noise ratios (S/N) of 5 for point
sources corresponding to m$_{AB}$=26.1, 25.5, 23.5, and 23.4 at 3.6,
4.5, 5.8, and 8.0 microns \citep{dah10}.

For the first version of our catalogue, only the first epoch of IRAC
observations of GOODS-S were available, in which the GMASS area was
covered by data at 4.5 and 8.0\,$\mu$m.  Sources were detected in the
4.5\,$\mu$m channel with SExtractor \citep{ber96}, using a Gaussian
detection kernel.  After careful inspection of blended and unblended
sources, we found that the projected distance between sources detected
in IRAC images and their counterparts in the $K$ band indicates
whether a source is blended in the IRAC image.  Empirically, we
found that the criterion of $<$\,0\farcs5 separation, applied by
ourselves, is efficient at discarding the vast majority of
substantially blended sources.  It was found that approximately 25\%
of the sources to the m(4.5) $<$ 23.0 limit were blended.  After the
second epoch of IRAC observations, data at 3.6 and 5.8\,$\mu$m
covering the GMASS area became available.  A new catalogue was generated
of sources detected in a summed image of channel one and two, after
applying a \emph{Mexican hat} kernel.  The higher deblending
efficiency of this kernel resulted in only $\sim10$\% of the
sources being blended \citep[see also][]{dad07a}.  Monte Carlo
simulations were performed by the GOODS Team (in particular H.\
Ferguson), by placing point sources at random in the IRAC images and
using an empirical PSF created by the \emph{Spitzer} Science Center.
The simulations confirm the empirical conclusion because about 10\% of
the simulated galaxies were detected further than 0\farcs5 from their
original position, at $m(4.5)=23.0$, for the \emph{Mexican hat}
kernel (and a significantly larger fraction for the Gaussian kernel).
The simulations also show that for sources unresolved at the IRAC
resolution (such as distant galaxies), we recover about 90\% of the
simulated sources at $m(4.5)=23.0$.

Galaxy photometry in the IRAC bands was performed using 4\arcsec\
diameter apertures.  Monte Carlo simulations were developed to measure
photometric aperture corrections to total magnitudes. The resulting
aperture corrections were 0.316, 0.355, 0.548, and 0.681 magnitudes for
the four IRAC channels.

The GMASS sample was extracted from the public IRAC 4.5\,$\mu$m image
of GOODS-South adopting a limiting magnitude of $m(4.5) \leq 23.0$ (AB
system), corresponding to a limiting flux of 2.3\,$\mu$Jy. In this
respect, the GMASS sample is a pure flux--limited sample with no
additional colour selection criteria.  The choices of 4.5$\mu$m band
and the cut of $m(4.5) \leq 23.0$ are the result of several
considerations related to the scientific aim of the project, the
survey design, and the spectroscopic multiplexing.  The main reasons
can be summarised as follows:

\begin{figure}
  \centering
  \includegraphics[width=\columnwidth]{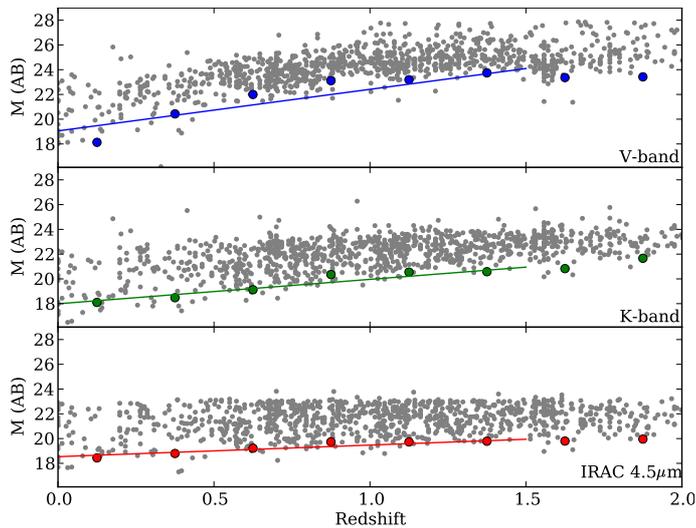}
  \caption{Magnitudes as a function of photometric redshift for the
    GMASS sample, in the ACS F606W band ({\it top}), ISAAC K$_s$ band
    (\it middle), and IRAC 4.5\,$\mu$m ({\it bottom}).  A straight
    line is fit to the median of the brightest 5\% in each $\Delta$z=0.25
    bin, up to $z$=1.5, that has a slope of 3.4, 2.0, and 0.9, respectively.
    This shows the relative strength of the K-correction for these
    bands, which is strong in $V$-band, and much weaker in
    the infrared IRAC band.}
  \label{fig:k-correction}
\end{figure}

(1) At the time the initial GMASS catalogue was developed, only the
4.5 and 8\,$\mu$m images were available for the CDFS field.  A severe
problem that occurs with this type of data is the blending of sources
due to the combination of low spatial resolution and high sensitivity.
The background confusion limit is therefore relatively quickly reached
in channel four, while the 4.5$\mu$m band is the optimal compromise
among the IRAC bands in terms of sensitivity, PSF, and image quality,
and has minor blending problems.  Moreover, it samples the rest-frame
near-infrared up to $z\sim 3$ (i.e. the expected upper redshift
envelope of the GMASS sample), thus allowing a selection that is most
sensitive to stellar mass.  In addition, the 4.5$\mu$m band detects
the redshifted rest-frame 1.6\,$\mu$m peak of the stellar SEDs for
$z>1.5$, which is consistent with the cut applied to photometric
redshifts.  The gradual shift of the 1.6\,$\mu$m peak in the
4.5\,$\mu$m band for $z>1.5$ is also responsible for a negative
k--correction effect, as illustrated in Fig.~\ref{fig:k-correction},
similar to that occurring in the submillimetre for dusty galaxies
\citep{bla93}.

(2) The limiting flux of $m(4.5) \leq 23.0$ was dictated partly by the
observational constraints imposed by the FORS2 mask exchange unit
(MXU) multiplexing, i.e. by the number of available slits with respect
to the surface density of targets at $z>1.4$ available in the
field. We carried out several tests by varying the limiting magnitude,
extracting the corresponding samples of galaxies with $z>1.4$, and
checking whether an appropriate number of targets was available to
maximise the number of targets and slits for both the blue and red
grism spectroscopy.  The cut $m(4.5) \leq 23.0$ represented the best
compromise.

(3) The photometric completeness at $m(4.5) \leq 23.0$ is 90\%.

(4) At magnitudes fainter than $m(4.5) = 23.0$, the fraction of
objects affected by blending increases significantly (e.g., from 10\%
at $m(4.5) = 23.0$ to 50\% at $m(4.5) = 25.0$).

(5) At $m(4.5) \leq 23.0$, the selection is sensitive to stellar
masses down to $\log(M/M_\odot) \approx 10.5$ for all redshifts ($0 <
z < 3$), using a Chabrier initial mass function (IMF). In particular,
the limiting mass sensitivities are $\log$(M/M$_\odot) \approx$ 9.8,
10.1, and 10.5 for $z=1.4$, 2, and 3, respectively.  This ensures that
it is possible to investigate the evolution of the galaxy mass
assembly within a mass range extending from the possible precursors of
massive galaxies (e.g., individual galaxies with $\log(M/M_\odot)
\approx 10$ that merge to form a more massive system) to the most
massive objects available at $z>1.5$.

\subsection{Optical and NIR observations and photometry}

Our optical and NIR data set consists of publicly available images
provided by several institutes.  The ground--based data includes
observations in the $U'$, $U$, $B$, $V$, $R$, $I$, $J$, $H$, and $Ks$
bands, some provided by ESO as part of its participation in the GOODS
project.

The $U'$ and $U$ band observations (PI J.~Krautter\footnote{ESO
  Programmes 164.O-0561 and 169.A-0725.}) were conducted at the
ESO/MPG 2.2\,m telescope at La Silla using the Wide-Field Imager
\citep[WFI, ][]{baa99}.  The data, which cover the full CDFS field,
have a seeing of 1\farcs1 and 1\farcs0 and reach a 5$\sigma$ limiting
magnitude, as measured within a 2 $\times$ FWHM aperture, of 26.0 and
25.7 for $U'$ and $U$, respectively \citep{arn01}.  We used release
DPS\_2.0 (7 Mar 2001\footnote{See
  http://www.eso.org/science/eis/old\_eis/eis\_rel/dps/dps\_rel.html.}),
which had been reduced by the \emph{ESO Imaging Survey} \citep[EIS,
][]{ren97} Team.  Deeper $U$ and $R$ band data obtained with the
\emph{VIsible Multi-Object Spectrograph} \citep[VIMOS, ][]{lef03} at
the VLT became available after we had constructed our catalogue
\citep{non09}.

The $B$, $V$, $R$, and $I$ band observations\footnote{ESO Programme
  64.O-0621(A).} were conducted at the ESO/VLT 8.2\,m telescope, using
FORS1.  The images have a seeing of $\sim$0\farcs7 and cover only part
of the GMASS field, their top edge being at DEC =
-27$^\circ$42\arcmin45\farcs49 (J2000).  For a description of the
data, we refer to \citet{gia01}, \citet{ros02}, and \citet{szo04}.

The $J$, $H$, and $K_s$ band observations\footnote{ESO Programme
  168.A-0485(A).} were conducted at the ESO/VLT 8.2\,m telescope,
using the Infrared Spectrometer And Array Camera \citep[ISAAC,
][]{moo98}.  At the time the GMASS catalogue was constructed, only the
$J$ and $K_s$ bands were available (GOODS/EIS release v1.0, 30 April
2004\footnote{See
  http://www.eso.org/science/goods/releases/20040430.}).  The $H$ band
(from release v1.5, 30 September 2005\footnote{See
  http://www.eso.org/science/goods/releases/20050930.}) data were
later added, but not used for the photometric redshift determination
described in Sec.~\ref{sec:photometric_redshifts}.  The individual
ISAAC pointings are assembled to form a mosaic covering the entire
GMASS field (and more). The seeing in the individual tiles varies from
0\farcs4 to 0\farcs6, and the 1$\sigma$ sky background limit in a
circular aperture of 0\farcs7 diameter from 27.4 to 27.8, from 26.6 to
27.4, and from 26.6 to 27.2, for $J$, $H$, and $K_s$ bands,
respectively \citep{ret10}.

The space--based data includes optical observations taken with the
Advanced Camera for Surveys (ACS) and NIR observations taken with the
Near Infrared Camera and Multi Object Spectrometer (NICMOS), both
aboard the Hubble Space Telescope (HST).  The GOODS ACS images
\citep{gia04} were taken with the Wide Field Channel (WFC), in four
broad, non-overlapping filters: F435W ($B_{435}$), F606W ($V_{606}$),
F775W ($i_{775}$), and F850LP ($z_{850}$) with exposure times of 3,
2.5, 2.5, and 5 orbits per filter, respectively.  The resulting
10$\sigma$ point-source sensitivities within an aperture diameter of
0\farcs2 are 27.8, 27.8, 27.1, and 26.6, respectively.  The values
reported are medians over the area covered by the HST/ACS imaging.  We
used the images reduced by the GOODS team, which was released as
version 1.0 (29 August 2003\footnote{See
  http://archive.stsci.edu/prepds/goods.}).  The Ultra Deep Field
(UDF) is located partly within the GMASS field.  The ACS UDF
observations consist of a single pointing at R.A.\ =
3$^h$32$^m$39$^s$0 and DEC = -27$^\circ$47\arcmin29\farcs1 (J2000)
imaged through the same four ACS filters as used in GOODS but for a
longer time, i.e.\ for 56, 56, 144, and 144 orbits, respectively.  The
expected 10$\sigma$ limiting magnitudes in an aperture of 0.2 square
arcsec are 28.7, 29.0, 29.0, and 28.4, respectively.  The UDF is also
(almost completely) covered by a 3$\times$3 mosaic of NICMOS pointings
through the two broad--band filters F110W ($Y_{110}$) and F160W
($H_{160}$), each filter being exposed during 8 orbits.  This resulted
in a 5$\sigma$ S/N of magnitude 27.7 through a 0\farcs6 diameter
aperture at both 1.1 and 1.6\,$\mu$m \citep{thom05}.  We used the ACS
and NICMOS data released as version 1.0 (9 March 2004\footnote{See
  http://www.stsci.edu/hst/udf/release.}).  We note that the CDFS
field will also be covered by the Cosmic Assembly Near-IR Deep
Extragalactic Legacy Survey (CANDELS), using the Wide Field Camera 3
on \emph{HST} \citep{koe11,gro11}, providing much more sensitive NIR
images than the NICMOS images used by us, and covering all of the
GMASS field.

We decided to use the $K_s$ image as the basis of our multi-band image
stack, that is, we cropped the $K_s$ image to produce the smallest
size image that still encompassed the GMASS field (whose orientation
is not such that north is up).  Since the $J$ and $H$ band images have
the same pixel scale, we performed the same procedure, after matching
their positions to the $K_s$ band image using the accurate astrometry
from the header.  The other images have different pixel scales and
were therefore transformed to match the geometry of the $K_s$ band
image using at least 200 detected objects per image, except for the
smaller NICMOS mosaic where 162 objects were used, and the shallower
$U'$ and $U$ band images, where 75 objects were used.  The RMS
deviations resulting from a surface fit to the matched object data
were about 0\farcs08 for the HST images, 0\farcs2 for the $U'$, $U$
band WFI images, 0\farcs04 for the $V$, $I$ band images, and 0\farcs1
for the $B$, $R$ images. The same procedures were followed for the
associated weight maps.

Subsequently, SExtractor \citep{ber96} was used in two-image mode
repeatedly on all images and their associated weight maps to carry out
matched-aperture photometry, using the $K_s$ image as a detection
image.  For the photometry, 1\farcs5 diameter circular apertures were
used.  The transformation of the images described above causes the
noise to be correlated between the pixels.  SExtractor, however,
assumes the noise to be uncorrelated between pixels for its noise
computation.  We therefore corrected the noise output from SExtractor
by a factor determined from independent noise measurements with
IRAF\footnote{IRAF is distributed by the National Optical Astronomy
  Observatories, which are operated by the Association of Universities
  for Research in Astronomy, Inc., under cooperative agreement with
  the National Science Foundation.}.  This factor depends on the band
but was typically below 2.0.  Objects had to have five contiguous
pixels with a S/N of $>$$1.5\sigma$ to be detected, resulting in 2609
$>$$3.3\sigma$ detections, the faintest having $K_s$ = 25.8.  We also
created a multi-band catalogue with objects that have five contiguous
pixels with a S/N of $>$$1.0\sigma$, resulting in 6207 $>$$2.2\sigma$
detections, the faintest having $K_s$ = 26.6, where faintest is in
this case defined as having an error in their magnitude smaller than
0.1.  We call this the \emph{faint} catalogue.  This catalogue was
only used to find four counterparts to IRAC detections not present in
the main $K_s$ based catalogue (see Sec.\
\ref{ssec:irac_opt_combination}).

We determined absolute magnitudes by applying a correction to each
band separately.  The correction per band was determined by comparing
the circular aperture magnitudes with SExtractor's BEST magnitude (the
Kron magnitudes for unblended cases and the isophotal one for blended
cases), for those objects deemed to be unresolved (i.e.\ SExtractor's
stellarity $>$ 0.90 and S/N $> 10$).  The correction factors correlate
quite well with the seeing on the images, i.e.\ the ACS images have
corrections in the range 0.03--0.04, the NICMOS images 0.06--0.08, the
ISAAC images 0.11--0.13, the FORS images 0.18--0.24, and the WFI
images 0.37--0.40.

\begin{figure}
  \centering
  \includegraphics[width=\columnwidth]{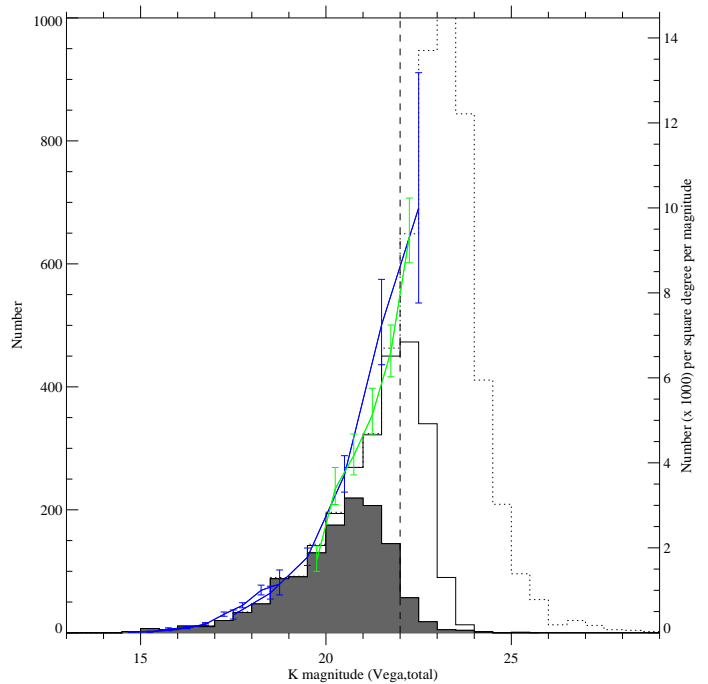}
  \caption{$K$ band counts for the 49.75 arcmin$^2$ GMASS field. Also
    indicated on the right-hand axis are counts/mag/arcmin$^2$.  The solid
    histogram is for the 3.3$\sigma$ $K$ band catalogue, while the dashed
    histogram is for the (faint) 2.2$\sigma$ catalogue.  The filled histogram
    indicated which objects are in the GMASS catalogue (i.e., are counterparts
    of an unblended 4.5\,$\mu$m source).  The vertical dashed line indicated
    the completeness limit for the 3.3$\sigma$ catalogue.  Overplotted are
    counts from \citet[][ green/light line]{gar93} and \citet[][ blue/dark
    line]{sar01}.}
  \label{fig:k_band_counts}
\end{figure}

As object detection was performed twice, once in the $K_s$ band and
once in the 4.5$\mu$m band, it is useful to know the completeness of
the $K_s$ band catalogue obtained.  We therefore compared the number
counts of the $K_s$ band catalogue and the faint $K_s$ band catalogue
with literature number counts obtained from \citet{gar93} and
\citet{sar01}\footnote{From the very useful galaxy counts webpage
  maintained in Durham:
  http://star-www.dur.ac.uk/nm/pubhtml/counts/counts.html}.  As shown
in Fig.~\ref{fig:k_band_counts}, the number counts (up to $K_{\rm
  Vega} = 22.5$) from our catalogue agree well with the literature
counts.  The number of sources detected in the $K_s$ band catalogue
deviates significantly and abruptly from the faint catalogue at
$K_{\rm Vega} > 22.0$ (or $K_{\rm AB} > 24.0$), which we therefore
accept as the completeness limit.

\subsection{Combination of IRAC and optical-NIR catalogues}\label{ssec:irac_opt_combination}

To construct the final multi-band catalogue, i.e., the GMASS catalogue,
covering wavelengths from the UV to the MIR, the optical-NIR catalogue
and the IRAC catalogue were combined by matching both catalogues.  This
matching was done by searching for counterparts to the IRAC
4.5\,$\mu$m detections in the $K_s$ band catalogue at a distance of
$\le$ 1\arcsec\ or less, using the centroid celestial coordinates.
Since the spatial resolution of the IRAC channel 2 image is not as
good as that of the $K_s$ band, some IRAC detections have two or even
three possible NIR counterparts.  All multiple counterpart cases were
checked by eye and if an unambiguous counterpart could be allocated by
eye, it was added to the GMASS catalogue.  In some cases, only a likely
counterpart could be identified, which was also added to the catalogue
but flagged as ambiguous.  At the time when only the first IRAC
catalogue was available (epoch 1, channels 2 and 4), this process
resulted in a list of 1202 objects, from which the spectroscopic
sample was selected.  We later repeated the process, using the second
IRAC catalogue (epoch 1+2, all channels), adding 70 new objects. 

For almost all IRAC sources, we found counterparts in the main $K_s$
based catalogue.  To find the remaining missing optical-NIR
counterparts, we checked the faint $K_s$ based catalogue and added
four objects from this catalogue.  Two, apparently very red,
4.5\,$\mu$m detections remained completely without a counterpart and
were added to the GMASS catalogue without optical-NIR information.  We
note that 52 of the original 1202 4.5\,$\mu$m detections are not
present in the second IRAC catalogue but were retained in the GMASS
catalogue as two had already been included in the first two GMASS
spectroscopy masks. These were mostly faint sources that, although
just being below the cut-off magnitude in the original catalogue
($m(4.5) \leq 23.0$), had a magnitude just above the cut-off in the
second IRAC catalogue. In addition, some sources that were within the
1\arcsec search radius in the original catalogue had moved outside
this radius in the second. Four bright sources were not present in the
second catalogue as they were either blended with other nearby sources
or close to a region containing artifacts from a bright star. After we
determined spectroscopic redshifts, we added 28 more objects to the
catalogue. These were included in the masks as fillers or
serendipitously. The final GMASS catalogue contains 1305 objects.

The internal consistency of the photometry in the GMASS catalogue was
tested by comparing optical, NIR, and IR photometry of stars in the
field using models by \citet{lej97} and \citet{bru03}, and IRAC
observations by \citet{eis04}.  As some discrepancy with the IRAC
photometry could not be ruled out to better than 10\%, we added this
uncertainty in quadrature to the measurement errors in the IRAC
photometry.

Only 7\% of the sources in the GMASS catalogue have magnitudes fainter
than $K_{\rm Vega} = 22.0$, where the $K_s$ band catalogue is
incomplete.  The number of 4.5$\mu$m counterparts in the $K_s$ band
begins to deviate slightly from the total number of $K_s$ band counts
at $K_{\rm Vega} = 19.5$ and significantly at $K_{\rm Vega} = 20.5$,
indicating that many $K_s$ band sources fainter than this limit do not
have an IRAC counterpart with $\mfpf < 23.0$, either because these
counterparts are indeed fainter than $\mfpf = 23.0$ or these
counterparts are blended with other sources in the IRAC image.

Within the catalogue, there are several objects that have been found
in other papers to be peculiar.  Using the deep ACS and NICMOS images
in the UDF, \citet{che04} identified nine galaxies at (their)
photometric redshift $z > 2.8$ that exhibit a pronounced discontinuity
between the F110W and F160W bandpasses.  These discontinuities are
consistent with redshifted 4000\,\AA\ breaks in E/S0 and Sab galaxy
model templates.  After some additional analysis of these nine
galaxies, they concluded that five of them have stellar masses
comparable to the present-day M$_*$ and are at least 1.6\,Gyr old.
\citet{yan04} used the same data in addition to IRAC observations, to
select objects with $f_\nu(3.6\mu m)/f_\nu(z_{850}) > 20$, called
IEROs for IRAC-selected extremely red objects.  After discarding 58
objects, whose IRAC photometry may be inaccurate because of nearby
objects, they retain a sample of seventeen bona-fide IEROs.  The SEDs
of these objects are best explained by the presence of an old
($\sim$1.5--2.5\,Gyr) stellar population in galaxies at $1.6 < z <
2.9$ with stellar masses of 0.1--1.6 $\times 10^{11}$ M$_\odot$.  All
nine objects from \citeauthor{che04} and 14 objects from
\citeauthor{yan04} are included in the GMASS catalogue.  Four of these
are common between the two papers.

\subsection{Photometric redshift determination}
\label{sec:photometric_redshifts}\label{sec:photz}

Photometric redshifts for the objects in the GMASS catalogue were
estimated by applying the \emph{HyperZ} software\footnote{See
  http://webast.ast.obs-mip.fr/hyperz.}, version 1.1 \citep{bol00}.
This photometric redshift code is based on the fitting of given
spectral energy distributions (SEDs) to the observed data.  Using a
range of redshifts and reddening vectors, the sum of the squared
difference between the observed and template flux divided by their
uncertainty, is minimised.  Redshifts were computed between $z = 0$
and $z=5$ in steps of $\Delta z = 0.044$.  A range of reddening was
also applied, using Calzetti's reddening law \citep{cal00} with $A_V$
between 0 and 1 magnitude and steps of 0.1 magnitude. These parameter
ranges are very broad and we therefore assume they represent flat
priors.

\begin{figure*}
  \centering
  \includegraphics[width=\linewidth]{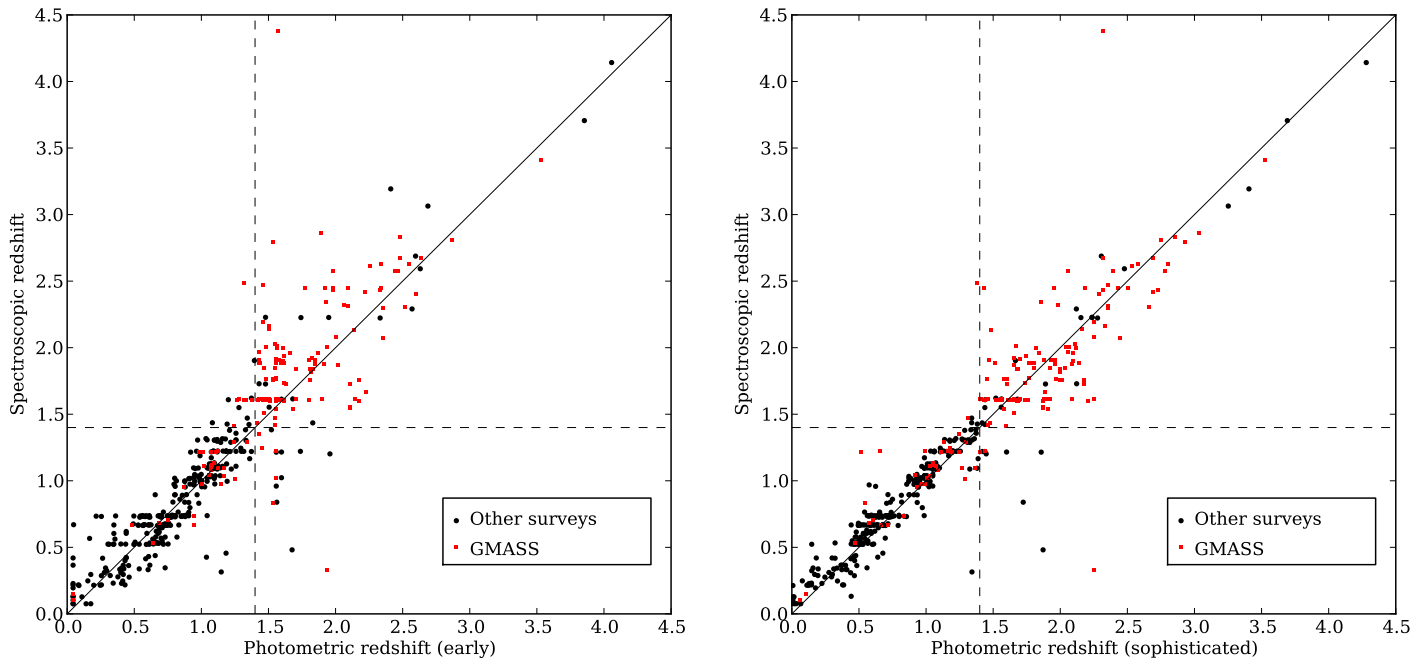}
  \caption{Plots of spectroscopic versus photometric redshifts for 309
    galaxies in the GMASS field with secure redshifts determined in
    other surveys (dark [black] circles) and 160 galaxies with secure
    redshifts determined by GMASS spectroscopy (bright [red] squares).
    The latter set of redshifts are shown here only to illustrate the
    quality of the photometric redshifts.  They were not used to
    optimise the photometric redshifts itself.  The plot on the left
    shows the \emph{early} photometric redshifts used to select the
    spectroscopic sample (based on the partly available IRAC images
    available at that time), while the plot on the right shows the
    photometric redshifts determined later, using all four IRAC bands,
    but based on the same set of 309 spectroscopic redshifts.}
  \label{fig:photzspecz}
\end{figure*}

The resulting photometric redshifts were compared to 309 secure
spectroscopic redshift values in the GMASS field available from the
literature (at that time, see Sec.~\ref{ssec:other_spec_redshifts}).
Assessing the difference between the fitted template flux and the
observed flux for all objects in the catalogue revealed systematic
offsets for some bands, which indicates that the colour terms are
caused mainly by the incorrect relative flux-calibration between the
bands, at least for the aperture photometry used here.  After
correcting these offsets, the photometric code was run again to see
whether photometric redshifts closer to the known spectroscopic
redshifts could be obtained, at which point the last two steps could
be repeated again, a process called \emph{tuning}.  We performed a
large number of tuning steps, which also involved including different
template spectra and excluding some observing bands (as some
ground-based bands overlap with some space-based bands).  After 28
runs, we concluded that we had obtained optimal results, given the
data at hand. The tuning resulted in zero-point offsets of -0.35,-0.33
for the U' and U bands, and values between -0.15 and 0.19 for the
other bands, except for the offsets for the IRAC 3.5 and 4.5 $\mu$m
bands, which were 0.18 and -0.21, respectively. These offsets most
likely represent corrections needed for imperfect PSF matching, and
possibly by partly inadequate template SEDs.  The zero-point offsets
were only applied to the photometric catalogue used as input for the
determination of photometric redshifts, not to the catalogue used to
construct plots in the remainder of the paper.  The mean difference
divided by $(1+z)$ and its standard deviation (RMS) between these
photometric redshifts and 309 secure redshifts from the literature
were $\Delta(z) = -0.0002$ and $\sigma(\Delta(z)) = 0.014$.  The final
SED of templates used consisted of four empirical templates and two
model templates.  The empirical templates, which were provided with
the HyperZ software, were constructed by taking the mean spectra of
local galaxies from \citet[CWW SEDs, ][]{col80} and extending these to
both the UV and IR regions using Bruzual \& Charlot models
\citep[BC93, ][]{bru93} with parameters (SFR and age) selected to
match the observed spectra \citep{bol00}.  These four templates
represent average E/S0, Sbc, Scd, and Im galaxies, but cannot
reproduce the very blue SEDs found for some high redshift galaxies.
To alleviate this problem, two model SEDs were added, representing
very young galaxies of 100\,Myr and 1\,Gyr old, generated with the
BC03 spectral synthesis code.  As the ISAAC $H$ band and the IRAC
channel 1 and 3 bands were unavailable at the time we estimated the
photometric redshifts, they were not used in this run.  In addition,
the FORS $I$ band was excluded because it is shallower than the other
available $I$ bands.  For the $B$ band, the UDF ACS band was used if
available, or otherwise the FORS $B$ band, and in places where neither
of those were available the GOODS ACS band.  The optical/NIR
magnitudes were \emph{boosted} per object to match the IRAC
magnitudes, by the difference between SExtractor's BEST magnitude and
the corrected aperture magnitude in the $K_s$ band for that object.
We checked by eye the observed SEDs and the fits made by HyperZ for
objects with $\zphot > 2.5$, all of which seemed to be fine.

After the selection of targets for spectroscopy, imaging data of the
CDFS for all four IRAC bands became available, and a more
sophisticated photometric-redshift determination was then attempted,
resulting in $\Delta(z) = 0.013$ and $\sigma(\Delta(z)) = 0.010$.
These later set of redshifts are used in the analyses in this
publication. In Fig.~\ref{fig:photzspecz}, we plot the photometric
versus spectroscopic redshifts for the 309 galaxies with known
spectroscopic redshifts, for both the \emph{early}
photometric-redshift determination and the later, more sophisticated,
determination.  We also plot the redshifts determined by GMASS, which
illustrate that the scatter at high redshifts (where few redshifts
were formerly known) is smaller for the second set of photometric
redshifts.

\begin{figure}
  \centering
  \includegraphics[width=\columnwidth]{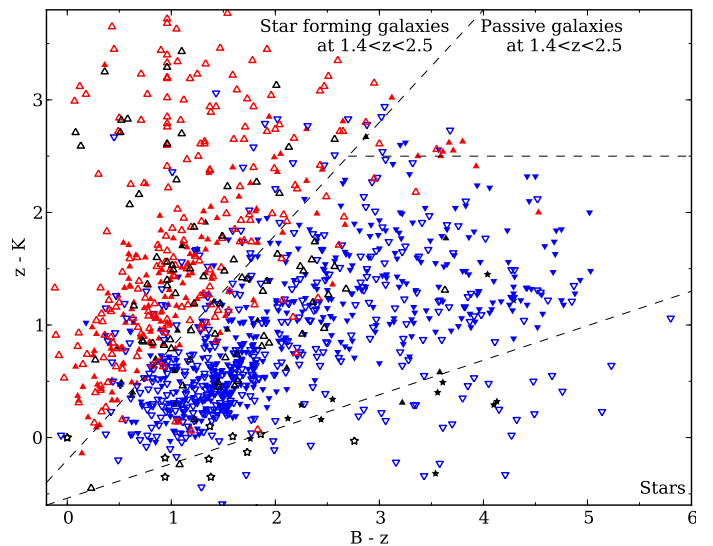}
  \caption{$B-z$ vs $z-K$ plot of all GMASS objects.  Closed symbols
    represent objects with secure spectroscopic redshifts.  For the
    remaining (open) symbols, the photometric redshift is used.
    Upward pointing triangles indicate $z > 1.4$; these are grey [red]
    for the range $1.4<z\leq2.5$ and black for $z>2.5$. Downward
    pointing (blue) triangles $z \leq 1.4$ and (black) stars indicate
    $z = 0$ (i.e., stars).  No special effort has been undertaken to
    identify stars based on their spatial appearance, hence most
    downward pointing symbols in the stellar region will indeed be
    stars.  The limits of the $BzK$ selection are indicated by dashed
    lines. }
  \label{fig:bzk_plot}
\end{figure}

Apart from photometric redshifts based on many photometric bands,
certain colour-colour selections can also give a good indication of
the redshift for particular redshift intervals.  A colour-colour
selection of special interest to our purposes, i.e.\ the selection of
galaxies at $z > 1.4$, is the $BzK$ selection \citep{dad04b}.
Applying the criterion $BzK=(z-K)-(B-z) > -0.2$ allows us to select
actively star-forming galaxies at $1.4\la z \la 2.5$, independent of
their dust reddening, while objects with $BzK<-0.2$ and $(z-K)>2.5$
colours include passively evolving galaxies in the same redshift
range.  A plot (see Fig.~\ref{fig:bzk_plot}) of the $B-z$ and $z-K$
colours of the objects in the GMASS catalogue, with the colour
selection superimposed, shows that the photometric redshifts and the
$BzK$ selection method are indeed consistent.  Of the 1275 objects
with photometric redshifts, 429 have $z_{\rm phot}>1.4$ and 865
$z_{\rm phot} \leq 1.4$.  Of the former (latter), 349 (92) fall in the
region allocated to $z>1.4$ by the BzK method.  The $BzK$ selection
method, however, should be most effective for the selection of
galaxies at $1.4<z<2.5$ \citep{dad04b}.  The number of objects with
$1.4<z_{\rm phot}\leq1.4$ is 343, while 934 have $z_{\rm phot} \leq
1.4$ or $z_{\rm phot}>2.5$.  Of the former (latter), 303 (138) fall in
the region allocated to $1.4\la z \la 2.5$ by the BzK method.  On the
basis of photometric redshifts, the $BzK$ selection seems therefore to
select $1.4<z<2.5$ galaxies with an efficiency of 69\% and to suffer
21\% contamination by $z<1.4$ galaxies.

\subsection{Other spectroscopic surveys in the CDFS field}
\label{ssec:other_spec_redshifts}
In addition to deep imaging, the CDFS field has extensive
optical spectroscopic coverage.  In particular, ESO has carried out
spectroscopic observations of all galaxies in both GOODS fields down
to a magnitude of 24--25, with this limiting magnitude being in the
$B$ and $V$ bands for objects observed with VIMOS and in the $z$
band for objects targeted with FORS2, both mounted at the VLT.  At the
time that the GMASS spectroscopic sample was defined,
\citet{van05,van06} reported that the first two FORS2 releases
together contained 930 observed sources and 724 redshift
determinations.  They used five categories of target selection, one of
which (partly) overlaps with the GMASS selection, namely, the
photometric--redshift selection based on the redshifts determined by
\citet{mob04} of galaxies at $1 < \zphot < 2$.  We note, however, that
the GMASS target selection also includes significantly fainter
objects. The full dataset of the ESO/GOODS FORS2 campaign was
presented by \citet{van08} and contains a total of 887 redshift
determinations (obtained from 1715 spectra of 1225 individual
targets). In addition, spectroscopic identifications for 114
additional galaxies were obtained in this field by
\citet{van09}. These, however, are $B$, $V$, and $i$-band dropouts, at
mean redshifts $z \sim 4$, 5, and 6.  The VIMOS spectroscopy in the
CDFS is part of the larger \emph{VIMOS VLT Deep Survey} \citep[VVDS,
][]{lef98} and targets galaxies as faint as $I \sim 24$.  At the time of
the GMASS spectroscopic sample definition, \citet{lef04} reported that
784 redshifts were determined within the GOODS field.  The redshift
distribution was peaked at a median redshift $z = 0.73$, but also
contained some redshifts at $z > 1.4$, up to $z \sim 4$.  The full
GOODS/VIMOS spectroscopic campaign \citep{pop09,bal10} produced 3218
redshifts, obtained from 5052 spectra.  These were observed with
either the \emph{blue} or \emph{orange} grism, targeting galaxies at
$1.8<z<3.5$, and $z<1 \cap z>3.5$, respectively. Another large VIMOS
survey, aimed at intermediate mass galaxies at $z \leq 1$ by
\citet{rav07}, provided an additional 531 redshifts in the CDFS.

The optical counterparts of X-ray sources found by \emph{Chandra} in
the CDFS were observed by \citet{szo04}, who presented spectroscopic
redshifts for 168 sources, mostly with magnitudes $R < 24$.  Another,
smaller, quasi-stellar object (QSO) survey based on optical and NIR
photometry was carried out by \citet{cro01}, resulting in 14 measured
redshifts.  In addition, the K20 survey was carried out in the CDFS
\citep{cim02c}.  This survey was designed to obtain optical and NIR
spectral information and redshifts of a complete sample of 545 objects
to $K_{s,{\rm Vega}} \leq 20.0$ over two independent fields, one of
which is the CDFS.  The reported redshift identification completeness
is very high ($> 92$\%, and has been increased to an even higher
percentage by the current work, see Sec.~\ref{ssec:K20}).

During our target selection, we excluded all the targets with secure
redshifts that were known at the time that the GMASS spectroscopic
sample was defined and available from the surveys mentioned above.  At
the time of the GMASS spectroscopic target selection, not all of the
above surveys had been finished.  In these cases, we avoided all
galaxies targeted by these surveys, as derived from the target lists
provided to us by the authors (e.g., Vanzella et al., private
communication).  In addition, we excluded the 29 distant supernova
(SN) host galaxies with secure spectroscopic redshifts found by
\citet{str04} in the CDFS.  The redshifts known in the CDFS were
collected by \citet{bal10} in a master catalogue that we extend
  with those obtained in the GMASS survey\footnote{CDFS master
    catalogue v2.0 by I.\ Balestra (2010) contains 7336 redshifts from
    16 observing programmes and can be found at
    http://www.eso.org/sci/activities/projects/goods/MasterSpectroscopy.html. We
    extend this catalogue, to v3.0, with 210 new entries
    (including 42 entries for galaxies that were already present
      but with lower quality redshifts, and 33 that were already
      present and had similar quality redshifts) (v3.0, see
    Sec.~\ref{sec:public_release}).}.

\subsection{Selection of targets for ultra-deep spectroscopy}

The goal of our spectroscopic campaign was to study a mass-selected
sample of galaxies at high redshift.  The mass selection is taken care
of by the use of IRAC photometry ($\mfpf \le 23.0$), while the high
redshift selection is guaranteed by the photometric redshift estimates
($\zphot \ge 1.4$).  We did not select sources on the basis of their
observed magnitudes, but we did set magnitude limits to assure that
spectroscopy was possible, of $I \le 26.0$ and $B \le 26.0$ for the
red and blue masks, respectively.  For P74, the $B$ band limit was set
to $B \le 26.5$ because we had seen from a first assessment of the
blue P73 mask that the S/N of the faintest objects in the mask was
still sufficient.  In addition, we divided the selected objects into
two samples: those most suitable for inclusion in the red masks being
red and at intermediate redshift, such that $z-K_s \ge 2.3$, $\zphot
\le 2.5$ and those most suitable for inclusion in the blue masks,
being blue with $z-K_s < 2.3$ or having $\zphot > 2.5$ such that UV
absorption lines are redshifted in the optical domain.

The target selection was done separately for P73 and, after a first
assessment of the results of P73, for P74.  For P73, using the
constraints given above, 128 and 32 objects were selected for
inclusion in the blue and red masks, respectively.  After a visual
assessment of the IRAC detections, some objects were removed from this
selection, as they seemed to have inaccurate photometry because of
blending (they were near bright objects), leaving 122 and 30 objects
in the blue and red parts of the spectroscopic target list,
respectively.

For P74, we excluded objects that already had been targeted in the P73
masks.  For the blue masks, we found 95 targets, using the fainter $B$
band limit.  This would have been 146, if the objects in the P73 masks
had not been excluded.  For the red mask, we used some extra
constraints to set priorities.  As highest priority targets (16), we
selected objects in the upper left part of the BzK diagram, i.e.\ with
$BzK < -0.2$ and $z - K > 2.5$, in the upper right, i.e.\ with $BzK >
-0.2$ and $z -K > 2.7$, objects with $\zphot > 4.0$, and
\emph{HyEROs}, i.e.\ with $J - K > 3$ (Vega magnitudes). As second
priority objects (25), we selected galaxies that had not already been
included in the steps above, which are faint in blue, but not bright
enough in red, i.e.\ $B > 26.5$.  As third priority objects (18), we
selected objects that had already been included in the red P73 mask,
but were very faint and/or not observed through optimal slits, and
objects close to the upper left part of the BzK diagram, i.e.\ $BzK <
-0.2$ and $2.2 < z - K < 2.5$, without selecting on the basis of
photometric redshift.

An additional 24 objects would have satisfied the constraint for
inclusion in the spectroscopic target list, had they not already
secure spectroscopic redshifts obtained in other surveys.

\begin{table*}
\caption{Spectroscopic sample selection criteria and redshift determination statistics}
\label{table:samples}
\centering
\begin{tabular}{rrrrrrrrrrrrrrrr}
\hline\hline
   &     &      &        & \multicolumn{3}{c}{Selection criteria$^{\mathrm{a}}$} &\multicolumn{5}{c}{Targets actually observed} & \multicolumn{2}{c}{$z_{\rm spec}$$\ge$0} & \multicolumn{2}{c}{$z_{\rm spec}$$>$1.4}\\
ID & P$^{\mathrm{b}}$\hspace{-1ex} & Prio$^{\mathrm{c}}$\hspace{-1ex} & M(AB)$^{\mathrm{d}}$\hspace{-1ex} & $z_{\rm phot}$ & other               & \#  & Tot$^{\mathrm{e}}$\hspace{-1ex} & Red$^{\mathrm{f}}$\hspace{-1ex} &Blue$^{\mathrm{f}}$\hspace{-1ex} & P73$^{\mathrm{g}}$\hspace{-1ex} & P74$^{\mathrm{g}}$\hspace{-1ex} & q=1$^{\mathrm{h}}$\hspace{-1ex} & q=0$^{\mathrm{h}}$\hspace{-1ex} & q=1$^{\mathrm{h}}$\hspace{-1ex} & q=0$^{\mathrm{h}}$\hspace{-1ex} \\
\hline
S2 & P73 & B1 & B$<$26.0 & 1.4$<$$z$$<$2.5& $z$$-$K$<$2.3      & 122 & 103 & 31 & 91 & 39 & 71 &  95 &   4 & 90 &  4 \\
S5b& P73 & B2 & B$<$26.0 & $z$$>$2.5      &                    &   5 &   5 &  1 &  5 &  5 &  2 &   5 &   0 &  5 &  0 \\
S1 & P73 & R1 & I$<$26.0 & 1.4$<$$z$$<$2.5& $z$$-$K$>$2.3      &  30 &  24 & 23 &  2 & 17 & 12 &  12 &   4 & 12 &  3 \\
S5 & P73 & R2 & I$<$26.0 & $z$$>$2.5      &                    &  23 &  15 & 10 &  7 &  5 & 12 &   7 &   3 &  6 &  3 \\
S6 & P73 & R3 & I$<$26.0 & 1.4$<$$z$$<$2.5& 1.8$<$$z$$-$K$<$2.3&  22 &  15 & 13 &  4 &  6 & 11 &   9 &   2 &  9 &  2 \\
S7 & P73 & R4 & I$<$26.0 & 1.4$<$$z$$<$2.5& 1.6$<$$z$$-$K$<$1.8&  18 &  15 &  8 &  9 &  9 &  8 &  10 &   3 & 10 &  3 \\
S8 & P73 & R5 & I$<$26.0 & 1.4$<$$z$$<$2.5& 1.4$<$$z$$-$K$<$1.6&  24 &  17 &  7 & 12 &  7 & 14 &  13 &   0 & 12 &  0 \\
   & P73 &\multicolumn{4}{l}{Total unique targets in P73 samples (S2-S8)}      & 202 &  86 & 61 &  34 & 44 &  57 &  51 & 12 & 49 & 11 \\
\hline
S21& P74 & B1 & B$<$26.5 & $z$$>$1.4      & $I$$>$26.0         &   1 &   1 &  1 &  1 &  0 &  1 &   1 &   0 &  1 &  0 \\
S22& P74 & B2 & B$<$26.5 & $z$$>$1.4      & $I$$<$26.0         &  94 &  71 & 19 & 70 &  1 & 70 &  64 &   2 & 59 &  2 \\
S25& P74 & R1 & I$<$26.0 & - &
$B$$z$$K^{\mathrm{i}}$, $z$$-$$K$$>$2.5                         &   8 &   8 &  8 &  0 &  0 &  8 &   7 &   1 &  5 &  0 \\
S27& P74 & R2 & I$<$26.0 & $z$$>$4.0      &                    &   4 &   4 &  4 &  0 &  0 &  4 &   1 &   2 &  0 &  2 \\
S28& P74 & R3 & I$<$26.0 & $z$$>$1.4      & $B$$<$26.5         &  25 &  12 & 12 &  0 &  0 & 12 &   5 &   3 &  5 &  3 \\
S29& P74 & R4 & I$<$26.0 & - &
\hspace{-2ex}$B$$z$$K^{\mathrm{i}}$, 2.2$<$$z$$-$$K$$<$2.5      &  11 &   9 &  9 &  0 &  0 &  9 &   7 &   0 &  1 &  0 \\
S30& P74 & R5 & I$<$26.0 & -              &$J$$-$$K$$>$3 (Vega)&   3 &   3 &  3 &  0 &  0 &  3 &   0 &   1 &  0 &  1 \\
S31& P74 & R6 & \multicolumn{3}{l}
                         {Promising faint targets in P73 masks}&   7 &   7 &  7 &  0 &  7 &  6 &   3 &   1 &  3 &  1 \\
   & P74 &\multicolumn{4}{l}{Total unique targets in P74 samples (S21-S31)}     & 144 & 114 & 62 &  71 &  8 & 112 &  88 & 10 &  74 &  9 \\
\hline
\multicolumn{2}{r}{P73+P74}
         & \multicolumn{4}{l}{Total unique targets in red or blue sample}                            
 & 221 & 174 & 92 & 105 & 66 & 125 & 135 & 15 & 120 & 13 \\
\multicolumn{2}{r}{P73+P74}
         & \multicolumn{4}{l}{Total unique targets in red sample}
 & 135 & 102 & 77 &  34 & 44 &  73 &  64 & 14 &  54 & 12 \\
\multicolumn{2}{r}{P73+P74}
         & \multicolumn{4}{l}{Total unique targets in blue sample}
 & 140 & 115 & 34 & 103 & 44 &  80 & 104 &  4 &  98 &  4 \\
\multicolumn{2}{r}{P73+P74}
         & \multicolumn{4}{l}{Total fillers from catalogue}      
 &     &  40 & 17 &  23 & 19 &  21 &  33 &  0 &   5 &  0 \\ 
\multicolumn{2}{r}{P73+P74}
         & \multicolumn{4}{l}{Total fillers not in catalogue}   
 &     &  41 & 25 &  16 & 15 &  16 &  26 &  3 &   8 &  2 \\ 
\hline
\end{tabular}
\begin{list}{}{}
\item[$^{\mathrm{a}}$] Additional criteria valid for all samples are m(4.5)$<$23.5, crfl$\le$2, and no secure known redshift from earlier spectroscopic surveys. For the P74 samples, targets already included in the P73 masks were excluded (except in sample S31). Note that the sample selection criteria are not mutually exclusive, i.e., objects can appear in more than one sample.
\item[$^{\mathrm{b}}$]Sample constructed for mask design in observing period 73 (P73) or 74 (P74).
\item[$^{\mathrm{c}}$]Priority during mask design for inclusion in a mask.
\item[$^{\mathrm{d}}$]Magnitude limit in B or I (AB).
\item[$^{\mathrm{e}}$]Total number of targets from this sample actually included in a spectroscopic mask.
\item[$^{\mathrm{f}}$]Observed in either a red or blue mask.
\item[$^{\mathrm{g}}$]Observed in either period 73 (P73) or period 74 (P74).
\item[$^{\mathrm{h}}$]Redshift determination quality flag, for either a (1) secure or (0) tentative determination.
\item[$^{\mathrm{i}}$] $B$$z$$K$ here indicates ($z$$-$$K$)$-$($B$$-$$z$)$<$$-0.2$
\end{list}
\end{table*}

\section{Spectroscopic observations}

\subsection{Spectroscopic strategy}

Spectroscopic observations were carried out in service mode in three
periods (ESO periods P73, P74, P75, and P76 from August 2004 until
November 2005) with FORS2 at ESO's 8.2m VLT ANTU (UT1).  The FORS2
spectrograph is equipped with a MXU, which contains laser-cut
multi-object spectroscopy masks.  It also has a range of available
grisms.  We chose to use the blue 300V grism without an order
separation filter and the red 300I grism with the order separation
filter OG590, both providing a dispersion of 1.7\,\AA\ per pixel.  The
exact wavelength range covered depends on the slit position, but for
central slits the coverages are 3300--6500\,\AA, and 6000--11000\,\AA,
respectively.  A relatively low resolution was chosen because of the
large wavelength coverage that it provides, given the wide range of
redshifts we wished to survey and the broad range of spectroscopic
features we wished to detect.  The resolution is, however, high enough
to resolve enough spectral features to permit a redshift
determination.  The field of view of FORS2 is imaged by two
backside-illuminated, 2048$\times$4096 pixel CCDs.  For readout, we
used the standard spectroscopic mode of 2$\times$2 binning, 100\,kHz
speed, and high gain.

The allocated 145 hours -- which included overheads -- were
distributed over six masks, three observed through the 300V grism (the
blue masks) and three observed through the 300I grism (the red masks).
In P73, 30 hours of observing time were allocated to test our
assumption that the stability of the instrument would allow us to
combine many one-hour exposures.  In this period, we therefore
observed both a red mask (referred to as r1, from now on) for 12 hours
and a blue mask (b2) for 11 hours of pure integration time.  The
results confirmed our assumptions, allowing for even longer co-added
exposures in P74: two blue masks (b3, b4) and two red masks (r5, r6)
for 15, 15, 32, and 30 hours, respectively.  These included the
longest integration time ever executed for spectroscopic VLT
observations.  As some targets were included in two or even three
masks (in some cases both blue and red, in other cases one colour
only), the total integration time for individual targets can be up to
77 hours (see Sec.~\ref{ssec:deepest}).

The objects in the red masks have stronger continuum emission in the
red than those in the blue masks, but their spectral features are more
challenging to identify as they most probably do not possess emission
lines.  In addition, the sky emission lines at wavelengths above
7200\,\AA\ cause extra noise.  We have used on-sky dithering to avoid
integrating complete spectra on bad pixels.  For the red masks, we
also used the dithering to permit background subtraction in a way
similar to NIR observing methods.  The blue masks were therefore
dithered to two positions at a distance of 2\farcs0, while the red
masks were dithered to four positions with a 1\farcs5 distance.  To
include at least 1\arcsec\ of sky on both sides of the assumed
1\arcsec\ sized target, we had to choose a minimum slit length of
9\arcsec\ for the red masks, while for the blue masks we chose a
minimum slit length of 8\arcsec\ to be able to measure enough of the
background to perform subtraction of the sky background.  The actual
wavelength coverage for a certain slit depends, apart from the grism
and order separation filter, on the position in the mask in the
dispersion direction.  We constrained the slits to be inside an area
where the coverage would be 3500--6500\,\AA\ and 6000--9700\,\AA, for
the blue and red masks, respectively, covering about 72\% and 66\% of
the field of view available for spectroscopy, respectively.

All slits were 1\arcsec\ wide.  To ensure a correct on-sky
positioning, three 2\arcsec$\times$2\arcsec\ openings were added to
the masks centred on stars bright enough to be seen during the
acquisition.  In addition, one slit with a 8\arcsec\ length was
centred on a relatively bright point-like object to track the on-sky
dithering and seeing.

\begin{table*}
\caption{Masks: slits and exposure times}
\label{table:masks}
\centering
\begin{tabular}{llrrrrrrrrr}
\hline\hline
M$^{\mathrm{a}}$ & Grating & Slits$^{\mathrm{b}}$ & Blue$^{\mathrm{c}}$ & Red$^{\mathrm{c}}$ & P73$^{\mathrm{d}}$ & P74$^{\mathrm{d}}$ & P75$^{\mathrm{d}}$ & P76$^{\mathrm{d}}$ & Total & Used$^{\mathrm{e}}$ \\
  &         &       &      &     & [h] & [h] & [h] & [h] &  [h]  &  [h] \\
\hline
1 & 300I& 41 &  1 & 33 & 12.75 &  1.00 &-   & -   & 13.75 & 12\\
2 & 300V& 45 & 32 &  - &  4.00 &  8.50 &-   & -   & 12.50 & 11\\
3 & 300V& 43 & 39 &  1 &  -    & 15.00 &-   & -   & 15.00 & 14\\
4 & 300V& 45 & 36 &  - &  -    & 16.00 &-   & -   & 16.00 & 15\\
5 & 300I& 42 &  8 & 26 &  -    & 34.00 &-   & -   & 34.00 & 32\\
6 & 300I& 39 & 14 & 19 &  -    &  3.00 &9.75&21.00& 33.75 & 30\\
\hline
\multicolumn{2}{l}{Total}& 255$^{\mathrm{f}}$ & 130 & 79
                   & 16.75 & 77.50 &9.75&21.00&123.00 &114 \\
\end{tabular}
\begin{list}{}{}
\item[$^{\mathrm{a}}$] Mask number
\item[$^{\mathrm{b}}$] The remaining slits contained fillers  (i.e., \#fillers = \#slits - \#blue - \#red).
\item[$^{\mathrm{c}}$] If a slit contained another target in addition, this is not counted here.
\item[$^{\mathrm{d}}$] Amount of exposure time (i.e., not including any kind of overheads) obtained during ESO periods 73, 74, 75, or 76, corresponding to Apr-Sep 2004, Oct-Mar 2004/5, Apr-Sep 2005, and Oct-Mar 2005/6, respectively.  This includes time during conditions worse than specified for the service mode observations, and aborted observing blocks.
\item[$^{\mathrm{e}}$] Exposure time actually used in the reduction of the spectra.
\item[$^{\mathrm{f}}$] The total number of slits is not equal to the total number of observed targets as some targets were observed in more than one mask. In addition, a few slits contained more than one target.
\end{list}
\end{table*}

\subsection{Mask preparation}

In March 2004, a twenty-minute $I$ band image was obtained with FORS2,
consisting of six exposures of 3m24s.  This image served as a
pre-image on which spectroscopic masks were designed to ensure the
correct positioning of the slits, and to avoid having to correct for
instrument distortions.  This shallow image is not deep enough to show
the positions of all spectroscopic targets, which includes targets as
faint as $I = 26.0 (26.5)$ and $B = 25.0 (26.0)$ for P73 (P74).  To
design the masks, it is however necessary to visually identify the
targets.  We therefore constructed two \emph{pseudo pre-images}, one
for each grism.  The red pre-image was constructed by co-adding the
FORS $I$ band and the ACS GOODS $i_{775}$ and $z_{850}$ band images,
while for the blue-image we used the ACS GOODS $V_{606}$ band.  The
images were transformed to the pre-image geometry before the
co-addition.

We used dedicated software to find the optimal mask position-angle
based on the spectroscopic targets selected for the blue and red
masks.  The masks were subsequently prepared with ESO's FORS
Instrumental Mask Simulator (FIMS) software using the pseudo
pre-images described above.  In each mask, we included as many
spectroscopic targets as possible, given the constraints on slit
length and wavelength coverage.  To fill the remaining spaces, we
placed slits on additional targets of secondary interest (i.e.,
fillers).  We first positioned slits on objects from the spectroscopic
target list, using slits that slightly violated our constraints, i.e.,
were in a position without the full required wavelength coverage
and/or had slit lengths shorter than required.  Second, we included
spectroscopic targets that had also been included in other masks (of
the same or the other colour), with slits fulfilling or not fulfilling
the constraints (in this mask).  Third, we included objects that
almost fulfilled our constraints for inclusion in the spectroscopic
target list, i.e.\ with photometric redshifts slightly below 1.4.  If
none of these secondary targets were available, we put the slit on a
random object in the GMASS catalogue.  If even such an object was
unavailable, we placed the slit on an object visible in the pseudo
pre-image but not present in the GMASS catalogue (i.e.\ with $\mfpf >
23.0$ and without a determined photometric redshift).  In some cases,
more than one object was present in a slit.  As the GMASS field has
the size of the field of view of the FORS instrument, the central
positions of the masks were very close to each other, while the
position angles were 290, 90, 303, 28, 278, and 353$^\circ$,
respectively for r1, b2, b3, b4, r5, and r6, where the FIMS convention
is followed, i.e.\ north through west, where 0$^\circ$ means pointing
north.

In the blue mask for P73, 32 objects from the P73 blue spectroscopic
target list were included, four of which had incomplete wavelength
coverage and, two of which had also been included in the red mask.
For the shallower P73 mask, we gave higher priority to objects with $B
< 25.0$, of which 17 were included.  In addition, 14 objects from the
GMASS catalogue that did not fulfil the constraints for inclusion in
the spectroscopic target lists were included as fillers.

In the red mask for P73, we were able to include 17 objects
from the spectroscopic target list. In addition, 16 objects were
included that were a bit less red (down to $z-K_s \ge 1.4$, two of
which had incomplete wavelength coverage) and one object with $\zphot
> 2.5$.  All of these are also in the spectroscopic target list, but
might have been more suitable for the blue mask.  In addition, one
object already included in the blue mask was included in this red
mask.  Finally, five objects from the GMASS catalogue that did not
fulfil the constraints for inclusion in the spectroscopic target lists
were included as fillers.

In the two blue masks for P74, 71 objects from the P74 blue
spectroscopic target list were included, 17 of which had incomplete
wavelength coverage or were close to the edge of the slit.  In
addition, seven objects were included that had also been included in
another blue mask.  In addition, nine objects from the GMASS catalogue
that did not fulfil the constraints for inclusion in the spectroscopic
target lists were included as fillers.

In the two red masks for P74, 42 objects from the P74 red
spectroscopic target list were included, eight of which had incomplete
wavelength coverage, as well as one target from the P74 blue
spectroscopic target list.  Four and twenty objects that had already
been included in other red or blue masks, respectively, were also
included.  In addition, 12 objects from the GMASS catalogue that did
not fulfil the constraints for inclusion in the spectroscopic target
lists were included as fillers.

This led to a total of 170 targets being included in the masks, out of
the 221 objects in the merged spectroscopic target lists for P73 and
P74. In addition, 46 objects in the GMASS catalogue that were not in the
spectroscopic target list were observed.  For these filler objects
that were not in the spectroscopic target lists, we preferred to select
objects that had no known spectroscopic redshift.  A small number of
other objects were included in the slits serendipitously, but are not
in the GMASS catalogue.

\begin{figure}
  \centering
  \includegraphics[width=\columnwidth]{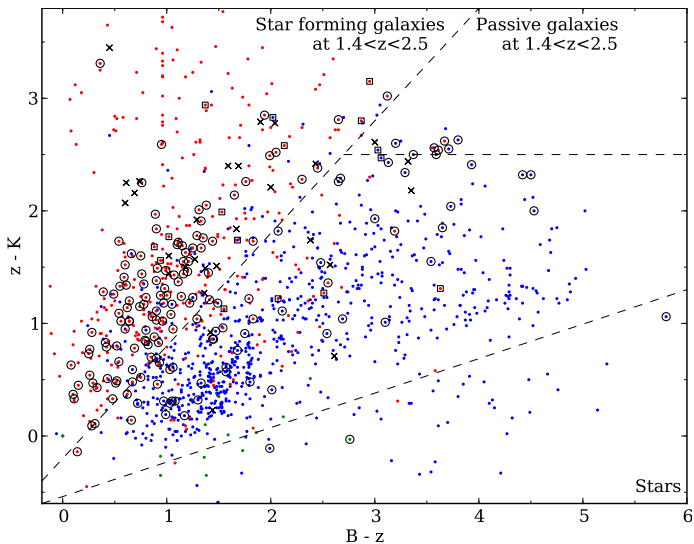}
  \caption{$B-z$ vs $z-K$ plot of all GMASS objects, showing which
    objects were included in the mask for spectroscopy.  Colours
    indicate photometric redshift either above $z=1.4$ (red) or below $z=1.4$
    (blue).  Objects included for spectroscopy are marked by crosses
    (no redshift obtained), boxes (tentative redshift obtained), and
    circles (secure redshift obtained).  Note that this includes mask
    fillers (at low redshift) and stars (for monitoring of the seeing
    and telescope pointing).  The limits of the $BzK$ selection are
    indicated by dashed lines. }
  \label{fig:bzk_spec}
\end{figure}

In total, 170 out of 221 objects from the spectroscopic selection
could be included in the masks, 36 of which were included in two
different masks (but not in three), and 5 in three different masks.
Table~\ref{table:samples} gives an overview of the samples, selection
criteria, and number of targets observed. In Table~\ref{table:masks},
we indicate the number of slits cut for targets from either the blue
or red samples (or fillers) in each mask.  In Fig.~\ref{fig:bzk_spec},
we show, in a BzK diagram, the distribution of targets actually
observed.

\subsection{Observations}

The observation blocks (OBs), including either two 30 minute (blue) or
four 15 minute (red) exposures, were carried out in service mode under
the following conditions: airmass $<$ 1.3 (blue) or 1.6 (red), lunar
illumination $<$ 0.1 (blue) or 0.4 (red), distance to moon $>$
60$^\circ$ (blue) or 120$^\circ$ (red), seeing $<$ 1\farcs0, and clear
sky.  Most of the OBs were carried out under photometric conditions.
For each mask, at least one standard star was observed, under
photometric conditions, before or after the science observations,
using a long slit but otherwise the same set-up as the science
observations.  The overheads amounted to about 25\% of the observing
time, mainly because of the time spent during the acquisition
procedure and on observations of the standard stars. In
Table~\ref{table:masks}, we provide a precise account of the exposure
times for each mask.

\section{Reduction}

The reduction of the 115 hour exposure of the six masks (and a few minutes of
standard stars) was carried out using IRAF and IDL\footnote{IDL, the
  \emph{Interactive Data Language}, is commercial software distributed by ITT
  Visual Information Solutions.}.  We note that the CCDs were read out using
on-chip pixel binning, resulting in images of 1024$\times$2048 pixels.  When
we refer to pixels in this section, we mean the latter (binned) pixels.  The
dispersion in the raw frames is therefore $\sim 3.4$\,\AA\ per pixel and the
spatial scale 2\farcs5 per pixel in this section.

Since the blue and red masks were affected by different sky backgrounds
and had different dithering patterns, the reduction differed in some
ways between them, but the first few steps were equivalent.

First, an assessment of the data quality of each observed OB was done,
including those that had been rejected by ESO.  We used some of these
rejected OBs.  These were OBs taken under conditions slightly worse than
requested (e.g., bad seeing).  Adding these improved the quality of the
co-added data, especially because we found several accepted OBs that were
also taken under slightly worse conditions than requested.

As FORS2 is equipped with two CCDs, all of the reduction steps
described below for the full frames were carried out for both CCDs.
The spectral dispersion direction is along the horizontal direction on
the CCDs.

\subsection{Flat fields}

We first treated the dome flat fields. These were taken for each night
that science observations had been carried out.  Between 5 and 20
flat-field frames were produced for each night.  As the flat fields
were very stable, we combined them all, making one flat field per
mask.  Bias values were subtracted by using the overscan region.
Using IRAF's \verb=response= task, a 75th order cubic spline was fit
interactively to the average of the lines of each separate slit.  Each
slit in the flat was then divided by its fit to form the normalized
response function.

\subsection{Wavelength calibration}

Secondly, we treated the wavelength calibration frames.  These were
taken at the same time as the flats and enabled us to also check the
instrumental stability.  As they turned out to be stable too, we used
the wavelength calibration frame for one night only for each mask.
After bias subtraction, trimming, flat fielding, and the construction
of a list of 24 (blue) and 20 (red) unblended lines out of the HeHgAr
and HeAr line lists provided by IRAF, the observed lamp lines were
identified interactively using IRAF's \verb=identify= and
\verb=reidentify= tasks.  Starting from the bottom of the CCD, three
lines were averaged, emission lines were identified, and a tenth-order
Legendre polynomial was fit to obtain a dispersion solution.  This
procedure was repeated for each set of three lines, re-using the last
dispersion solution obtained as long as the same slit was concerned,
until the top of the CCD was reached.  Depending on the position of
the slit and therefore the actual wavelength coverage, typically fewer
lines than the number of entries in the line list could be identified.
The order of the polynomial fit was decreased for slits with fewer
than 17 identified lines to ensure a plausible solution.  An
IDL procedure was written to divide the resulting database into
separate parts for each slit, removing the first and last records,
i.e.\ dismissing the first and last three lines of a slit as these
were typically contaminated by emission lines from the neighbouring
slit.

\subsection{From masks to slits}

Thirdly, we treated the science frames.  These were bias subtracted
and trimmed.  Shifts in the dispersion direction between the frames
were determined using three sky lines in three different slits (i.e.\
nine lines per frame).  The shifts were of the order of one pixel.  As
the wavelength calibration is more accurate than one pixel, these
shifts had to be corrected.  The shifts in the spatial direction were
determined using the bright object observed in the slit for dithering
tracing.  Apart from the dithering, shifts of up to several pixels
were measured.  These also had to be taken into account before the
frames could be combined.  As any non-integer pixel shifts involve
interpolation that degrades the quality of the data, we preferred to
carry out only one such step in the entire reduction process.  This
means that the distortion correction and the positional corrections
had to be done at the same time.  Interpolation of data containing
cosmic rays leads to spreading of the cosmic rays over several pixels,
which is much more difficult to remove than the cosmic rays in the
original data.  We experimented with several methods of cosmic ray
detection and removal and found the method designed by \citet{vdo01},
based on a variation of Laplacian edge detection, to work best.  This
method works by first removing the sky lines using a low-order
polynomial fit to the CCD columns and then identifies cosmic rays by
subsampling the image and convolving with the appropriate kernel.
This only works for single spectra, so we extracted the individual
two-dimensional slit spectra from each frame before applying the
procedure.  We note that a mask with an average of 40 slits, observed
for 30 hours at four dither positions, is represented by 4800 single
files (called slits from now on) at this stage.  The resulting
cosmic-ray mask is kept for later use.  The slits were subsequently
flat fielded.

The rectification transformation was determined with IRAF's
\verb=fitcoords=\footnote{We reported a (confirmed) bug in this task
  causing the displayed rms to represent the rms using the present fit
  but including also values not used (i.e.\ deleted) for the fit.}
from the fits to the arc lamp lines made earlier using a
two-dimensional Legendre polynomial of sixth order in the dispersion
direction and second order in the spatial direction (note that an
individual slit has typically only about 30 lines).  In some
cases, a fifth or seventh order was used in the dispersion direction,
depending on the number of emission lines fit.  Using the resulting
rectification transformation solutions, the slits were interpolated to
a linear wavelength scale with a dispersion of 2.5\,\AA\ per pixel,
while at the same time the shifts in the spatial and dispersion
directions were corrected so that the resulting rectified slits could
be co-added without further corrections.  Using eight unblended sky
lines, the dispersion in the rectified slits was checked and found to
be correct to within 0.5 pixels.

\subsection{Co-addition and extraction of one-dimensional spectra}

Before combining the individual frames, we computed the average
airmass of all the frames together.  First the airmasses at the
beginning, middle, and end of each exposure were computed using the
date and time, hour angle, and declination values obtained from the
FITS header. The average airmass (AM) for an exposure was then
computed by taking (AM(start) + 4$\times$AM(middle) + AM(end)) / 6.
Finally, the airmasses of all frames were averaged to obtain the
average airmass for the combined frame.

At this point, the individual files can be combined to form one file
representing an exposure of up to 32 hours.  There are different
methods for performing this.  We used different methods for the blue and
red masks, following the different dithering strategies used.  We
experimented with various methods to compare the results, trying three
methods for the blue masks and eight for the red masks, after which we
decided which method to use for the final reduction.  In the
following, we describe the methods used for the blue and red masks.

\subsubsection{Blue masks}

For the blue masks, we carried out the following steps for each slit.
First, all frames were averaged per dither position, without rejection
(as cosmic rays had already been removed).  Flux calibration,
extinction correction, and telluric absorption correction were then
applied to the two-dimensional frames.  To remove the sky lines,
IRAF's \verb=background= task was used, fitting a second-order
Legendre polynomial (with four iterations to exclude deviant pixels
from the fit) to all lines in a column.  From a visual assessment of
the background-subtracted two-dimensional image, the position(s) of
the spectrum or spectra was (were) determined and the background
subtraction was repeated on the original image using this information
to exclude the lines containing the spectrum from the column fits.  In
principle, the two two-dimensional frames could now be averaged to
form the final two-dimensional spectrum, but in almost all cases there
were defects that had to be corrected by hand at this stage.  These
included the residuals of cosmic rays, CCD defects, bright sky lines
from neighbouring slits, and slit edges.  The latter two are
particularly common in slits at the outer edges of the field of view.
The applied distortion correction corrects only along the lines, which
means that the strong distortion in FORS2 causes straight slits to be
projected onto the CCD as curved stripes.  As we created individual
slit images using a fixed number of CCD lines per slit, unexposed
pixels from regions outside the slit became visible at either short or
long wavelengths, for some of the slits.  One way to resolve this
problem is to reduce the vertical size of the region on the image
allocated to slits, but this would have reduced the area from which
the background signal can be measured significantly as the deformation
can lead to a difference in the vertical position of up to 1\arcsec\
between the blue and red edges of a slit.  The latter two effects and
the first two when occurring outside the location of the spectrum of
interest, caused undesired offsets in the background estimates.  These
were corrected by replacing the affected part of the column(s) by an
unaffected part of the column(s), typically three pixels, and redoing
the background subtraction, or excluding the affected lines from the
background fit (for certain columns).  If a defect occurred inside the
region where the actual spectrum was located, it could not be replaced
by another part of the column.  In that case, it was replaced by the
same two-dimensional region at the other dither position, reducing the
S/N in this region in the final image by a factor of $\sqrt 2$.
Finally, one-dimensional spectra were extracted from the
two-dimensional ones using unweighted summing over a 6 pixel (=
1\farcs5) aperture, unless there was clear evidence of a spatially
extended source, in which case the aperture was broadened.

\subsubsection{Red masks}

For the red masks, we used a method similar to the one applied in the
NIR, starting with the frames where cosmic rays were removed and flat
fielding was carried out.  In the following, the four dither positions
are called A, B, C, and D.  First, three dither positions (BCD, CDA,
DAB, and ABC) were median-combined without shifting to form a
representation of the sky background.  These median frames were
subtracted from the position that was not part of the median (A, B, C,
and D, respectively).  This should have taken care of the sky
background removal, but owing to temporal variations in the strength
of the sky lines some residuals remain.  The frames were subsequently
transformed to correct for the distortion.  To remove the sky line
residuals as well as possible, we used IRAF's \verb=background= task
to fit a first-order Legendre polynomial (i.e.\ a line) to the columns
and subtract this fit.  As for the blue masks, this step was repeated
once after the location of spectra in the two-dimensional frame had
been determined, avoiding the lines containing the spectra.  Finally,
the sky-subtracted frames were averaged using a sigma-clipping
rejection method and applying the appropriate shifts to obtain the
final two-dimensional spectra.  The two-dimensional spectra were
flux-calibrated and corrected for extinction and telluric absorption.
One-dimensional spectra were extracted from the two-dimensional ones
using unweighted summing over a seven-pixel (= 1\farcs75) aperture,
unless there was clear evidence of a spatially extended source, in
which case the aperture was broadened.

\subsection{Spectrophotometric calibration}\label{sec:spec_cal}

Standard stars (LTT1788 and LTT3218) were observed during some
(photometric) nights through 5\arcsec\ slits.  The observations of the
standard stars were bias subtracted using the overscan regions and
flat fielded using the flat fields taken for the 5\arcsec\ slit.
Distortion correction was also carried out using the dispersion
solution obtained from the wavelength calibration frames for the
5\arcsec\ slit.  The observations of one standard star repeated over
several nights were combined and a 14 pixel (or 3\farcs5) wide
aperture was used to extract the spectrum, using a third- to
fifth-order Legendre polynomial to trace the spectrum position.  A
response curve was determined using a standard star for each period
(P73 or P74), by fitting a 15th order cubic spline to the parts of the
standard-star observation unaffected by telluric absorption, while
those parts affected were used to create a curve representing the
telluric absorption.  We attempted to create a telluric absorption
curve from the brighter spectra in a science mask, but this turned out
to be impossible as the S/N is insufficiently high.

\begin{table}[h]
\caption{\label{table:recal_per_mask}
Average magnitude and flux difference between spectral and imaging photometry}
\centering\begin{tabular}{ccr@{$\pm$}lr@{$\pm$}l}
Mask & Grism & \multicolumn{2}{c}{$\Delta$mag\tablefootmark{a}} & \multicolumn{2}{c}{$\Delta$Flux\tablefootmark{a}}\\
\hline\hline
 1&300I& 0.23& 0.24& 1.2& 0.3\\
 2&300V& 0.59& 0.20& 1.7& 0.3\\
 3&300V& 0.58& 0.25& 1.7& 0.4\\
 4&300V& 0.48& 0.25& 1.6& 0.4\\
 5&300I& 0.13& 0.37& 1.2& 0.4\\
 6&300I& 0.49& 0.28& 1.6& 0.4\\
\end{tabular}
\tablefoot{
\tablefoottext{a}{The values are 3$\sigma$-clipped averaged over all objects in the respective mask and all three filters ($B_{435}$, $V_{606}$, $i_{775}$) that were covered by spectroscopy.}
}
\end{table}

We computed synthetic spectral magnitudes by convolving the spectra
with the HST/ACS $B_{435}$, $V_{606}$ and $i_{775}$ filters, wherever
the spectra covered the full wavelength range of these filters. Five
spectra cover only part of the $i_{775}$ filter and none of the
others. Here we convolved only the appropriate part of the $i_{775}$
filter (which was more than 50\% of its full width in all five cases).
As expected, due to the more severe flux loss -- caused by the finite
width of the slits -- for extended galaxies than for the unresolved
standard stars, the synthetic magnitudes are in almost all cases
higher than the imaging magnitudes, by 0.4 on average. The
3$\sigma$-clipped average offsets differ per mask and are listed in
Table \ref{table:recal_per_mask}, together with the corresponding
average ratio in flux. The obtained average values differ less than
0.1 mag from the median values (which were not $\sigma$-clipped).  We
excluded twelve outliers manually beforehand that were mostly
serendipitous objects not centered in the slits and therefore suffer
from additional slit losses. Also excluded were 14 objects for which
imaging photometry is not available. The average offset is slightly
higher for the bluer filters, by 0.1--0.2 mag. In Table
\ref{table:gmass_galaxies}, we list the flux ratio per spectrum
(averaged over multiple filters if available), that is also the
multiplicative factor needed to normalise the spectra in order to
obtain fluxes consistent with the imaging magnitudes.

\subsection{Galaxies observed in multiple masks}

To combine spectra of the targets that had been observed in multiple
masks, we scaled the one-dimensional spectra using their common
wavelength range, after which the common part of the spectra was
averaged.  Since the spectra had been flux-calibrated, we neither scaled nor
weighted the spectra during the combination.

\section{Spectroscopy results}

\subsection{One-dimensional spectra}

One-dimensional spectra of galaxies for which we were able to
determine a redshift (either secure or tentative, see below) are
presented in Appendix~\ref{sec:app_spectra}.  All objects present in
the GMASS catalogue that were observed spectroscopically are listed,
together with their redshifts, photometry, and the Table in
Appendix~\ref{sec:table}.

\subsection{Redshifts}

\subsubsection{Determination}

Redshifts were principally determined by finding and identifying
absorption and emission features in the galaxy spectra.  In addition,
once a sufficient number of redshifts had been determined for a mask,
an average de-redshifted SED was constructed using the spectra with
known redshifts and subsequently used to determine the redshift of
galaxies for which the first method did not result in a redshift.
Using the second method, only a few more redshifts were found and
subsequently confirmed by identifying several spectral features that
had not been noticed before.  The quality of the redshifts determined
was assessed, taking into account the number of features used and the
S/N of these, resulting in three quality flags: (1) secure redshifts;
(0) tentative redshifts, often based on only one spectral feature,
very low S/N features, or discontinuities in the observed SED; and
(-1) where no redshift could be determined.

\begin{figure}
  \centering
  \includegraphics[width=\linewidth]{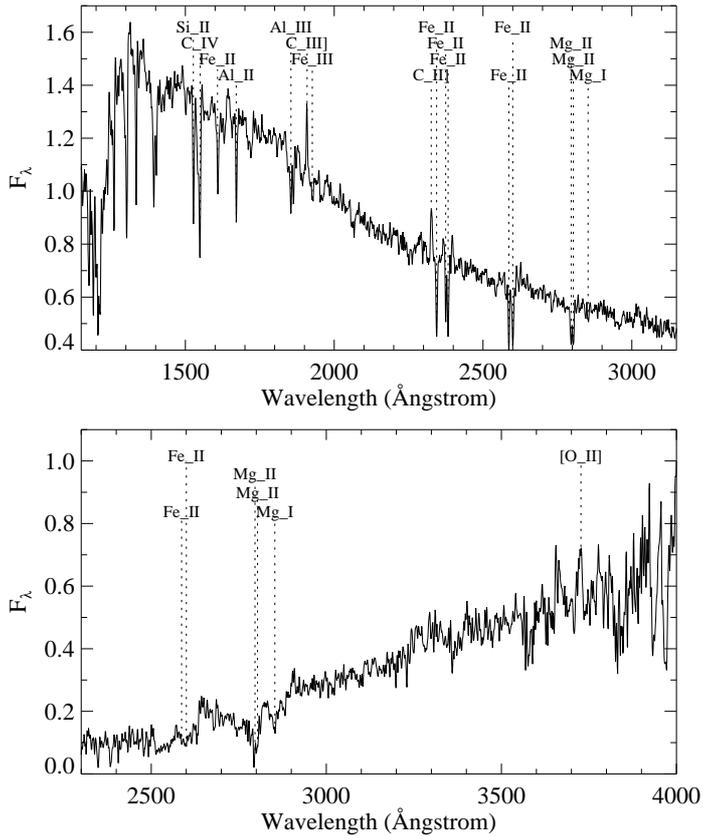}
  \caption{Rest-frame composite spectra of galaxies in the blue
    (\emph{top}) and red (\emph{bottom}) masks. The most important
    absorption lines and one emission line
    ([\ion{O}{II}]$\lambda$3727) are indicated. Note the clear
    difference in slope and strength of the FeII and MgII,MgI
    absorption lines.}
  \label{fig:bluered}
\end{figure}

In Fig.~\ref{fig:bluered}, we show composite spectra of blue (obtained
in all masks) and red (obtained in the red masks only) galaxies,
similar (but of higher S/N) to those used to determine the redshifts
of individual galaxy spectra.  During the co-addition to produce the
composites, each spectrum was shifted to its rest-frame, rebinned to
1~\AA\ bins, and normalized in the 3000-3500~\AA\ (2000-2500~\AA)
wavelength range, which is always present in the observed
spectroscopic window of the galaxies observed in the red (blue) masks.

\subsubsection{Redshift determination success}

\begin{figure*}
  \centering
  \includegraphics[width=\linewidth]{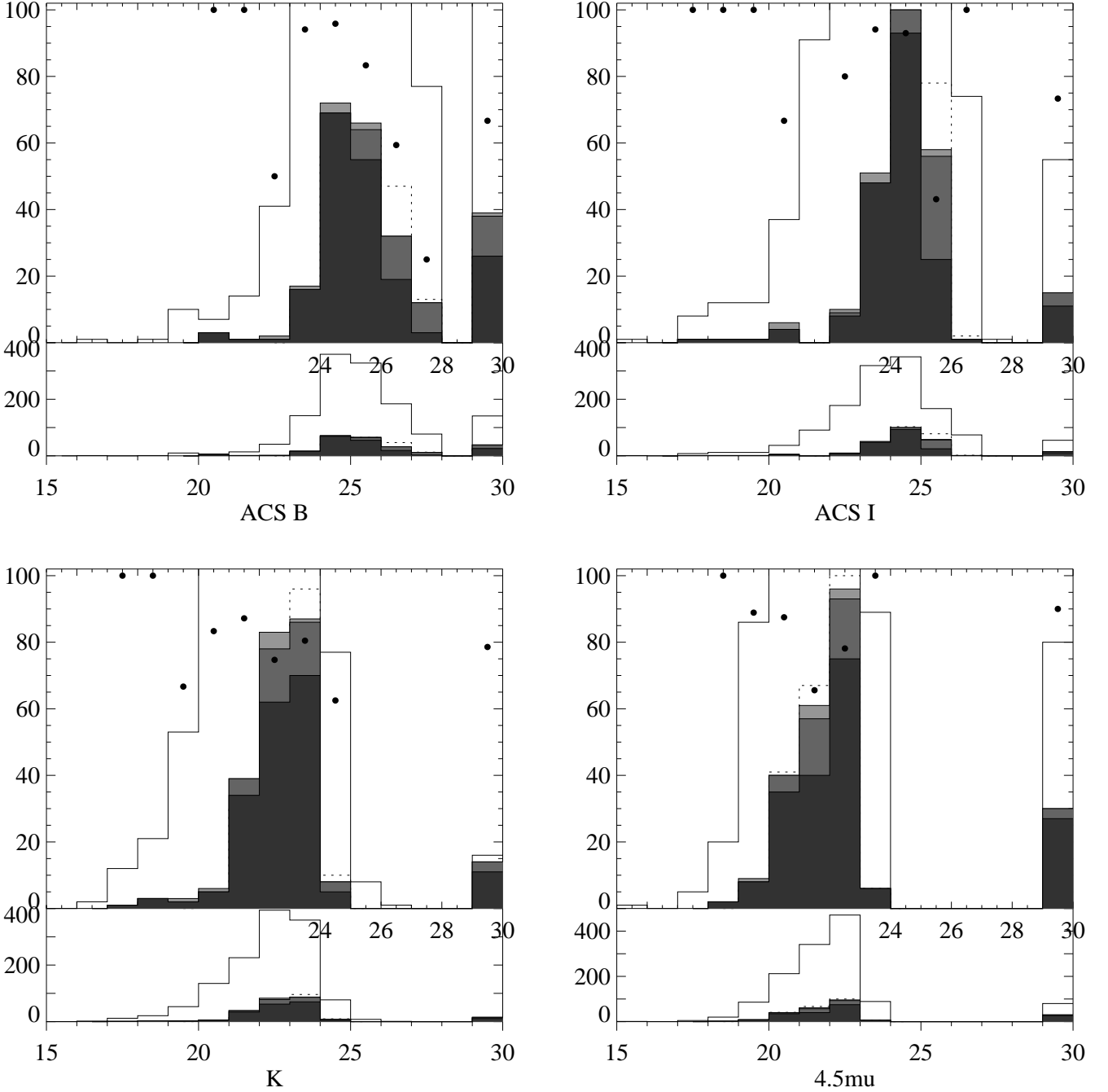}
  \caption{Magnitude histograms of the 1305 objects in the GMASS catalogue for the
    following bands: ACS $B$, ACS $I$, ISAAC $K_s$, and IRAC 4,5\,$\mu$m (from
    left to right and top to bottom).  The smaller histograms represent
    objects in the spectroscopic sample (dashed), those observed spectroscopically
    (light grey), those resulting in redshifts (grey), and those with secure
    redshifts (dark grey).  The objects in the bin at magnitude 29.5 were not
    detected in the respective bands.  Each panel contains the full histogram
    at the bottom, while the top histogram is a zoomed image of the spectroscopic
    sample and includes filled circles representing the secure spectroscopic
    redshift determination success rate per magnitude bin (in percentages).
    Note that the spectroscopic sample does not include objects with secure
    redshifts published in the literature. }
  \label{fig:mag_zspec_histograms}
\end{figure*}

In total, we were able to determine 135 secure redshifts (quality flag
1) for the 174 objects belonging to the spectroscopic target list and
observed in at least one of the GMASS masks.  In addition, 15 of these
objects have less secure redshifts (quality flag 0).  For 22 objects,
the extracted spectra did not provide clues about their redshift or
result in conflicting redshift determinations.  Finally, two objects
turned out to be too faint to allow extraction of their spectra.
Among the objects with newly determined secure redshifts, 22 had been
observed in previous surveys, but did not yet have secure redshifts.
The success rate of redshift determination by GMASS is therefore 76\%
(86\%, including less secure redshifts) for the full spectroscopic
sample observed, 63\% (76\%) for the red sample, and 90\% (94\%) for
the blue sample.  We note that these rates would have been even
higher, had we not excluded the (\emph{easier}) targets in the GMASS
field, for which redshifts had been previously determined in other
surveys.  The efficiency of selecting galaxies at $z>1.4$, i.e., the
fraction for which $z_{\rm spec}>1.4$ among those with determined
secure (tentative) redshifts is 89\% (89\%) for the complete sample,
84\% (85\%) for the red, and 94\% (94\%) for the blue sample.  In
Table \ref{table:samples}, we list some more statistics, including the
number of redshifts determined (at $z \ge 0$ and $z > 1.4$) per
sample.

Among the targets used to fill empty places in the masks, 40 objects
were in the GMASS photometric catalogue, but had not been classified
as spectroscopic targets.  For 33 of these, we managed to obtain a
secure redshift, 5 of these being at $z_{\rm spec}>1.4$.  In addition,
we extracted 41 spectra of sources not present in the GMASS
photometric catalogue, most of these being serendipitously included in
slits placed on other targets.  We were able to determine 26 secure
and 3 tentative redshifts for these fillers, 8 and 2 at $z_{\rm
  spec}>1.4$, respectively.  These are also listed in Table
\ref{table:samples}.

In Fig.~\ref{fig:mag_zspec_histograms}, we have plotted the number of
redshifts determined as a function of magnitude, for several bands, in the
form of histograms.  The spectroscopic redshift determination success is
relatively independent of the $K$ and 4.5\,$\mu$m magnitudes, but is, as
expected, a strong function of magnitude in the $B$ and $I$ bands, decreasing
from $>$90\% for $B, I < 25$ to 25\%, and 43\% for $B > 27$ and $I > 25$,
respectively.

\begin{figure}
  \centering
  \includegraphics[width=\linewidth]{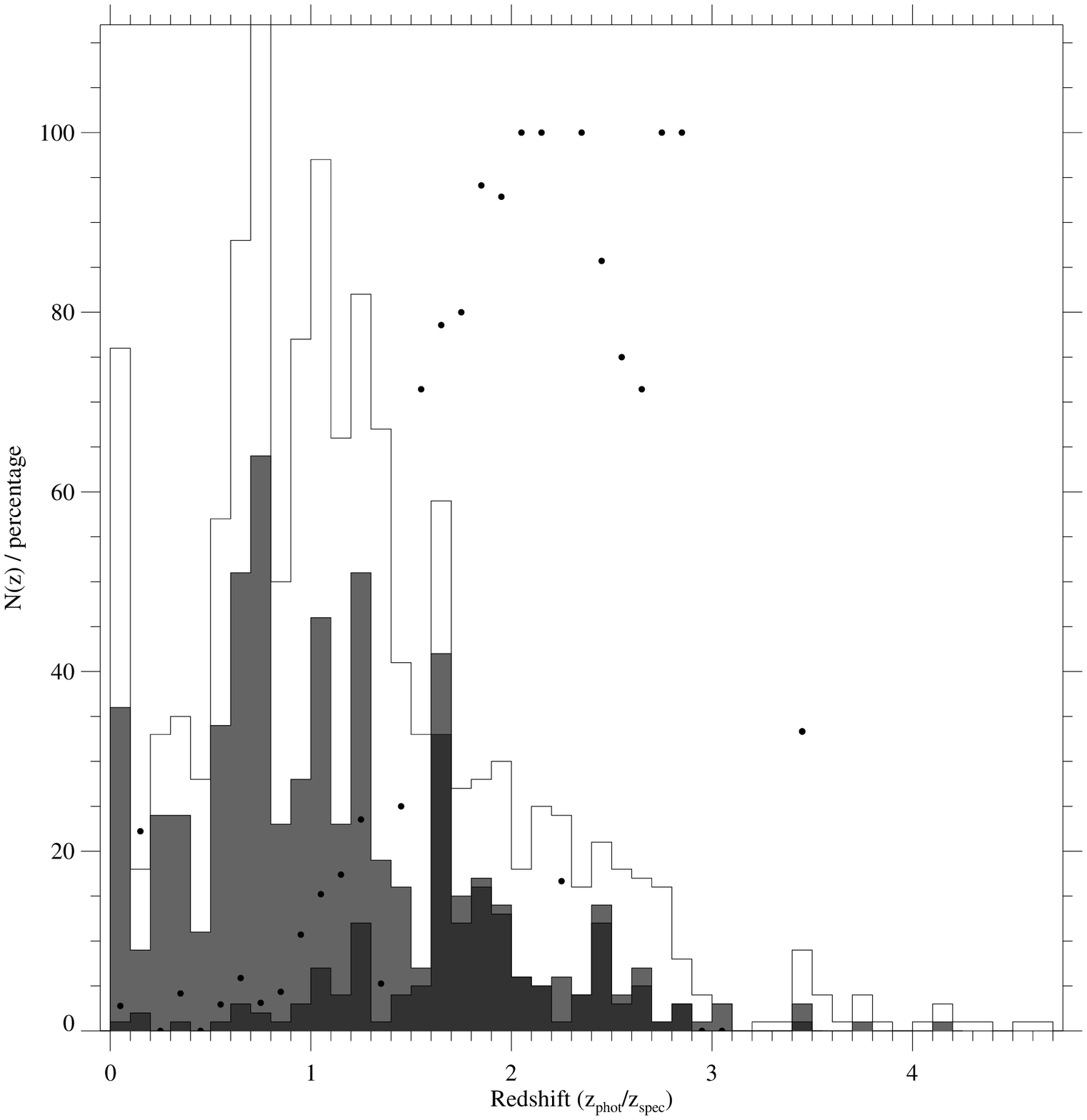} %
  \caption{Histogram of redshifts in the GMASS catalogue.  The open histogram
    represents photometric redshifts, while the grey (dark) histogram
    represents secure (GMASS only) spectroscopic redshifts.  The dots indicate
    the percentage of secure redshifts determined by the GMASS survey.}
  \label{fig:zspec_histogram}
\end{figure}

In Fig.~\ref{fig:zspec_histogram} (left panel), we have plotted a
histogram of photometric and secure spectroscopic redshifts, both
derived from GMASS observations and other surveys.  We also indicate
the ratio of spectroscopic redshifts derived from GMASS to the total
number. It is clear that, within the GMASS field, at $z > 1.5$ most
redshift information comes only from GMASS, namely 120 out of the 152
(or 80\%) spectroscopic redshifts, and, in the range $1.5 < z < 2.9$,
119 out of 145 (or 79\%) redshifts.  The redshift distribution is
inhomogeneous: several peaks are visible in the histogram.  The
properties of the most significant high-redshift overdensity at $z =
1.6$ are described in \citet{kur09}.

\subsubsection{Comparison with photometric redshifts and BzK selection}

In Fig.~\ref{fig:photzspecz}, we compare the newly obtained
spectroscopic redshifts with both the \emph{early} photometric
redshifts (see Sec.~\ref{sec:photz}) used for the sample selection and
the more sophisticated photometric redshifts obtained later.  We also
show with dashed lines for both $z_{\rm phot}$ and $z_{\rm spec}$
equal to 1.4 the lower limit to our photometric redshift selection.
The deviation of spectroscopic from the early (sophisticated)
photometric redshifts for the new GMASS redshifts (bright [red]
squares in the figure) is $\Delta(z) = 0.021$ (-0.005) and
$\sigma(\Delta(z)) = 0.041$ (0.021).  This is a factor of two higher
than the deviation from the training set of more than 300
spectroscopic redshifts in this field, most (92\%) at $z\le1.4$.
Indeed, the fraction of new $z_{\rm spec} > 1.4$ that have $z_{\rm
  phot,early} (z_{\rm phot,soph.}) > 1.4$ is 94\% (98\%), while the
fraction of new $z_{\rm spec} \le 1.4$ that have $z_{\rm phot,early}
(z_{\rm phot,soph.}) \le 1.4$ is 81\% (92\%).

As described in Sec.~\ref{sec:photz}, the $BzK$ diagram can also be
used to select galaxies at $1.4\la z \la 2.5$ with high efficiency and
low contamination.  The sample of known spectroscopic redshifts in the
GMASS field (produced by GMASS and other surveys) confirms this: of
the 570 objects with secure redshifts, 142 have $1.4<z_{\rm
  phot}\leq2.5$ (110 of these, i.e.\ 77\% have redshifts measured by
GMASS) and 428 $z_{\rm phot} \leq 1.4$ or $z_{\rm phot}>2.5$.  Of the
former (latter) 124 (27) fall in the region allocated to $1.4\la z \la
2.5$ by the $BzK$ method (see also Fig.~\ref{fig:bzk_plot}).  The
$BzK$ selection therefore seems to be efficient (82\%) and to suffer
only low contamination (18\%).  If the $BzK$ criteria are used to
select $z>1.4$ galaxies, the contamination is only 11\%.  These
percentages compare favourably to those computed for the larger sample
of 1275 photometric redshifts, which were 69\% and 31\% (or 21\% if we
are not concerned about \emph{contamination} by $z>2.5$ galaxies),
respectively.  The larger contamination among photometric redshifts
may therefore be due to inaccuracies in the photometric redshift
determination rather than due to $1.4\la z \la 2.5$ galaxies with
colours inconsistent with the $BzK$ criteria.  We note, however, that
the galaxies with secure redshifts are a sub-sample of all galaxies
within the $BzK$ region, which are probably biased towards brighter
galaxies and/or with emission lines.

\subsection{SED-derived properties of the GMASS samples}
\begin{figure}
  \centering
  \includegraphics[width=\linewidth]{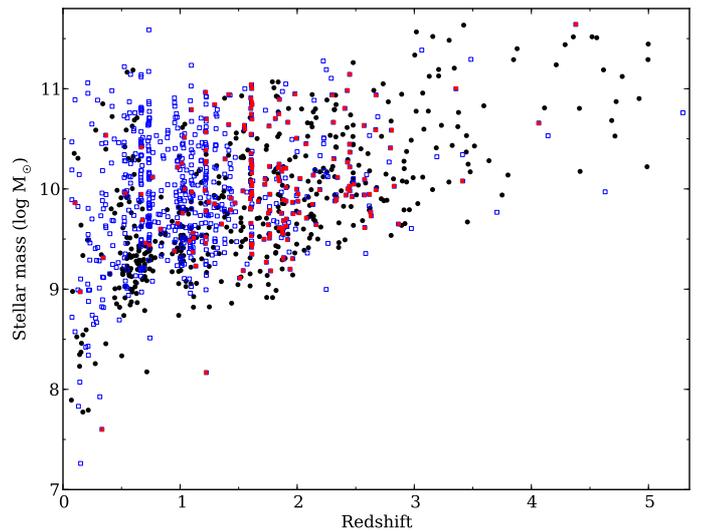}
  \caption{Stellar mass as a function of redshift in the GMASS
    catalogue.  Galaxies with spectroscopic redshifts are identified by
    open [blue] squares.  Those determined in the course of the GMASS
    survey are shown as filled [red] squares.  Photometric redshifts
    are identified by the dark [black] filled circles.}
  \label{fig:massvsz}
\end{figure}

Once the spectroscopic redshifts had been determined, new SED fits
were obtained for the whole GMASS photometric catalogue, this time
fixing the redshift parameter to the 609 spectroscopic redshifts (both
secure and tentative) known, and leaving it as a free parameter for
the other objects (resulting in a photometric redshift).  During the
fitting, observed magnitudes were used only up to rest-frame
wavelength $\lambda_0$ = 2.5 $\mu$m to avoid the influence of dust
emission (which was not included in the models used) and to minimise
the effect that different stellar population synthesis models would
have.  No photometric shifts were applied in this fitting procedure,
owing to uncertainties regarding their origin.  We employed
exponentially declining star formation histories, with characteristic
times $\tau$ = 0.1, 0.3, 1, 2, 3, 5, 10, 15, and 30 Gyr, plus a model
with a constant rate of star formation.  A minimum age of 0.09\,Gyr
was imposed.  We used only solar metallicities as we had found that
introducing a choice of metallicities did not lead to a substantial
improvement in the quality of the best-fits and produced differences
in the best-fit stellar masses $\la 0.1$\,dex, compared to solar
metallicity SEDs, at the cost of introducing an additional parameter.
We used values of extinction covering the range $0 < A_V < 4$.
Moreover, we applied a \emph{prior} in the choice of the best-fit
models, similar to the one used by \citet{fon04} and \citet{bol10},
that is, to exclude models with $A_V > 0.6$ and age/$\tau > 4$ (i.e.,
old galaxies must have a moderate dust extinction) and models with
$\tau < 0.6$ Gyr and ages for which $z_{\rm form}$ is $< 1$ (to obtain
a better estimate of the ages of early-type galaxies typically fitted
by these low-$\tau$ models). In addition to the canonical stellar
population models provided by \citet[][BC03]{bru03}, we computed
masses using the stellar population models of \citet{mar05}, with the
Kroupa initial mass function (IMF) \citep{kro01}, similar to the
Chabrier IMF \citep{cha03} used in Bruzual \& Charlot models, and
those by Charlot \& Bruzual \citep{bru07a,bru07b}, both of which
include the thermally pulsing asymptotic giant branch phase of stellar
evolution.  For intermediate age stellar populations, this phase can
contribute up to $\sim $50\% to the total bolometric light, radiated
mostly in the NIR \citep[e.g., ][]{mar05}.  The use of different SED
models implies different mass estimates: for instance the change in
the IMF from Chabrier to Salpeter produces higher estimates of the
stellar masses (0.23 dex, i.e., a factor $\sim$1.7). In the remainder
of this paper, when we refer to stellar masses, we refer to those
computed using the BC03 models as these provide the best fits among
the three models, and for consistency with previous works.

The stellar masses of observed galaxies and other galaxies in the GMASS
photometric catalogue are shown in Fig.~\ref{fig:massvsz}.  The range of
derived stellar masses for the galaxies with redshifts from GMASS
observations is between 7.5 and 11.6 in log(M$_\odot$), with most
(96\%) galaxies being between 9.0 and 11.0 in log(M$_\odot$).

\begin{figure}
  \centering
  \includegraphics[width=\linewidth]{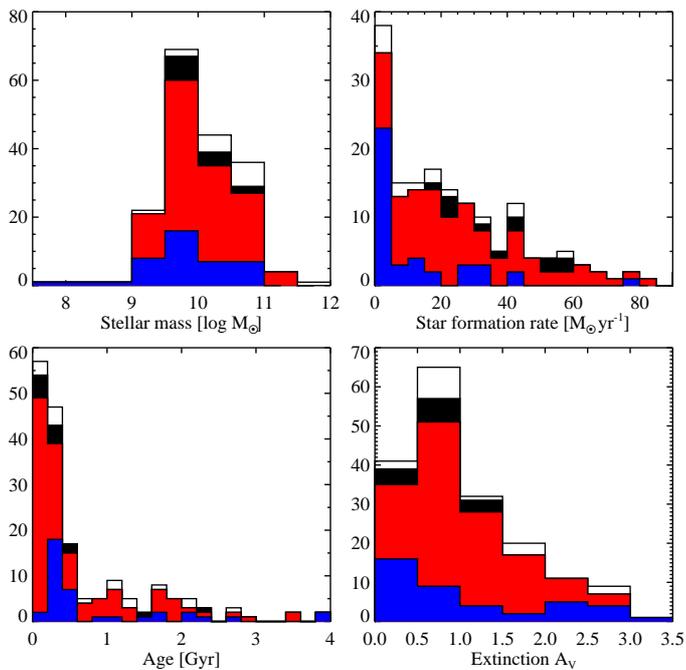}
  \caption{Histograms of SED-derived stellar masses, star formation
    rates, ages, and extinction (in A$_{\rm V}$) for the sample with
    spectroscopic redshifts derived by GMASS (open histograms), secure
    spectroscopic redshifts derived by GMASS (filled histograms), for
    all redshifts (black), redshifts $z<2.0$ (red), and redshifts
    $z<1.4$ (blue). These histograms are plotted on top of each other,
    i.e., they are not cumulative.}
  \label{fig:histograms}
\end{figure}

In Fig.~\ref{fig:histograms}, we show histograms of SED-derived
stellar masses, star formation rates, ages, and extinction (in A$_{\rm
  V}$) for the sample with spectroscopic redshifts derived by GMASS
(open histograms), secure spectroscopic redshifts derived by GMASS
(filled histograms), for all redshifts (black), redshifts $z<2.0$
(red), and redshifts $z<1.4$ (blue). There are no obvious redshift
differences in the distributions of mass and extinction. For star
formation rates, the high-redshift galaxies have the highest star
formation rates, while the lowest bin (SFR $<$ 5 M$_\odot$ yr$^{-1}$)
is dominated by the galaxies at $z<1.4$. As the GMASS catalogue was
selected on MIR magnitude, this difference in SFR can only be partly
explained by selection effects. The highest redshift galaxies, have,
as expected the lowest ages (most are younger than 0.4 Gyr).

\subsection{Extending the CDFS spectroscopic catalogue}

\begin{figure*}
  \centering
  \includegraphics[width=\linewidth]{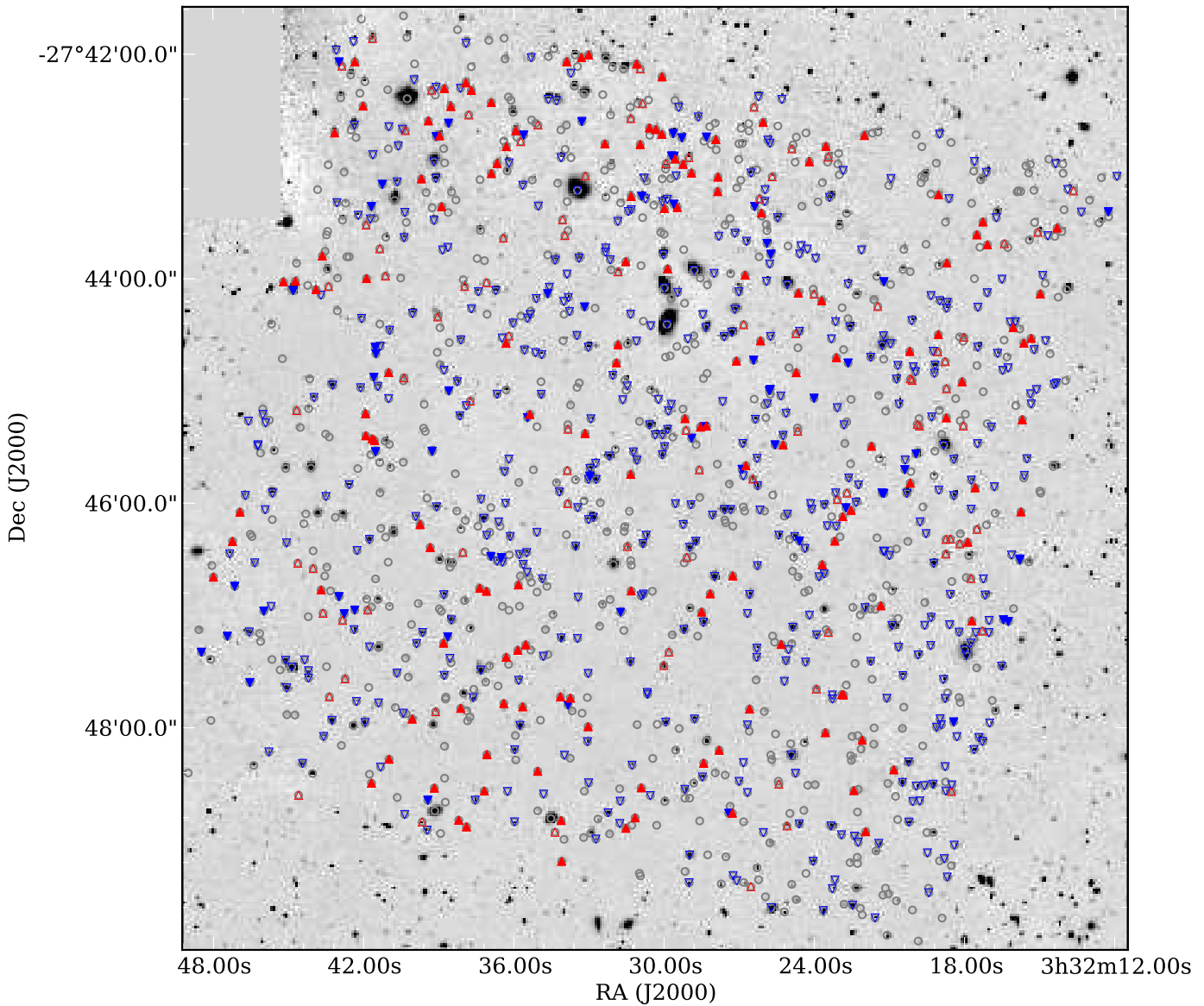}
  \caption{$K_s$-band image of the GMASS field in CDFS. Grey circles
    indicate the 1275 objects in the GMASS photometric catalogue. Blue
    downward pointing triangles indicate galaxies with spectroscopic
    redshifts $z<1.4$, while red upward pointing triangles indicate
    galaxies with spectroscopic redshifts $z\geq1.4$.  The triangles
    are filled if the redshift was determined by spectroscopy from the
    GMASS survey.}
  \label{fig:gmass_pos}
\end{figure*}

\begin{figure}
  \centering
  \includegraphics[width=\linewidth]{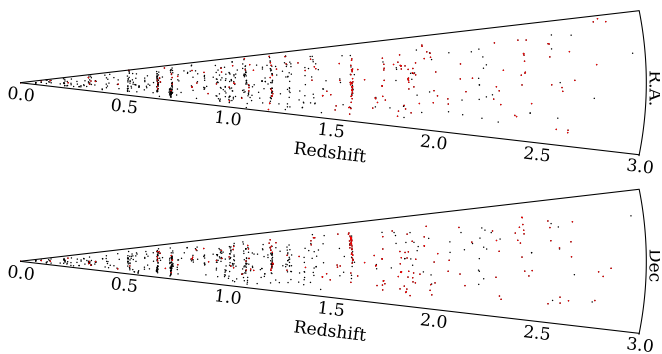}
  \caption{Cone plots showing the projection in R.A.\ in the \emph{top}
    panel and declination in the \emph{bottom} panel of the spatial
    distribution of galaxies in the GMASS field.  Only galaxies with
    spectroscopic redshifts up to $z=3.0$ are shown.  Red symbols
    indicate redshifts determined by spectroscopy from the GMASS
    survey. The angle of each cones was stretched by a factor of six to
    help visualisation.}
  \label{fig:gmass_cone}
\end{figure}

As described in Sec.~\ref{ssec:other_spec_redshifts}, the CDFS, in
which the GMASS field is located, is the focus of many spectroscopic
campaigns, providing thousands of galaxy redshifts. The GMASS survey
provides additional spectra and redshifts, which fill an important
niche in parameter space: its resulting redshifts are preferentially
in the former \emph{redshift desert} at $1.4 < z < 2.5$, and the
galaxies are up to two magnitudes fainter ($B,I < 26.0$) than those
targeted in most other surveys, which explains the extremely long
integration times needed ($\sim$ 30 hours). \citet{bal10} compiled a
\emph{master} catalogue of spectroscopic redshifts obtained by 16
authors with 7332 entries. We extend this catalogue\footnote{Version
  3.0 of the GOODS/CDFS spectroscopy master catalogue is available
  from the ESO website at
  http://www.eso.org/sci/activities/projects/goods/MasterSpectroscopy.html}
by 210 entries obtained by the GMASS project. Some of these (42)
  concern galaxies that had tentative redshifts from other surveys
  that are now replaced by more secure GMASS redshifts.  In
Fig.~\ref{fig:gmass_pos}, we show the positions in the GMASS field of
the galaxies with redshifts obtained by GMASS and other surveys. In
Fig.~\ref{fig:gmass_cone}, a redshift cone of galaxies in the GMASS
field is displayed.

\section{Notes on individual objects}

Here we provide notes on individual objects, in particular those that have
been detected at other wavelengths or observed by other surveys.

\subsection{Radio sources}

The CDFS was observed at 1.4\,GHz using the Australian Telescope
Compact Array down to a 1$\sigma$ limiting sensitivity of
$\sim$14\,$\mu$Jy (Koekemoer et al., in preparation).  Within the area
covered by the ACS observations, a total of 64 radio sources are found
with 1.4\,GHz fluxes between 63\,$\mu$Jy and 20\,mJy
\citep{afo06}. \citet{afo06} identified these radio sources with
objects detected on the ACS $z_{850}$ band using a likelihood method.
Identifications were inspected visually to check for cases where the
likelihood method might not apply.  Seven of the radio sources were
not identified with an optical source.  We cross-correlated the GMASS
catalogue with the catalogue published by \citet{afo06}, using the
coordinates of the ACS counterpart, except for the seven cases without
a counterpart, for which we used the coordinates of the radio source.
All objects at distances smaller than 1\farcs0 were considered
matches.  Fourteen of these were found, all at distances $\le$
0\farcs3 and with optical identifications by \citeauthor{afo06}.
Twelve of these already had secure spectroscopic redshifts determined
by \citet{szo04}, one has no spectroscopy at all, and one was observed
in a GMASS mask.  This last object -- GMASS\,2113, No.~24 in
\citeauthor{afo06} -- has a spectroscopic redshift of 1.613.  Our
spectrum of this galaxy displays a narrow [\ion{O}{II}] emission line,
but there is no evidence of broad lines.  It has an extended,
irregular morphology, with a colour gradient.  As this galaxy is part
of the redshift spike at $z=1.61$ in the GOODS-S field, its spectrum
and HST image can be found in \citet{kur09}, where this spike is
described in detail.  This galaxy was also observed with SINFONI, the
VLT's NIR integral field spectrograph, displaying evidence of a
merging system \citep{for09}, the clearest interacting system among
the 63 galaxies detected in the SINS survey. We note that
\citet{kel08} later also published deep radio observations of the
CDFS, performed with the Very Large Array (VLA) at 20 and 6 cm,
containing 266 sources. An even deeper VLA survey of Extended-CDFS is
presented in Miller et al. (in prep.), containing 883 sources that are
identified with optical/mid-IR sources and their spectroscopic and
photometric redshifts by Bonzini et al. (submitted).

\subsection{X-ray sources}\label{ssec:xraysources}

\citet{szo04} carried out a spectroscopic survey of optical
counterparts to X-ray sources in the CDFS, as observed by
\emph{Chandra} for 942\,ks \citep{gia02,ros02}.  They used FORS2 with
the 150I grism with typical exposure times of two to four hours. To
check whether any of the new redshifts determined by the GMASS survey
correspond to X-ray emitting sources, we cross-correlated the GMASS
catalogue with the 1\,Ms X-ray catalogue from \citet{gia02}, using a
distance of 1\farcs5 to match the coordinates, following
\citet{szo04}.  The latter authors note that within this error circle,
0.15 field galaxies are expected to fall.  That is, one false
candidate is expected for every seventh X-ray source at $R < 26$ (Vega
magnitude).

There are four (possible) X-ray counterparts with either new or now
confirmed formerly tentative redshifts:

\noindent {\bf GMASS\,2443, 2043} These objects, at distances of
1\farcs40 and 1\farcs35 from their X-ray counterparts
(\citeauthor{gia02}, Nos.\ 148, 231, resp.) were also observed by
\citet[][ Nos.\ 241, 23, resp.]{szo04} but they were unable to
determine redshifts for these objects. We observed GMASS\,2443 for a
total of 44\,h in two red masks and GMASS\,2043 for 15\,h in a blue
mask, resulting in secure redshifts of $z = 2.298$$\pm$0.004, and $z =
2.576$$\pm$0.002, respectively. GMASS\,2043 exhibits a broad emission
line and has a compact morphology, suggesting that it is a QSO.

\noindent {\bf GMASS\,1084} This object at 0\farcs96 distance from its X-ray
counterpart (\citeauthor{gia02}, No.\ 227) was observed by \citet[][ No.\ 26,
no redshift obtained]{szo04} and \citet[][ No.\ 736, tentative redshift of $z
= 1.552$]{van06}. We observed it for 15, 32, and 30\,h in one blue and two red
masks, respectively, resulting in a secure redshift of $z =
1.552$$\pm$0.004.

\noindent {\bf GMASS\,253} This object was also described by \citet[][ No.\ 7,
or 1446 in the UDF publicly available catalogue]{dad05a} as a high redshift
elliptical with probable redshift of $z = 2.47$.  We observed this object
for 30\,h in a red mask, but were unable to secure its redshift, although
our best estimate is $z = 2.670$$\pm$0.001, close to that proposed by
\citeauthor{dad05a}.  It is 1\farcs45 away from its X-ray counterpart
(\citeauthor{gia02}, No.\ 224).

\noindent In addition, {\bf GMASS\,1155} at $z=1.727$ has broad
emission lines and a compact morphology, reminiscent of a QSO.  It is
listed as No.\ 145 without a redshift by \citet{szo04}. \citet{lef04}
report a redshift of $z=1.730$ for this source.

\subsection{IEROS}

Three IEROs \citep{yan04} were included in the masks, but
unfortunately their emission was too faint to determine unambiguous
redshifts. One of these is GMASS\,253, which was also described by
\citet{che04}. It has a tentative GMASS redshift estimate of 2.670
(see Sec.~\ref{ssec:xraysources}), which is almost consistent with the
photometric redshift of $>$2.8 determined by \citet{che04} and is
consistent with the redshift range $1.6 < z < 2.9$ given by
\citet{yan04}.

\subsection{Objects from the K20 survey}\label{ssec:K20}

As the GMASS field partly overlaps with that of the K20 survey
\citep{cim02c}, eight objects previously selected for the K20 survey
were included in the GMASS masks, one already having a secure
redshift.  For all K20 targets, secure redshifts could be determined
from the GMASS observations, raising the number of K20 objects with
secure redshifts from 501 to 508 and the spectroscopic redshift
completeness of the K20 survey from 92\% to 93\%
\citep[508/545,][]{mig05}.

\subsubsection{The deepest spectra}\label{ssec:deepest}

Five targets were observed in three different masks (none in four or
more masks): GMASS\,1084, 1314, 1380, 1788, and 2454.  These objects
were observed for 77, 62, 53, 55, and 77\,h, respectively, resulting
in secure redshifts of $z = 1.552$$\pm$0.004, 2.007$\pm$0.002,
1.612$\pm$0.003, 3.413$\pm$0.003, and 1.602$\pm$0.002, respectively.
Apart from GMASS\,1084 described in Sec.~\ref{ssec:xraysources}, only
GMASS\,1380 had been observed before by \citet{van06}, who derived a
tentative redshift of $z = 1.611$ that we confirm ($z =
1.612$$\pm$0.003).  In addition, there are three targets observed in
both masks r5 and r6, resulting in a total exposure time of 62\,h:
GMASS\,1030, 1901, and 2239.  None of these targets previously had
spectroscopic data.  Our observations resulted in a tentative redshift
of $z = 2.447$$\pm$0.003 for GMASS\,1030, and secure redshifts of $z =
0.1032$$\pm$0.0002, 1.415$\pm$0.002, respectively, for the others.  We
note that GMASS\,1901 is a bright object with a point--like core that
was used to position both masks and has data of the highest S/N among
our spectra and quite likely the deepest spectrum ever taken for a $z
\sim 0.1$ galaxy.

We are unaware of any other galaxy spectra, published in the
literature, with exposure times of 60 hours or more and the data
presented above are therefore likely the deepest galaxy spectra ever
taken, with GMASS\,1084 and GMASS\,2454 being the record holders.

\subsubsection{The faintest object with a secure redshift}

The faintest object in the $K$ band with a secure redshift and
m(4.5\,$\mu$m)$<$23.0 is GMASS\,2032 at $z = 1.962$$\pm$0.005 and $K =
24.1$, while the faintest object in the $I$ band is GMASS\,1667 at $z
= 1.613$$\pm$0.002.  The least luminous objects with secure redshifts
are GMASS\,365 and GMASS\,408 for the $K$ and $I$ bands, at $z =
1.609$$\pm$0.002, and 1.508$\pm$0.003 and $M_K = -21.0$ and $M_I =
-20.7$, respectively.

\subsubsection{The highest redshift objects}

The most distant object for which a secure redshift could be derived
from our observations is GMASS\,1788 at $z=3.413$$\pm$0.003, described
in Sec.~\ref{ssec:deepest}, followed by GMASS\,1160 at $z =
2.865$$\pm$0.002, which forms the top of a more continuous redshift
distribution down to $z = 0$ (see also
Fig.~\ref{fig:zspec_histogram}).  We note that none of the forty
targets with secure redshifts $z > 2.00$ derived from GMASS
observations had previously published secure (or even tentative)
redshifts.

The highest redshift object with a tentative redshift is GMASS\,2467
at $z = 4.379$$\pm$0.006, observed for 32\,h in a red mask.  This
redshift is based on the presence of two emission lines identified
with Ly$\alpha$ and \ion{C}{IV}, the latter being blue-shifted by 4500
km\,s$^{-1}$ w.r.t. to Ly$\alpha$.  In the spectrum, there is a hint
of continuum emission blueward of the tentative Ly$\alpha$ line, which
would be uncommon for this redshift.  In addition, in the $U$-band
image, there seems to be emission from this galaxy, although a nearby
much brighter object prevents a firm detection.  If the bluest
emission line were to be identified with \ion{C}{II}, the redshift would
be $z=2.384$, which is rather consistent with its photometric redshift
of $z=2.3$.  The second, redder, line should then be considered
spurious, however, despite its S/N similar to that of the bluer
line. We therefore prefer to list the two-line identification as
tentative, but note $z=2.384$ as an alternative solution.

\section{Public release}\label{sec:public_release}

As originally stated in the proposal of this Large Programme, we make
available for the general public the fully calibrated reduced spectra,
both two- and one-dimensional, the corresponding two-dimensional fully
calibrated sky background, and the GMASS catalogue.  The GMASS
catalogue contains both the photometric information on which the
photometric redshifts are based as well as spectroscopic information,
such as the redshift and its quality.  The data can be accessed on a
web page dedicated to GMASS\footnote{currently at
http://www.mpe.mpg.de/$\sim$kurk/gmass, soon also at
http://www.astronomia.unibo.it/Astronomia/default.htm
}.  We have also updated the
compilation of GOODS/CDFS spectroscopy master catalogue (now v3.0),
available from the ESO website\footnote{
  http://www.eso.org/sci/activities/projects/goods/MasterSpectroscopy.html}.

\section{Summary}

We have undertaken a spectroscopic survey of galaxies in CDFS,
targeted specifically at galaxies selected in terms of mass at
$z>1.4$. This field is one of most intensively imaged, from radio to
X-ray wavelengths, and also the focus of extraordinary spectroscopic
efforts. Nevertheless, the number of spectroscopic redshifts known in
the range $1.5<z<2.5$ is relatively low, especially for galaxies
that do not exhibit strong features related to on-going star
formation. The spectra of these galaxies do not reveal sufficient
detail in the typical exposures of a few hours to determine their
redshift or other fundamental properties. We therefore carried out a
spectroscopic survey, using one of the most sensitive optical
multi-object spectrographs available at 8\,m-class telescopes, FORS2
at the VLT, employing exceptionally long exposure times of 12 to 32
hours.

The galaxies targeted were all detected at m(4.5\,$\mu$m)$<$23.0 to
ensure that these were relatively massive (log($M/M_\odot)\ga10.5$ for
$0<z<3$). The first release of the \emph{Spitzer/IRAC} public imaging
of the CDFS had just become available at the time we performed the
target selection.  The 4.5\,$\mu$m photometry allows a more accurate
estimate of the stellar mass than optical or NIR photometry and this
channel also provides the best compromise amongst the four IRAC
channels, in terms of sensitivity, PSF, image quality, and blending
problems.  Additional $U',U,B,V,R,I,J,H$, and $K_s$ band photometry
allowed us to determine accurate photometric redshifts.  In addition
to the 4.5\,$\mu$m criterium, we selected the targets for spectroscopy
based on (our own) photometric redshift $z_{\rm phot}>1.4$ and
magnitude limits $B<26.5$ or $I<26.0$ for galaxies observed with the
blue-sensitive 300V or red-sensitive 300I grisms, respectively. We
excluded objects with known spectroscopic redshifts and those targeted
by other spectroscopic surveys, but not yet observed at that time.
These criteria left a sample of 221 targets, of which we could include
174 in three {\it blue} and {\it three} red spectroscopic masks. In
addition, more than 46 objects were included to fill available spaces
in the masks.  We obtained exposure times from 11 to 32 hours per mask
over the years from 2004 to 2006.

We reduced the spectra, ensuring that interpolation for wavelength
calibration and rectification was performed only once. Background
subtraction for the blue masks was performed in the way usually
applied to optical spectra, while for the red masks we used a method
similar to one applied in the NIR, taking advantage of having observed
the spectra at four dither positions along the slits.  Galaxy
redshifts were determined by identifying absorption and emission
features, and also by cross-correlating with composite galaxy spectra
obtained from the sample of galaxies with known redshifts in the
respective masks.  Among the 244 objects for which we extracted
one-dimensional spectra, we were able to determine 210 redshifts, of
which 145 are at $z > 1.4$, and 192 are securely identified.  Among
the 174 high-redshift galaxies targeted, we obtained redshifts for
150, of which 133 are at $z > 1.4$, and 135 are securely identified.
Within the field covered by GMASS, 80\% of the known redshifts $z>1.5$
originate from our work, while the remainder is provided by the other
numerous spectroscopic surveys within the CDFS.  We extened the
spectroscopic catalogue of the CDFS by 210 entries, 135 of which are
new, 44 are more secure than determined in previous spectroscopy, and
31 are equal to existing entries.  The redshift distribution has
several noticeable peaks, the highest of which represents an
overdensity of galaxies at $z=1.6$.

We have used the newly determined spectroscopic redshifts to assess
the \emph{BzK} selection criteria for selecting $z>1.4$ galaxies and
found that these are efficient (82\%) and suffer low contamination
(11\%).  We used the GMASS spectra and extensive photometry in the
CDFS to perform several studies, including that of quiescent superdense
galaxies at $z>1.4$ \citep{cim08,cap09}, the stellar metallicity and
outflows of star-forming $z\sim2$ galaxies \citep[][Talia et al.,
submitted]{hal08}, the evolution of the rest-frame colour distribution
and dust properties of high redshift galaxies \citep{cas08,nol09}, and
the properties of galaxies in, and inflow of cold gas into, the galaxy
overdensity at $z=1.6$ \citep{kur09,gia11}.  The public release of the
GMASS spectra will facilitate further studies of the distant galaxies
targeted by our survey.

\begin{acknowledgements}
  JK acknowledges the \emph{Deut\-sche For\-schungs\-ge\-mein\-schaft}
  (DFG) for support via grant SFB-439, and via German-Israeli Project
  Cooperation grant STE1869/1-1.GE625/15-1.  ED acknowledges funding
  support from ERC-StG grant UPGAL 240039 and ANR-08-JCJC- 0008.  This
  work is based [in part] on observations made with the Spitzer Space
  Telescope, which is operated by the Jet Propulsion Laboratory,
  California Institute of Technology under a contract with NASA.
\end{acknowledgements}

\bibliographystyle{aa} 
\bibliography{kurk_gmass} 

\begin{thebibliography}{111}
\expandafter\ifx\csname natexlab\endcsname\relax\def\natexlab#1{#1}\fi

\bibitem[{{Afonso} {et~al.}(2006){Afonso}, {Mobasher}, {Koekemoer}, {Norris},
  \& {Cram}}]{afo06}
{Afonso}, J., {Mobasher}, B., {Koekemoer}, A., {Norris}, R.~P., \& {Cram}, L.
  2006, \aj, 131, 1216

\bibitem[{{Appenzeller} {et~al.}(1998){Appenzeller}, {Fricke}, {Furtig},
  {Gassler}, {Hafner}, {Harkl}, {Hess}, {Hummel}, {Jurgens}, {Kudritzki},
  {Mantel}, {Meisl}, {Muschielok}, {Nicklas}, {Rupprecht}, {Seifert}, {Stahl},
  {Szeifert}, \& {Tarantik}}]{app98}
{Appenzeller}, I., {Fricke}, K., {Furtig}, W., {et~al.} 1998, The Messenger,
  94, 1

\bibitem[{{Arnouts} {et~al.}(2001){Arnouts}, {Vandame}, {Benoist},
  {Groenewegen}, {da Costa}, {Schirmer}, {Mignani}, {Slijkhuis},
  {Hatziminaoglou}, {Hook}, {Madejsky}, {Rit{\'e}}, \& {Wicenec}}]{arn01}
{Arnouts}, S., {Vandame}, B., {Benoist}, C., {et~al.} 2001, \aap, 379, 740

\bibitem[{{Baade} {et~al.}(1999){Baade}, {Meisenheimer}, {Iwert}, {Alonso},
  {Augusteijn}, {Beletic}, {Bellemann}, {Benesch}, {Boehm}, {Boehnhardt},
  {Brewer}, {Deiries}, {Delabre}, {Donaldson}, {Dupuy}, {Franke}, {Gerdes},
  {Gilliotte}, {Grimm}, {Haddad}, {Hess}, {Ihle}, {Klein}, {Lenzen}, {Lizon},
  {Mancini}, {Muench}, {Pizarro}, {Prado}, {Rahmer}, {Reyes}, {Richardson},
  {Robledo}, {Sanchez}, {Silber}, {Sinclaire}, {Wackermann}, \&
  {Zaggia}}]{baa99}
{Baade}, D., {Meisenheimer}, K., {Iwert}, O., {et~al.} 1999, The Messenger, 95,
  15

\bibitem[{{Balestra} {et~al.}(2010){Balestra}, {Mainieri}, {Popesso},
  {Dickinson}, {Nonino}, {Rosati}, {Teimoorinia}, {Vanzella}, {Cristiani},
  {Cesarsky}, {Fosbury}, {Kuntschner}, \& {Rettura}}]{bal10}
{Balestra}, I., {Mainieri}, V., {Popesso}, P., {et~al.} 2010, \aap, 512, A12+

\bibitem[{{Bertin} \& {Arnouts}(1996)}]{ber96}
{Bertin}, E. \& {Arnouts}, S. 1996, \aaps, 117, 393

\bibitem[{{Blain} \& {Longair}(1993)}]{bla93}
{Blain}, A.~W. \& {Longair}, M.~S. 1993, \mnras, 264, 509

\bibitem[{{Bolzonella} {et~al.}(2010){Bolzonella}, {Kova{\v c}}, {Pozzetti},
  {Zucca}, {Cucciati}, {Lilly}, {Peng}, {Iovino}, {Zamorani}, {Vergani},
  {Tasca}, {Lamareille}, {Oesch}, {Caputi}, {Kampczyk}, {Bardelli}, {Maier},
  {Abbas}, {Knobel}, {Scodeggio}, {Carollo}, {Contini}, {Kneib}, {Le
  F{\`e}vre}, {Mainieri}, {Renzini}, {Bongiorno}, {Coppa}, {de la Torre}, {de
  Ravel}, {Franzetti}, {Garilli}, {Le Borgne}, {Le Brun}, {Mignoli},
  {Pell{\'o}}, {Perez-Montero}, {Ricciardelli}, {Silverman}, {Tanaka},
  {Tresse}, {Bottini}, {Cappi}, {Cassata}, {Cimatti}, {Guzzo}, {Koekemoer},
  {Leauthaud}, {Maccagni}, {Marinoni}, {McCracken}, {Memeo}, {Meneux},
  {Porciani}, {Scaramella}, {Aussel}, {Capak}, {Halliday}, {Ilbert},
  {Kartaltepe}, {Salvato}, {Sanders}, {Scarlata}, {Scoville}, {Taniguchi}, \&
  {Thompson}}]{bol10}
{Bolzonella}, M., {Kova{\v c}}, K., {Pozzetti}, L., {et~al.} 2010, \aap, 524,
  A76+

\bibitem[{{Bolzonella} {et~al.}(2000){Bolzonella}, {Miralles}, \&
  {Pell{\'o}}}]{bol00}
{Bolzonella}, M., {Miralles}, J.-M., \& {Pell{\'o}}, R. 2000, \aap, 363, 476

\bibitem[{{Bruzual} \& {Charlot}(2003)}]{bru03}
{Bruzual}, G. \& {Charlot}, S. 2003, \mnras, 344, 1000

\bibitem[{{Bruzual}(2007{\natexlab{a}})}]{bru07a}
{Bruzual}, A., G. 2007{\natexlab{a}}, ArXiv Astrophysics e-prints,
  astro-ph/0703052

\bibitem[{{Bruzual}(2007{\natexlab{b}})}]{bru07b}
{Bruzual}, A., G. 2007{\natexlab{b}}, ArXiv Astrophysics e-prints,
  astro-ph/0702091

\bibitem[{{Bruzual A.} \& {Charlot}(1993)}]{bru93}
{Bruzual A.}, G. \& {Charlot}, S. 1993, \apj, 405, 538

\bibitem[{{Bundy} {et~al.}(2006){Bundy}, {Ellis}, {Conselice}, {Taylor},
  {Cooper}, {Willmer}, {Weiner}, {Coil}, {Noeske}, \& {Eisenhardt}}]{bun06}
{Bundy}, K., {Ellis}, R.~S., {Conselice}, C.~J., {et~al.} 2006, \apj, 651, 120

\bibitem[{{Calzetti} {et~al.}(2000){Calzetti}, {Armus}, {Bohlin}, {Kinney},
  {Koornneef}, \& {Storchi-Bergmann}}]{cal00}
{Calzetti}, D., {Armus}, L., {Bohlin}, R.~C., {et~al.} 2000, \apj, 533, 682

\bibitem[{{Cappellari} {et~al.}(2009){Cappellari}, {di Serego Alighieri},
  {Cimatti}, {Daddi}, {Renzini}, {Kurk}, {Cassata}, {Dickinson},
  {Franceschini}, {Mignoli}, {Pozzetti}, {Rodighiero}, {Rosati}, \&
  {Zamorani}}]{cap09}
{Cappellari}, M., {di Serego Alighieri}, S., {Cimatti}, A., {et~al.} 2009,
  \apjl, 704, L34

\bibitem[{{Caputi} {et~al.}(2005){Caputi}, {Dunlop}, {McLure}, \&
  {Roche}}]{cap05}
{Caputi}, K.~I., {Dunlop}, J.~S., {McLure}, R.~J., \& {Roche}, N.~D. 2005,
  \mnras, 361, 607

\bibitem[{{Caputi} {et~al.}(2006){Caputi}, {McLure}, {Dunlop}, {Cirasuolo}, \&
  {Schael}}]{cap06}
{Caputi}, K.~I., {McLure}, R.~J., {Dunlop}, J.~S., {Cirasuolo}, M., \&
  {Schael}, A.~M. 2006, \mnras, 366, 609

\bibitem[{{Cassata} {et~al.}(2008){Cassata}, {Cimatti}, {Kurk}, {Rodighiero},
  {Pozzetti}, {Bolzonella}, {Daddi}, {Mignoli}, {Berta}, {Dickinson},
  {Franceschini}, {Halliday}, {Renzini}, {Rosati}, \& {Zamorani}}]{cas08}
{Cassata}, P., {Cimatti}, A., {Kurk}, J., {et~al.} 2008, \aap, 483, L39

\bibitem[{{Chabrier}(2003)}]{cha03}
{Chabrier}, G. 2003, \pasp, 115, 763

\bibitem[{{Chen} \& {Marzke}(2004)}]{che04}
{Chen}, H.-W. \& {Marzke}, R.~O. 2004, \apj, 615, 603

\bibitem[{{Cimatti} {et~al.}(2008){Cimatti}, {Cassata}, {Pozzetti}, {Kurk},
  {Mignoli}, {Renzini}, {Daddi}, {Bolzonella}, {Brusa}, {Rodighiero},
  {Dickinson}, {Franceschini}, {Zamorani}, {Berta}, {Rosati}, \&
  {Halliday}}]{cim08}
{Cimatti}, A., {Cassata}, P., {Pozzetti}, L., {et~al.} 2008, \aap, 482, 21

\bibitem[{{Cimatti} {et~al.}(2006){Cimatti}, {Daddi}, \& {Renzini}}]{cim06}
{Cimatti}, A., {Daddi}, E., \& {Renzini}, A. 2006, \aap, 453, L29

\bibitem[{{Cimatti} {et~al.}(2004){Cimatti}, {Daddi}, {Renzini}, {Cassata},
  {Vanzella}, {Pozzetti}, {Cristiani}, {Fontana}, {Rodighiero}, {Mignoli}, \&
  {Zamorani}}]{cim04}
{Cimatti}, A., {Daddi}, E., {Renzini}, A., {et~al.} 2004, \nat, 430, 184

\bibitem[{{Cimatti} {et~al.}(2002{\natexlab{a}}){Cimatti}, {Mignoli}, {Daddi},
  {Pozzetti}, {Fontana}, {Saracco}, {Poli}, {Renzini}, {Zamorani},
  {Broadhurst}, {Cristiani}, {D'Odorico}, {Giallongo}, {Gilmozzi}, \&
  {Menci}}]{cim02c}
{Cimatti}, A., {Mignoli}, M., {Daddi}, E., {et~al.} 2002{\natexlab{a}}, \aap,
  392, 395

\bibitem[{{Cimatti} {et~al.}(2002{\natexlab{b}}){Cimatti}, {Pozzetti},
  {Mignoli}, {Daddi}, {Menci}, {Poli}, {Fontana}, {Renzini}, {Zamorani},
  {Broadhurst}, {Cristiani}, {D'Odorico}, {Giallongo}, \& {Gilmozzi}}]{cim02b}
{Cimatti}, A., {Pozzetti}, L., {Mignoli}, M., {et~al.} 2002{\natexlab{b}},
  \aap, 391, L1

\bibitem[{{Coleman} {et~al.}(1980){Coleman}, {Wu}, \& {Weedman}}]{col80}
{Coleman}, G.~D., {Wu}, C.-C., \& {Weedman}, D.~W. 1980, \apjs, 43, 393

\bibitem[{{Comastri} {et~al.}(2011){Comastri}, {Ranalli}, {Iwasawa}, {Vignali},
  {Gilli}, {Georgantopoulos}, {Barcons}, {Brandt}, {Brunner}, {Brusa},
  {Cappelluti}, {Carrera}, {Civano}, {Fiore}, {Hasinger}, {Mainieri},
  {Merloni}, {Nicastro}, {Paolillo}, {Puccetti}, {Rosati}, {Silverman},
  {Tozzi}, {Zamorani}, {Balestra}, {Bauer}, {Luo}, \& {Xue}}]{com11}
{Comastri}, A., {Ranalli}, P., {Iwasawa}, K., {et~al.} 2011, \aap, 526, L9+

\bibitem[{{Cowie} {et~al.}(1996){Cowie}, {Songaila}, {Hu}, \& {Cohen}}]{cow96}
{Cowie}, L.~L., {Songaila}, A., {Hu}, E.~M., \& {Cohen}, J.~G. 1996, \aj, 112,
  839

\bibitem[{{Cresci} {et~al.}(2009){Cresci}, {Hicks}, {Genzel}, {Schreiber},
  {Davies}, {Bouch{\'e}}, {Buschkamp}, {Genel}, {Shapiro}, {Tacconi},
  {Sommer-Larsen}, {Burkert}, {Eisenhauer}, {Gerhard}, {Lutz}, {Naab},
  {Sternberg}, {Cimatti}, {Daddi}, {Erb}, {Kurk}, {Lilly}, {Renzini},
  {Shapley}, {Steidel}, \& {Caputi}}]{cre09}
{Cresci}, G., {Hicks}, E.~K.~S., {Genzel}, R., {et~al.} 2009, \apj, 697, 115

\bibitem[{{Croom} {et~al.}(2001){Croom}, {Warren}, \& {Glazebrook}}]{cro01}
{Croom}, S.~M., {Warren}, S.~J., \& {Glazebrook}, K. 2001, \mnras, 328, 150

\bibitem[{{Daddi} {et~al.}(2007{\natexlab{a}}){Daddi}, {Alexander},
  {Dickinson}, {Gilli}, {Renzini}, {Elbaz}, {Cimatti}, {Chary}, {Frayer},
  {Bauer}, {Brandt}, {Giavalisco}, {Grogin}, {Huynh}, {Kurk}, {Mignoli},
  {Morrison}, {Pope}, \& {Ravindranath}}]{dad07b}
{Daddi}, E., {Alexander}, D.~M., {Dickinson}, M., {et~al.} 2007{\natexlab{a}},
  \apj, 670, 173

\bibitem[{{Daddi} {et~al.}(2002){Daddi}, {Cimatti}, {Broadhurst}, {Renzini},
  {Zamorani}, {Mignoli}, {Saracco}, {Fontana}, {Pozzetti}, {Poli}, {Cristiani},
  {D'Odorico}, {Giallongo}, {Gilmozzi}, \& {Menci}}]{dad02}
{Daddi}, E., {Cimatti}, A., {Broadhurst}, T., {et~al.} 2002, \aap, 384, L1

\bibitem[{{Daddi} {et~al.}(2004){Daddi}, {Cimatti}, {Renzini}, {Fontana},
  {Mignoli}, {Pozzetti}, {Tozzi}, \& {Zamorani}}]{dad04b}
{Daddi}, E., {Cimatti}, A., {Renzini}, A., {et~al.} 2004, \apj, 617, 746

\bibitem[{{Daddi} {et~al.}(2007{\natexlab{b}}){Daddi}, {Dickinson}, {Morrison},
  {Chary}, {Cimatti}, {Elbaz}, {Frayer}, {Renzini}, {Pope}, {Alexander},
  {Bauer}, {Giavalisco}, {Huynh}, {Kurk}, \& {Mignoli}}]{dad07a}
{Daddi}, E., {Dickinson}, M., {Morrison}, G., {et~al.} 2007{\natexlab{b}},
  \apj, 670, 156

\bibitem[{{Daddi} {et~al.}(2005){Daddi}, {Renzini}, {Pirzkal}, {Cimatti},
  {Malhotra}, {Stiavelli}, {Xu}, {Pasquali}, {Rhoads}, {Brusa}, {di Serego
  Alighieri}, {Ferguson}, {Koekemoer}, {Moustakas}, {Panagia}, \&
  {Windhorst}}]{dad05a}
{Daddi}, E., {Renzini}, A., {Pirzkal}, N., {et~al.} 2005, \apj, 626, 680

\bibitem[{{Dahlen} {et~al.}(2010){Dahlen}, {Mobasher}, {Dickinson}, {Ferguson},
  {Giavalisco}, {Grogin}, {Guo}, {Koekemoer}, {Lee}, {Lee}, {Nonino}, {Riess},
  \& {Salimbeni}}]{dah10}
{Dahlen}, T., {Mobasher}, B., {Dickinson}, M., {et~al.} 2010, \apj, 724, 425

\bibitem[{{di Serego Alighieri} {et~al.}(2005){di Serego Alighieri}, {Vernet},
  {Cimatti}, {Lanzoni}, {Cassata}, {Ciotti}, {Daddi}, {Mignoli}, {Pignatelli},
  {Pozzetti}, {Renzini}, {Rettura}, \& {Zamorani}}]{ser05}
{di Serego Alighieri}, S., {Vernet}, J., {Cimatti}, A., {et~al.} 2005, \aap,
  442, 125

\bibitem[{{Doherty} {et~al.}(2005){Doherty}, {Bunker}, {Ellis}, \&
  {McCarthy}}]{doh05}
{Doherty}, M., {Bunker}, A.~J., {Ellis}, R.~S., \& {McCarthy}, P.~J. 2005,
  \mnras, 361, 525

\bibitem[{{Drory} {et~al.}(2005){Drory}, {Salvato}, {Gabasch}, {Bender},
  {Hopp}, {Feulner}, \& {Pannella}}]{dro05}
{Drory}, N., {Salvato}, M., {Gabasch}, A., {et~al.} 2005, \apjl, 619, L131

\bibitem[{{Eisenhardt} {et~al.}(2004){Eisenhardt}, {Stern}, {Brodwin}, {Fazio},
  {Rieke}, {Rieke}, {Werner}, {Wright}, {Allen}, {Arendt}, {Ashby}, {Barmby},
  {Forrest}, {Hora}, {Huang}, {Huchra}, {Pahre}, {Pipher}, {Reach}, {Smith},
  {Stauffer}, {Wang}, {Willner}, {Brown}, {Dey}, {Jannuzi}, \& {Tiede}}]{eis04}
{Eisenhardt}, P.~R., {Stern}, D., {Brodwin}, M., {et~al.} 2004, \apjs, 154, 48

\bibitem[{{Elbaz} {et~al.}(2011){Elbaz}, {Dickinson}, {Hwang}, {Diaz-Santos},
  {Magdis}, {Magnelli}, {Le Borgne}, {Galliano}, {Pannella}, {Chanial},
  {Armus}, {Charmandaris}, {Daddi}, {Aussel}, {Popesso}, {Kartaltepe},
  {Altieri}, {Valtchanov}, {Coia}, {Dannerbauer}, {Dasyra}, {Leiton},
  {Mazzarella}, {Buat}, {Burgarella}, {Chary}, {Gilli}, {Ivison}, {Juneau},
  {LeFloc'h}, {Lutz}, {Morrison}, {Mullaney}, {Murphy}, {Pope}, {Scott},
  {Alexander}, {Brodwin}, {Calzetti}, {Cesarsky}, {Charlot}, {Dole},
  {Eisenhardt}, {Ferguson}, {Foerster-Schreiber}, {Frayer}, {Giavalisco},
  {Huynh}, {Koekemoer}, {Papovich}, {Reddy}, {Surace}, {Teplitz}, {Yun}, \&
  {Wilson}}]{elb11}
{Elbaz}, D., {Dickinson}, M., {Hwang}, H.~S., {et~al.} 2011, ArXiv e-prints,
  1105.2537

\bibitem[{{Fazio} {et~al.}(2004){Fazio}, {Hora}, {Allen}, {Ashby}, {Barmby},
  {Deutsch}, {Huang}, {Kleiner}, {Marengo}, {Megeath}, {Melnick}, {Pahre},
  {Patten}, {Polizotti}, {Smith}, {Taylor}, {Wang}, {Willner}, {Hoffmann},
  {Pipher}, {Forrest}, {McMurty}, {McCreight}, {McKelvey}, {McMurray}, {Koch},
  {Moseley}, {Arendt}, {Mentzell}, {Marx}, {Losch}, {Mayman}, {Eichhorn},
  {Krebs}, {Jhabvala}, {Gezari}, {Fixsen}, {Flores}, {Shakoorzadeh}, {Jungo},
  {Hakun}, {Workman}, {Karpati}, {Kichak}, {Whitley}, {Mann}, {Tollestrup},
  {Eisenhardt}, {Stern}, {Gorjian}, {Bhattacharya}, {Carey}, {Nelson},
  {Glaccum}, {Lacy}, {Lowrance}, {Laine}, {Reach}, {Stauffer}, {Surace},
  {Wilson}, {Wright}, {Hoffman}, {Domingo}, \& {Cohen}}]{faz04}
{Fazio}, G.~G., {Hora}, J.~L., {Allen}, L.~E., {et~al.} 2004, \apjs, 154, 10

\bibitem[{{Feulner} {et~al.}(2003){Feulner}, {Bender}, {Drory}, {Hopp},
  {Snigula}, \& {Hill}}]{feu03}
{Feulner}, G., {Bender}, R., {Drory}, N., {et~al.} 2003, \mnras, 342, 605

\bibitem[{{Feulner} {et~al.}(2005){Feulner}, {Gabasch}, {Salvato}, {Drory},
  {Hopp}, \& {Bender}}]{feu05}
{Feulner}, G., {Gabasch}, A., {Salvato}, M., {et~al.} 2005, \apjl, 633, L9

\bibitem[{{Fontana} {et~al.}(2004){Fontana}, {Pozzetti}, {Donnarumma},
  {Renzini}, {Cimatti}, {Zamorani}, {Menci}, {Daddi}, {Giallongo}, {Mignoli},
  {Perna}, {Salimbeni}, {Saracco}, {Broadhurst}, {Cristiani}, {D'Odorico}, \&
  {Gilmozzi}}]{fon04}
{Fontana}, A., {Pozzetti}, L., {Donnarumma}, I., {et~al.} 2004, \aap, 424, 23

\bibitem[{{F{\"o}rster Schreiber} {et~al.}(2009){F{\"o}rster Schreiber},
  {Genzel}, {Bouch{\'e}}, {Cresci}, {Davies}, {Buschkamp}, {Shapiro},
  {Tacconi}, {Hicks}, {Genel}, {Shapley}, {Erb}, {Steidel}, {Lutz},
  {Eisenhauer}, {Gillessen}, {Sternberg}, {Renzini}, {Cimatti}, {Daddi},
  {Kurk}, {Lilly}, {Kong}, {Lehnert}, {Nesvadba}, {Verma}, {McCracken},
  {Arimoto}, {Mignoli}, \& {Onodera}}]{for09}
{F{\"o}rster Schreiber}, N.~M., {Genzel}, R., {Bouch{\'e}}, N., {et~al.} 2009,
  \apj, 706, 1364

\bibitem[{{Gabasch} {et~al.}(2006){Gabasch}, {Hopp}, {Feulner}, {Bender},
  {Seitz}, {Saglia}, {Snigula}, {Drory}, {Appenzeller}, {Heidt}, {Mehlert},
  {Noll}, {B{\"o}hm}, {J{\"a}ger}, \& {Ziegler}}]{gab06}
{Gabasch}, A., {Hopp}, U., {Feulner}, G., {et~al.} 2006, \aap, 448, 101

\bibitem[{{Gardner} {et~al.}(1993){Gardner}, {Cowie}, \& {Wainscoat}}]{gar93}
{Gardner}, J.~P., {Cowie}, L.~L., \& {Wainscoat}, R.~J. 1993, \apjl, 415, L9

\bibitem[{{Genzel} {et~al.}(2008){Genzel}, {Burkert}, {Bouch{\'e}}, {Cresci},
  {F{\"o}rster Schreiber}, {Shapley}, {Shapiro}, {Tacconi}, {Buschkamp},
  {Cimatti}, {Daddi}, {Davies}, {Eisenhauer}, {Erb}, {Genel}, {Gerhard},
  {Hicks}, {Lutz}, {Naab}, {Ott}, {Rabien}, {Renzini}, {Steidel}, {Sternberg},
  \& {Lilly}}]{gen08}
{Genzel}, R., {Burkert}, A., {Bouch{\'e}}, N., {et~al.} 2008, \apj, 687, 59

\bibitem[{{Giacconi} {et~al.}(2001){Giacconi}, {Rosati}, {Tozzi}, {Nonino},
  {Hasinger}, {Norman}, {Bergeron}, {Borgani}, {Gilli}, {Gilmozzi}, \&
  {Zheng}}]{gia01}
{Giacconi}, R., {Rosati}, P., {Tozzi}, P., {et~al.} 2001, \apj, 551, 624

\bibitem[{{Giacconi} {et~al.}(2002){Giacconi}, {Zirm}, {Wang}, {Rosati},
  {Nonino}, {Tozzi}, {Gilli}, {Mainieri}, {Hasinger}, {Kewley}, {Bergeron},
  {Borgani}, {Gilmozzi}, {Grogin}, {Koekemoer}, {Schreier}, {Zheng}, \&
  {Norman}}]{gia02}
{Giacconi}, R., {Zirm}, A., {Wang}, J., {et~al.} 2002, \apjs, 139, 369

\bibitem[{{Giavalisco} {et~al.}(2004){Giavalisco}, {Ferguson}, {Koekemoer},
  {Dickinson}, {Alexander}, {Bauer}, {Bergeron}, {Biagetti}, {Brandt},
  {Casertano}, {Cesarsky}, {Chatzichristou}, {Conselice}, {Cristiani}, {Da
  Costa}, {Dahlen}, {de Mello}, {Eisenhardt}, {Erben}, {Fall}, {Fassnacht},
  {Fosbury}, {Fruchter}, {Gardner}, {Grogin}, {Hook}, {Hornschemeier}, {Idzi},
  {Jogee}, {Kretchmer}, {Laidler}, {Lee}, {Livio}, {Lucas}, {Madau},
  {Mobasher}, {Moustakas}, {Nonino}, {Padovani}, {Papovich}, {Park},
  {Ravindranath}, {Renzini}, {Richardson}, {Riess}, {Rosati}, {Schirmer},
  {Schreier}, {Somerville}, {Spinrad}, {Stern}, {Stiavelli}, {Strolger},
  {Urry}, {Vandame}, {Williams}, \& {Wolf}}]{gia04}
{Giavalisco}, M., {Ferguson}, H.~C., {Koekemoer}, A.~M., {et~al.} 2004, \apjl,
  600, L93

\bibitem[{{Giavalisco} {et~al.}(2011){Giavalisco}, {Vanzella}, {Salimbeni},
  {Tripp}, {Dickinson}, {Cassata}, {Renzini}, {Guo}, {Ferguson}, {Nonino},
  {Cimatti}, {Kurk}, {Mignoli}, {Tang}, \& {.}}]{gia11}
{Giavalisco}, M., {Vanzella}, E., {Salimbeni}, S., {et~al.} 2011, ArXiv
  e-prints, 1106.1205

\bibitem[{{Glazebrook} {et~al.}(2004){Glazebrook}, {Tober}, {Thomson},
  {Bland-Hawthorn}, \& {Abraham}}]{gla04}
{Glazebrook}, K., {Tober}, J., {Thomson}, S., {Bland-Hawthorn}, J., \&
  {Abraham}, R. 2004, \aj, 128, 2652

\bibitem[{{Grogin} {et~al.}(2011){Grogin}, {Kocevski}, {Faber}, {Ferguson},
  {Koekemoer}, {Riess}, {Acquaviva}, {Alexander}, {Almaini}, {Ashby}, {Barden},
  {Bell}, {Bournaud}, {Brown}, {Caputi}, {Casertano}, {Cassata}, {Challis},
  {Chary}, {Cheung}, {Cirasuolo}, {Conselice}, {Roshan Cooray}, {Croton},
  {Daddi}, {Dahlen}, {Dav{\'e}}, {de Mello}, {Dekel}, {Dickinson}, {Dolch},
  {Donley}, {Dunlop}, {Dutton}, {Elbaz}, {Fazio}, {Filippenko}, {Finkelstein},
  {Fontana}, {Gardner}, {Garnavich}, {Gawiser}, {Giavalisco}, {Grazian}, {Guo},
  {Hathi}, {H{\"a}ussler}, {Hopkins}, {Huang}, {Huang}, {Jha}, {Kartaltepe},
  {Kirshner}, {Koo}, {Lai}, {Lee}, {Li}, {Lotz}, {Lucas}, {Madau}, {McCarthy},
  {McGrath}, {McIntosh}, {McLure}, {Mobasher}, {Moustakas}, {Mozena}, {Nandra},
  {Newman}, {Niemi}, {Noeske}, {Papovich}, {Pentericci}, {Pope}, {Primack},
  {Rajan}, {Ravindranath}, {Reddy}, {Renzini}, {Rix}, {Robaina}, {Rodney},
  {Rosario}, {Rosati}, {Salimbeni}, {Scarlata}, {Siana}, {Simard}, {Smidt},
  {Somerville}, {Spinrad}, {Straughn}, {Strolger}, {Telford}, {Teplitz},
  {Trump}, {van der Wel}, {Villforth}, {Wechsler}, {Weiner}, {Wiklind}, {Wild},
  {Wilson}, {Wuyts}, {Yan}, \& {Yun}}]{gro11}
{Grogin}, N.~A., {Kocevski}, D.~D., {Faber}, S.~M., {et~al.} 2011, ArXiv
  e-prints, 1105.3753

\bibitem[{{Halliday} {et~al.}(2008){Halliday}, {Daddi}, {Cimatti}, {Kurk},
  {Renzini}, {Mignoli}, {Bolzonella}, {Pozzetti}, {Dickinson}, {Zamorani},
  {Berta}, {Franceschini}, {Cassata}, {Rodighiero}, \& {Rosati}}]{hal08}
{Halliday}, C., {Daddi}, E., {Cimatti}, A., {et~al.} 2008, \aap, 479, 417

\bibitem[{{Hopkins} \& {Beacom}(2006)}]{hop06}
{Hopkins}, A.~M. \& {Beacom}, J.~F. 2006, \apj, 651, 142

\bibitem[{{Juneau} {et~al.}(2005){Juneau}, {Glazebrook}, {Crampton},
  {McCarthy}, {Savaglio}, {Abraham}, {Carlberg}, {Chen}, {Le Borgne}, {Marzke},
  {Roth}, {J{\o}rgensen}, {Hook}, \& {Murowinski}}]{jun05}
{Juneau}, S., {Glazebrook}, K., {Crampton}, D., {et~al.} 2005, \apjl, 619, L135

\bibitem[{{Karim} {et~al.}(2011){Karim}, {Schinnerer},
  {Mart{\'{\i}}nez-Sansigre}, {Sargent}, {van der Wel}, {Rix}, {Ilbert},
  {Smol{\v c}i{\'c}}, {Carilli}, {Pannella}, {Koekemoer}, {Bell}, \&
  {Salvato}}]{kar11}
{Karim}, A., {Schinnerer}, E., {Mart{\'{\i}}nez-Sansigre}, A., {et~al.} 2011,
  \apj, 730, 61

\bibitem[{{Kellermann} {et~al.}(2008){Kellermann}, {Fomalont}, {Mainieri},
  {Padovani}, {Rosati}, {Shaver}, {Tozzi}, \& {Miller}}]{kel08}
{Kellermann}, K.~I., {Fomalont}, E.~B., {Mainieri}, V., {et~al.} 2008, \apjs,
  179, 71

\bibitem[{{Koekemoer} {et~al.}(2011){Koekemoer}, {Faber}, {Ferguson}, {Grogin},
  {Kocevski}, {Koo}, {Lai}, {Lotz}, {Lucas}, {McGrath}, {Ogaz}, {Rajan},
  {Riess}, {Rodney}, {Strolger}, {Casertano}, {Dahlen}, {Dickinson}, {Dolch},
  {Fontana}, {Giavalisco}, {Grazian}, {Guo}, {Hathi}, {Huang}, {van der Wel},
  {Yan}, {Acquaviva}, {Almaini}, {Ashby}, {Barden}, {Bell}, {Bournaud},
  {Brown}, {Caputi}, {Cassata}, {Challis}, {Chary}, {Cheung}, {Cirasuolo},
  {Conselice}, {Roshan Cooray}, {Croton}, {Daddi}, {Dav{\'e}}, {de Mello}, {de
  Ravel}, {Dekel}, {Donley}, {Dunlop}, {Dutton}, {Elbaz}, {Fazio},
  {Filippenko}, {Finkelstein}, {Frazer}, {Gardner}, {Garnavich}, {Gawiser},
  {Gruetzbauch}, {Hartley}, {H{\"a}ussler}, {Herrington}, {Hopkins}, {Huang},
  {Jha}, {Johnson}, {Kartaltepe}, {Khostovan}, {Kirshner}, {Lani}, {Lee}, {Li},
  {Madau}, {McCarthy}, {McIntosh}, {McLure}, {McPartland}, {Mobasher},
  {Moreira}, {Mortlock}, {Moustakas}, {Mozena}, {Nandra}, {Newman}, {Nielsen},
  {Niemi}, {Noeske}, {Papovich}, {Pentericci}, {Pope}, {Primack},
  {Ravindranath}, {Reddy}, {Renzini}, {Rix}, {Robaina}, {Rosario}, {Rosati},
  {Salimbeni}, {Scarlata}, {Siana}, {Simard}, {Smidt}, {Snyder}, {Somerville},
  {Spinrad}, {Straughn}, {Telford}, {Teplitz}, {Trump}, {Vargas}, {Villforth},
  {Wagner}, {Wandro}, {Wechsler}, {Weiner}, {Wiklind}, {Wild}, {Wilson},
  {Wuyts}, \& {Yun}}]{koe11}
{Koekemoer}, A.~M., {Faber}, S.~M., {Ferguson}, H.~C., {et~al.} 2011, ArXiv
  e-prints, 1105.3754

\bibitem[{{Kong} {et~al.}(2006){Kong}, {Daddi}, {Arimoto}, {Renzini},
  {Broadhurst}, {Cimatti}, {Ikuta}, {Ohta}, {da Costa}, {Olsen}, {Onodera}, \&
  {Tamura}}]{kon06}
{Kong}, X., {Daddi}, E., {Arimoto}, N., {et~al.} 2006, \apj, 638, 72

\bibitem[{{Kroupa}(2001)}]{kro01}
{Kroupa}, P. 2001, \mnras, 322, 231

\bibitem[{{Kurk} {et~al.}(2009){Kurk}, {Cimatti}, {Zamorani}, {Halliday},
  {Mignoli}, {Pozzetti}, {Daddi}, {Rosati}, {Dickinson}, {Bolzonella},
  {Cassata}, {Renzini}, {Franceschini}, {Rodighiero}, \& {Berta}}]{kur09}
{Kurk}, J., {Cimatti}, A., {Zamorani}, G., {et~al.} 2009, \aap, 504, 331

\bibitem[{{Le F{\`e}vre} {et~al.}(2004){Le F{\`e}vre}, {Vettolani}, {Paltani},
  {Tresse}, {Zamorani}, {Le Brun}, {Moreau}, {Bottini}, {Maccagni}, {Picat},
  {Scaramella}, {Scodeggio}, {Zanichelli}, {Adami}, {Arnouts}, {Bardelli},
  {Bolzonella}, {Cappi}, {Charlot}, {Contini}, {Foucaud}, {Franzetti},
  {Garilli}, {Gavignaud}, {Guzzo}, {Ilbert}, {Iovino}, {McCracken}, {Mancini},
  {Marano}, {Marinoni}, {Mathez}, {Mazure}, {Meneux}, {Merighi}, {Pell{\`o}},
  {Pollo}, {Pozzetti}, {Radovich}, {Zucca}, {Arnaboldi}, {Bondi}, {Bongiorno},
  {Busarello}, {Ciliegi}, {Gregorini}, {Mellier}, {Merluzzi}, {Ripepi}, \&
  {Rizzo}}]{lef04}
{Le F{\`e}vre}, O., {Vettolani}, G., {Paltani}, S., {et~al.} 2004, \aap, 428,
  1043

\bibitem[{{Le Fevre} {et~al.}(1998){Le Fevre}, {Vettolani}, {Maccagni},
  {Mancini}, {Picat}, {Mellier}, {Mazure}, {Saisse}, {Cuby}, {Delabre},
  {Garilli}, {Hill}, {Prieto}, {Arnold}, {Conconi}, {Cascone}, {Mattaini}, \&
  {Voet}}]{lef98}
{Le Fevre}, O., {Vettolani}, G.~P., {Maccagni}, D., {et~al.} 1998, in Proc.
  SPIE Vol. 3355, p. 8-19, Optical Astronomical Instrumentation, Sandro
  D'Odorico; Ed., ed. S.~{D'Odorico}, 8--19

\bibitem[{{LeFevre} {et~al.}(2003){LeFevre}, {Saisse}, {Mancini}, {Brau-Nogue},
  {Caputi}, {Castinel}, {D'Odorico}, {Garilli}, {Kissler-Patig}, {Lucuix},
  {Mancini}, {Pauget}, {Sciarretta}, {Scodeggio}, {Tresse}, \&
  {Vettolani}}]{lef03}
{LeFevre}, O., {Saisse}, M., {Mancini}, D., {et~al.} 2003, in Instrument Design
  and Performance for Optical/Infrared Ground-based Telescopes. Edited by Iye,
  Masanori; Moorwood, Alan F. M. Proceedings of the SPIE, Volume 4841, pp.
  1670-1681 (2003)., ed. M.~{Iye} \& A.~F.~M. {Moorwood}, 1670--1681

\bibitem[{{Lejeune} {et~al.}(1997){Lejeune}, {Cuisinier}, \& {Buser}}]{lej97}
{Lejeune}, T., {Cuisinier}, F., \& {Buser}, R. 1997, \aaps, 125, 229

\bibitem[{{Lutz} {et~al.}(2011){Lutz}, {Poglitsch}, {Altieri}, {Andreani},
  {Aussel}, {Berta}, {Bongiovanni}, {Brisbin}, {Cava}, {Cepa}, {Cimatti},
  {Daddi}, {Dominguez-Sanchez}, {Elbaz}, {Forster Schreiber}, {Genzel},
  {Grazian}, {Gruppioni}, {Harwit}, {Le Floc'h}, {Magdis}, {Magnelli},
  {Maiolino}, {Nordon}, {Perez Garcia}, {Popesso}, {Pozzi}, {Riguccini},
  {Rodighiero}, {Saintonge}, {Sanchez Portal}, {Santini}, {Shao}, {Sturm},
  {Tacconi}, {Valtchanov}, {Wetzstein}, \& {Wieprecht}}]{lut11}
{Lutz}, D., {Poglitsch}, A., {Altieri}, B., {et~al.} 2011, ArXiv e-prints,
  1106.3285

\bibitem[{{Maraston}(2005)}]{mar05}
{Maraston}, C. 2005, \mnras, 362, 799

\bibitem[{{McCarthy} {et~al.}(2001){McCarthy}, {Carlberg}, {Chen}, {Marzke},
  {Firth}, {Ellis}, {Persson}, {McMahon}, {Lahav}, {Wilson}, {Martini},
  {Abraham}, {Sabbey}, {Oemler}, {Murphy}, {Somerville}, {Beckett}, {Lewis}, \&
  {MacKay}}]{mcc01}
{McCarthy}, P.~J., {Carlberg}, R.~G., {Chen}, H.-W., {et~al.} 2001, \apjl, 560,
  L131

\bibitem[{{McCarthy} {et~al.}(2004){McCarthy}, {Le Borgne}, {Crampton}, {Chen},
  {Abraham}, {Glazebrook}, {Savaglio}, {Carlberg}, {Marzke}, {Roth},
  {J{\o}rgensen}, {Hook}, {Murowinski}, \& {Juneau}}]{mcc04}
{McCarthy}, P.~J., {Le Borgne}, D., {Crampton}, D., {et~al.} 2004, \apjl, 614,
  L9

\bibitem[{{Mignoli} {et~al.}(2005){Mignoli}, {Cimatti}, {Zamorani}, {Pozzetti},
  {Daddi}, {Renzini}, {Broadhurst}, {Cristiani}, {D'Odorico}, {Fontana},
  {Giallongo}, {Gilmozzi}, {Menci}, \& {Saracco}}]{mig05}
{Mignoli}, M., {Cimatti}, A., {Zamorani}, G., {et~al.} 2005, \aap, 437, 883

\bibitem[{{Mobasher} {et~al.}(2004){Mobasher}, {Idzi}, {Ben{\'{\i}}tez},
  {Cimatti}, {Cristiani}, {Daddi}, {Dahlen}, {Dickinson}, {Erben}, {Ferguson},
  {Giavalisco}, {Grogin}, {Koekemoer}, {Mignoli}, {Moustakas}, {Nonino},
  {Rosati}, {Schirmer}, {Stern}, {Vanzella}, {Wolf}, \& {Zamorani}}]{mob04}
{Mobasher}, B., {Idzi}, R., {Ben{\'{\i}}tez}, N., {et~al.} 2004, \apjl, 600,
  L167

\bibitem[{{Moorwood} {et~al.}(1998){Moorwood}, {Cuby}, {Biereichel}, {Brynnel},
  {Delabre}, {Devillard}, {van Dijsseldonk}, {Finger}, {Gemperlein},
  {Gilmozzi}, {Herlin}, {Huster}, {Knudstrup}, {Lidman}, {Lizon}, {Mehrgan},
  {Meyer}, {Nicolini}, {Petr}, {Spyromilio}, \& {Stegmeier}}]{moo98}
{Moorwood}, A., {Cuby}, J.-G., {Biereichel}, P., {et~al.} 1998, The Messenger,
  94, 7

\bibitem[{{Noll} {et~al.}(2009){Noll}, {Pierini}, {Cimatti}, {Daddi}, {Kurk},
  {Bolzonella}, {Cassata}, {Halliday}, {Mignoli}, {Pozzetti}, {Renzini},
  {Berta}, {Dickinson}, {Franceschini}, {Rodighiero}, {Rosati}, \&
  {Zamorani}}]{nol09}
{Noll}, S., {Pierini}, D., {Cimatti}, A., {et~al.} 2009, \aap, 499, 69

\bibitem[{{Nonino} {et~al.}(2009){Nonino}, {Dickinson}, {Rosati}, {Grazian},
  {Reddy}, {Cristiani}, {Giavalisco}, {Kuntschner}, {Vanzella}, {Daddi},
  {Fosbury}, \& {Cesarsky}}]{non09}
{Nonino}, M., {Dickinson}, M., {Rosati}, P., {et~al.} 2009, \apjs, 183, 244

\bibitem[{{Oke}(1974)}]{oke74}
{Oke}, J.~B. 1974, \apjs, 27, 21

\bibitem[{{Pannella} {et~al.}(2009){Pannella}, {Carilli}, {Daddi}, {McCracken},
  {Owen}, {Renzini}, {Strazzullo}, {Civano}, {Koekemoer}, {Schinnerer},
  {Scoville}, {Smol{\v c}i{\'c}}, {Taniguchi}, {Aussel}, {Kneib}, {Ilbert},
  {Mellier}, {Salvato}, {Thompson}, \& {Willott}}]{pan09}
{Pannella}, M., {Carilli}, C.~L., {Daddi}, E., {et~al.} 2009, \apjl, 698, L116

\bibitem[{{Popesso} {et~al.}(2009){Popesso}, {Dickinson}, {Nonino}, {Vanzella},
  {Daddi}, {Fosbury}, {Kuntschner}, {Mainieri}, {Cristiani}, {Cesarsky},
  {Giavalisco}, {Renzini}, \& {GOODS Team}}]{pop09}
{Popesso}, P., {Dickinson}, M., {Nonino}, M., {et~al.} 2009, \aap, 494, 443

\bibitem[{{Pozzetti} {et~al.}(2003){Pozzetti}, {Cimatti}, {Zamorani}, {Daddi},
  {Menci}, {Fontana}, {Renzini}, {Mignoli}, {Poli}, {Saracco}, {Broadhurst},
  {Cristiani}, {D'Odorico}, {Giallongo}, \& {Gilmozzi}}]{poz03}
{Pozzetti}, L., {Cimatti}, A., {Zamorani}, G., {et~al.} 2003, \aap, 402, 837

\bibitem[{{Ravikumar} {et~al.}(2007){Ravikumar}, {Puech}, {Flores}, {Proust},
  {Hammer}, {Lehnert}, {Rawat}, {Amram}, {Balkowski}, {Burgarella}, {Cassata},
  {Cesarsky}, {Cimatti}, {Combes}, {Daddi}, {Dannerbauer}, {di Serego
  Alighieri}, {Elbaz}, {Guiderdoni}, {Kembhavi}, {Liang}, {Pozzetti},
  {Vergani}, {Vernet}, {Wozniak}, \& {Zheng}}]{rav07}
{Ravikumar}, C.~D., {Puech}, M., {Flores}, H., {et~al.} 2007, \aap, 465, 1099

\bibitem[{{Renzini}(2009)}]{ren09}
{Renzini}, A. 2009, \mnras, 398, L58

\bibitem[{{Renzini} \& {da Costa}(1997)}]{ren97}
{Renzini}, A. \& {da Costa}, L.~N. 1997, The Messenger, 87, 23

\bibitem[{{Retzlaff} {et~al.}(2010){Retzlaff}, {Rosati}, {Dickinson},
  {Vandame}, {Rit{\'e}}, {Nonino}, {Cesarsky}, \& {GOODS Team}}]{ret10}
{Retzlaff}, J., {Rosati}, P., {Dickinson}, M., {et~al.} 2010, \aap, 511, A50+

\bibitem[{{Rodighiero} {et~al.}(2010){Rodighiero}, {Cimatti}, {Gruppioni},
  {Popesso}, {Andreani}, {Altieri}, {Aussel}, {Berta}, {Bongiovanni},
  {Brisbin}, {Cava}, {Cepa}, {Daddi}, {Dominguez-Sanchez}, {Elbaz}, {Fontana},
  {F{\"o}rster Schreiber}, {Franceschini}, {Genzel}, {Grazian}, {Lutz},
  {Magdis}, {Magliocchetti}, {Magnelli}, {Maiolino}, {Mancini}, {Nordon},
  {Perez Garcia}, {Poglitsch}, {Santini}, {Sanchez-Portal}, {Pozzi},
  {Riguccini}, {Saintonge}, {Shao}, {Sturm}, {Tacconi}, {Valtchanov},
  {Wetzstein}, \& {Wieprecht}}]{rod10}
{Rodighiero}, G., {Cimatti}, A., {Gruppioni}, C., {et~al.} 2010, \aap, 518,
  L25+

\bibitem[{{Rodighiero} {et~al.}(2011){Rodighiero}, {Daddi}, {Baronchelli},
  {Cimatti}, {Renzini}, {Aussel}, {Popesso}, {Lutz}, {Andreani}, {Berta},
  {Cava}, {Elbaz}, {Feltre}, {Fontana}, {F{\"o}rster Schreiber},
  {Franceschini}, {Genzel}, {Grazian}, {Gruppioni}, {Ilbert}, {Le Floch},
  {Magdis}, {Magliocchetti}, {Magnelli}, {Maiolino}, {McCracken}, {Nordon},
  {Poglitsch}, {Santini}, {Pozzi}, {Riguccini}, {Tacconi}, {Wuyts}, \&
  {Zamorani}}]{rod11}
{Rodighiero}, G., {Daddi}, E., {Baronchelli}, I., {et~al.} 2011, \apjl, 739,
  L40

\bibitem[{{Rosati} {et~al.}(2002){Rosati}, {Tozzi}, {Giacconi}, {Gilli},
  {Hasinger}, {Kewley}, {Mainieri}, {Nonino}, {Norman}, {Szokoly}, {Wang},
  {Zirm}, {Bergeron}, {Borgani}, {Gilmozzi}, {Grogin}, {Koekemoer}, {Schreier},
  \& {Zheng}}]{ros02}
{Rosati}, P., {Tozzi}, P., {Giacconi}, R., {et~al.} 2002, \apj, 566, 667

\bibitem[{{Saracco} {et~al.}(2001){Saracco}, {Giallongo}, {Cristiani},
  {D'Odorico}, {Fontana}, {Iovino}, {Poli}, \& {Vanzella}}]{sar01}
{Saracco}, P., {Giallongo}, E., {Cristiani}, S., {et~al.} 2001, \aap, 375, 1

\bibitem[{{Saracco} {et~al.}(2005){Saracco}, {Longhetti}, {Severgnini}, {Della
  Ceca}, {Braito}, {Mannucci}, {Bender}, {Drory}, {Feulner}, {Hopp}, \&
  {Maraston}}]{sar05}
{Saracco}, P., {Longhetti}, M., {Severgnini}, P., {et~al.} 2005, \mnras, 357,
  L40

\bibitem[{{Scarlata} {et~al.}(2007){Scarlata}, {Carollo}, {Lilly}, {Sargent},
  {Feldmann}, {Kampczyk}, {Porciani}, {Koekemoer}, {Scoville}, {Kneib},
  {Leauthaud}, {Massey}, {Rhodes}, {Tasca}, {Capak}, {Maier}, {McCracken},
  {Mobasher}, {Renzini}, {Taniguchi}, {Thompson}, {Sheth}, {Ajiki}, {Aussel},
  {Murayama}, {Sanders}, {Sasaki}, {Shioya}, \& {Takahashi}}]{sca07}
{Scarlata}, C., {Carollo}, C.~M., {Lilly}, S., {et~al.} 2007, \apjs, 172, 406

\bibitem[{{Scott} {et~al.}(2010){Scott}, {Yun}, {Wilson}, {Austermann},
  {Aguilar}, {Aretxaga}, {Ezawa}, {Ferrusca}, {Hatsukade}, {Hughes}, {Iono},
  {Giavalisco}, {Kawabe}, {Kohno}, {Mauskopf}, {Oshima}, {Perera}, {Rand},
  {Tamura}, {Tosaki}, {Velazquez}, {Williams}, \& {Zeballos}}]{sco10}
{Scott}, K.~S., {Yun}, M.~S., {Wilson}, G.~W., {et~al.} 2010, \mnras, 405, 2260

\bibitem[{{Steidel} {et~al.}(2003){Steidel}, {Adelberger}, {Shapley},
  {Pettini}, {Dickinson}, \& {Giavalisco}}]{ste03}
{Steidel}, C.~C., {Adelberger}, K.~L., {Shapley}, A.~E., {et~al.} 2003, \apj,
  592, 728

\bibitem[{{Strolger} {et~al.}(2004){Strolger}, {Riess}, {Dahlen}, {Livio},
  {Panagia}, {Challis}, {Tonry}, {Filippenko}, {Chornock}, {Ferguson},
  {Koekemoer}, {Mobasher}, {Dickinson}, {Giavalisco}, {Casertano}, {Hook},
  {Blondin}, {Leibundgut}, {Nonino}, {Rosati}, {Spinrad}, {Steidel}, {Stern},
  {Garnavich}, {Matheson}, {Grogin}, {Hornschemeier}, {Kretchmer}, {Laidler},
  {Lee}, {Lucas}, {de Mello}, {Moustakas}, {Ravindranath}, {Richardson}, \&
  {Taylor}}]{str04}
{Strolger}, L.-G., {Riess}, A.~G., {Dahlen}, T., {et~al.} 2004, \apj, 613, 200

\bibitem[{{Szokoly} {et~al.}(2004){Szokoly}, {Bergeron}, {Hasinger}, {Lehmann},
  {Kewley}, {Mainieri}, {Nonino}, {Rosati}, {Giacconi}, {Gilli}, {Gilmozzi},
  {Norman}, {Romaniello}, {Schreier}, {Tozzi}, {Wang}, {Zheng}, \&
  {Zirm}}]{szo04}
{Szokoly}, G.~P., {Bergeron}, J., {Hasinger}, G., {et~al.} 2004, \apjs, 155,
  271

\bibitem[{{Talia} {et~al.}(2012){Talia}, {Mignoli}, {Cimatti}, {Kurk}, {Berta},
  {Bolzonella}, {Cassata}, {Daddi}, {Dickinson}, {Franceschini}, {Halliday},
  {Pozzetti}, {Renzini}, {Rodighiero}, {Rosati}, \& {Zamorani}}]{tal12}
{Talia}, M., {Mignoli}, M., {Cimatti}, A., {et~al.} 2012, \aap, 539, A61

\bibitem[{{Tanaka} {et~al.}(2004){Tanaka}, {Goto}, {Okamura}, {Shimasaku}, \&
  {Brinkmann}}]{tan04}
{Tanaka}, M., {Goto}, T., {Okamura}, S., {Shimasaku}, K., \& {Brinkmann}, J.
  2004, \aj, 128, 2677

\bibitem[{{Thomas} {et~al.}(2005){Thomas}, {Maraston}, {Bender}, \& {Mendes de
  Oliveira}}]{tho05}
{Thomas}, D., {Maraston}, C., {Bender}, R., \& {Mendes de Oliveira}, C. 2005,
  \apj, 621, 673

\bibitem[{{Thompson} {et~al.}(2005){Thompson}, {Illingworth}, {Bouwens},
  {Dickinson}, {Eisenstein}, {Fan}, {Franx}, {Riess}, {Rieke}, {Schneider},
  {Stobie}, {Toft}, \& {van Dokkum}}]{thom05}
{Thompson}, R.~I., {Illingworth}, G., {Bouwens}, R., {et~al.} 2005, \aj, 130, 1

\bibitem[{{Treu} {et~al.}(2005){Treu}, {Ellis}, {Liao}, \& {van
  Dokkum}}]{tre05}
{Treu}, T., {Ellis}, R.~S., {Liao}, T.~X., \& {van Dokkum}, P.~G. 2005, \apjl,
  622, L5

\bibitem[{{van der Wel} {et~al.}(2005){van der Wel}, {Franx}, {van Dokkum},
  {Rix}, {Illingworth}, \& {Rosati}}]{vdw05}
{van der Wel}, A., {Franx}, M., {van Dokkum}, P.~G., {et~al.} 2005, \apj, 631,
  145

\bibitem[{{van Dokkum}(2001)}]{vdo01}
{van Dokkum}, P.~G. 2001, \pasp, 113, 1420

\bibitem[{{Vanzella} {et~al.}(2008){Vanzella}, {Cristiani}, {Dickinson},
  {Giavalisco}, {Kuntschner}, {Haase}, {Nonino}, {Rosati}, {Cesarsky},
  {Ferguson}, {Fosbury}, {Grazian}, {Moustakas}, {Rettura}, {Popesso},
  {Renzini}, {Stern}, \& {GOODS Team}}]{van08}
{Vanzella}, E., {Cristiani}, S., {Dickinson}, M., {et~al.} 2008, \aap, 478, 83

\bibitem[{{Vanzella} {et~al.}(2005){Vanzella}, {Cristiani}, {Dickinson},
  {Kuntschner}, {Moustakas}, {Nonino}, {Rosati}, {Stern}, {Cesarsky}, {Ettori},
  {Ferguson}, {Fosbury}, {Giavalisco}, {Haase}, {Renzini}, {Rettura}, {Serra},
  \& {The Goods Team}}]{van05}
{Vanzella}, E., {Cristiani}, S., {Dickinson}, M., {et~al.} 2005, \aap, 434, 53

\bibitem[{{Vanzella} {et~al.}(2006){Vanzella}, {Cristiani}, {Dickinson},
  {Kuntschner}, {Nonino}, {Rettura}, {Rosati}, {Vernet}, {Cesarsky},
  {Ferguson}, {Fosbury}, {Giavalisco}, {Grazian}, {Haase}, {Moustakas},
  {Popesso}, {Renzini}, {Stern}, \& {The Goods Team}}]{van06}
{Vanzella}, E., {Cristiani}, S., {Dickinson}, M., {et~al.} 2006, \aap, 454, 423

\bibitem[{{Vanzella} {et~al.}(2009){Vanzella}, {Giavalisco}, {Dickinson},
  {Cristiani}, {Nonino}, {Kuntschner}, {Popesso}, {Rosati}, {Renzini}, {Stern},
  {Cesarsky}, {Ferguson}, \& {Fosbury}}]{van09}
{Vanzella}, E., {Giavalisco}, M., {Dickinson}, M., {et~al.} 2009, \apj, 695,
  1163

\bibitem[{{Wei{\ss}} {et~al.}(2009){Wei{\ss}}, {Kov{\'a}cs}, {Coppin}, {Greve},
  {Walter}, {Smail}, {Dunlop}, {Knudsen}, {Alexander}, {Bertoldi}, {Brandt},
  {Chapman}, {Cox}, {Dannerbauer}, {De Breuck}, {Gawiser}, {Ivison}, {Lutz},
  {Menten}, {Koekemoer}, {Kreysa}, {Kurczynski}, {Rix}, {Schinnerer}, \& {van
  der Werf}}]{wei09}
{Wei{\ss}}, A., {Kov{\'a}cs}, A., {Coppin}, K., {et~al.} 2009, \apj, 707, 1201

\bibitem[{{Werner} {et~al.}(2004){Werner}, {Roellig}, {Low}, {Rieke}, {Rieke},
  {Hoffmann}, {Young}, {Houck}, {Brandl}, {Fazio}, {Hora}, {Gehrz}, {Helou},
  {Soifer}, {Stauffer}, {Keene}, {Eisenhardt}, {Gallagher}, {Gautier}, {Irace},
  {Lawrence}, {Simmons}, {Van Cleve}, {Jura}, {Wright}, \&
  {Cruikshank}}]{wer04}
{Werner}, M.~W., {Roellig}, T.~L., {Low}, F.~J., {et~al.} 2004, \apjs, 154, 1

\bibitem[{{Xue} {et~al.}(2011){Xue}, {Luo}, {Brandt}, {Bauer}, {Lehmer},
  {Broos}, {Schneider}, {Alexander}, {Brusa}, {Comastri}, {Fabian}, {Gilli},
  {Hasinger}, {Hornschemeier}, {Koekemoer}, {Liu}, {Mainieri}, {Paolillo},
  {Rafferty}, {Rosati}, {Shemmer}, {Silverman}, {Smail}, {Tozzi}, \&
  {Vignali}}]{xue11}
{Xue}, Y.~Q., {Luo}, B., {Brandt}, W.~N., {et~al.} 2011, \apjs, 195, 10

\bibitem[{{Yan} {et~al.}(2004){Yan}, {Dickinson}, {Eisenhardt}, {Ferguson},
  {Grogin}, {Paolillo}, {Chary}, {Casertano}, {Stern}, {Reach}, {Moustakas}, \&
  {Fall}}]{yan04}
{Yan}, H., {Dickinson}, M., {Eisenhardt}, P.~R.~M., {et~al.} 2004, \apj, 616,
  63

\end{thebibliography}

\begin{appendix}
\section{Table of observed galaxies}\label{sec:table}
\addtocounter{table}{1}
\longtab{1}{
\begin{longtable}{rrrr@{$\pm$}lr@{$\pm$}lr@{$\pm$}lr@{$\pm$}lr@{$\pm$}l@{\hspace{0.5ex}}r@{\hspace{0.5ex}}r@{\hspace{0.5ex}}r@{\hspace{0.8ex}}r@{\hspace{0.8ex}}r}
\caption{\label{table:gmass_galaxies}
Galaxies observed in the GMASS masks: coordinates, photometry, redshifts, S/N, photometric normalisation factor, sample and mask(s)}\\
\hline\hline
ID & \multicolumn{1}{c}{R.A.} & \multicolumn{1}{c}{Dec.} & \multicolumn{2}{c}{B} & \multicolumn{2}{c}{I} & \multicolumn{2}{c}{K$_{\rm s}$} & \multicolumn{2}{c}{m$_{\rm4.5}$} & \multicolumn{2}{c}{\zspec} & q\tablefootmark{a} & S/N\tablefootmark{b} & Norm\tablefootmark{c} & S\tablefootmark{d} & M\tablefootmark{e} \\
\hline
\endfirsthead
\caption{continued}\\
\hline\hline
ID & \multicolumn{1}{c}{R.A.} & \multicolumn{1}{c}{Dec.} & \multicolumn{2}{c}{B} & \multicolumn{2}{c}{I} & \multicolumn{2}{c}{K$_{\rm s}$} & \multicolumn{2}{c}{m$_{\rm4.5}$} & \multicolumn{2}{c}{\zspec} & q\tablefootmark{a} & S/N\tablefootmark{b} & Norm\tablefootmark{c} & S\tablefootmark{d} & M\tablefootmark{e} \\
\hline
\endhead
\hline
\endfoot
2467& 3:32:38.96&-27:42:43.7&99.00&0.00&25.77&0.30&21.57&0.03&20.26&0.01&        4.3792&0.0064& 1&   1.1& 1.5&    1&    5\\
9111& 3:32:16.07&-27:44:26.1&\multicolumn{2}{c}{-}&\multicolumn{2}{c}{-}&\multicolumn{2}{c}{-}&\multicolumn{2}{c}{-}&        4.1365&0.0061& 0&   0.3& 0.0&     &    5\\
 418& 3:32:33.03&-27:47:59.5&99.00&0.00&25.04&0.17&23.77&0.13&22.66&0.02&        4.0627&0.0062& 0&   0.9& 2.2&  1,3&    6\\
1788& 3:32:36.31&-27:44:34.6&25.94&0.21&24.09&0.07&23.08&0.07&22.73&0.02&        3.4129&0.0030& 1&   4.2& 1.4&  1,2&1,2,5\\
1807& 3:32:19.05&-27:44:29.9&99.00&0.00&25.94&0.33&22.95&0.06&21.03&0.01&        3.3561&0.0012& 0&   0.6& 2.0&  1,3&    5\\
9101& 3:32:41.56&-27:45:26.5&\multicolumn{2}{c}{-}&\multicolumn{2}{c}{-}&\multicolumn{2}{c}{-}&\multicolumn{2}{c}{-}&        3.1000&0.0028& 1&   0.2& 0.0&     &    3\\
9102& 3:32:41.91&-27:45:24.0&\multicolumn{2}{c}{-}&\multicolumn{2}{c}{-}&\multicolumn{2}{c}{-}&\multicolumn{2}{c}{-}&        3.0711&0.0067& 1&   0.1& 0.0&     &    3\\
1160& 3:32:46.93&-27:46:04.7&25.40&0.14&24.43&0.10&23.86&0.18&22.86&0.02&        2.8645&0.0021& 1&   3.8& 1.3&  2,4&    4\\
 920& 3:32:47.99&-27:46:39.5&24.51&0.06&23.60&0.05&22.90&0.08&22.61&0.02&        2.8281&0.0013& 1&   8.8& 1.7&  2,4&    4\\
1048& 3:32:47.23&-27:46:20.4&25.81&0.20&24.67&0.13&23.01&0.11&22.09&0.01&        2.8054&0.0027& 1&   1.6& 2.7&  1,2&    2\\
 307& 3:32:35.05&-27:48:23.3&26.23&0.26&25.34&0.22&23.30&0.08&22.35&0.01&        2.7973&0.0040& 1&   1.6& 2.3&    1&    2\\
9103& 3:32:30.59&-27:42:39.5&\multicolumn{2}{c}{-}&\multicolumn{2}{c}{-}&\multicolumn{2}{c}{-}&\multicolumn{2}{c}{-}&        2.6891&0.0001& 1&   0.2& 0.0&     &    3\\
1980& 3:32:14.99&-27:44:08.2&25.50&0.13&24.75&0.11&23.49&0.11&22.79&0.03&        2.6734&0.0017& 1&   5.1& 1.1&  2,4&  4,5\\
1479& 3:32:15.70&-27:45:15.4&25.77&0.18&24.70&0.12&22.53&0.05&21.31&0.01&        2.6732&0.0068& 1&   3.0& 2.2&  1,2&  2,3\\
 253& 3:32:39.18&-27:48:32.3&27.08&0.51&25.87&0.34&22.78&0.06&22.02&0.01&        2.6697&0.0006& 0&   0.8& 1.7&  1,3&    6\\
 191& 3:32:27.28&-27:48:45.7&24.34&0.05&23.58&0.05&23.33&0.11&22.76&0.02&        2.6305&0.0033& 1&   8.1& 1.6&  1,2&    2\\
 330& 3:32:28.42&-27:48:19.0&24.66&0.07&24.08&0.07&23.68&0.12&22.96&0.02&        2.6274&0.0035& 1&   6.2& 1.6&    2&    2\\
1796& 3:32:15.64&-27:44:34.5&24.34&0.05&23.88&0.05&23.36&0.11&22.28&0.02&        2.6161&0.0037& 1&  11.6& 2.0&  2,4&    3\\
2161& 3:32:29.24&-27:42:58.9&26.46&0.24&25.24&0.16&23.97&0.24&22.86&0.03&        2.5769&0.0024& 1&   4.6& 1.9&    4&    3\\
2043& 3:32:41.88&-27:43:59.9&25.81&0.19&24.92&0.14&23.19&0.09&22.05&0.01&        2.5763&0.0021& 1&   2.4& 1.8&  2,4&    4\\
 167& 3:32:37.89&-27:48:53.0&24.51&0.06&23.79&0.05&23.20&0.10&22.48&0.01&        2.5729&0.0016& 1&   7.1& 2.0&  2,4&    4\\
2512& 3:32:42.01&-27:42:27.8&26.87&0.46&25.55&0.25&23.58&0.17&22.09&0.01&        2.5713&0.0030& 0&   1.6& 1.2&    3&    6\\
1049& 3:32:23.18&-27:46:20.3&24.74&0.07&24.12&0.07&23.10&0.08&22.56&0.02&        2.4832&0.0018& 1&   5.6& 1.9&     &    4\\
 885& 3:32:43.69&-27:46:46.4&25.01&0.14&24.79&0.19&23.35&0.10&23.01&0.02&        2.4677&0.0059& 1&   3.8& 1.4&    2&    2\\
2562& 3:32:33.30&-27:42:01.9&25.73&0.17&24.71&0.13&22.57&0.07&21.41&0.01&        2.4495&0.0020& 1&   3.9& 1.8&1,2,4&  3,6\\
2207& 3:32:36.89&-27:43:03.8&25.22&0.11&24.77&0.13&23.60&0.17&21.75&0.01&        2.4488&0.0018& 1&   7.9& 1.1&  2,4&    3\\
2303& 3:32:38.88&-27:43:21.5&24.58&0.07&24.02&0.07&23.34&0.14&22.80&0.03&        2.4487&0.0017& 1&  11.0& 1.6&  2,4&    3\\
2363& 3:32:39.41&-27:42:35.7&25.99&0.22&24.51&0.10&22.97&0.10&21.79&0.01&        2.4485&0.0005& 1&   3.0& 1.5&  2,4&    6\\
2578& 3:32:33.01&-27:42:00.5&25.55&0.15&24.70&0.13&22.00&0.05&20.88&0.01&        2.4481&0.0049& 1&   2.0& 1.2&    1&    1\\
1030& 3:32:39.34&-27:46:23.7&27.50&0.70&25.57&0.27&23.72&0.14&22.88&0.02&        2.4469&0.0027& 0&   1.5& 1.4&  1,3&  5,6\\
1489& 3:32:29.17&-27:45:14.8&25.44&0.14&24.58&0.09&23.25&0.08&22.68&0.02&        2.4334&0.0029& 1&   2.7& 1.6&    2&    2\\
2471& 3:32:32.36&-27:42:48.0&24.55&0.06&23.79&0.06&22.91&0.11&22.08&0.01&        2.4301&0.0024& 1&   9.3& 2.2&  2,4&  3,4\\
1989& 3:32:43.89&-27:44:05.8&24.46&0.06&23.74&0.05&21.67&0.02&20.83&0.01&        2.4286&0.0005& 0&   4.8& 1.2&  1,2&    1\\
2090& 3:32:18.72&-27:43:51.7&24.84&0.08&24.20&0.08&22.86&0.07&22.11&0.01&        2.4164&0.0011& 1&   6.0& 1.7&  2,4&    4\\
2252& 3:32:19.05&-27:43:15.2&25.52&0.15&24.82&0.13&22.66&0.07&21.47&0.01&        2.4065&0.0028& 1&   2.1& 2.4&  1,2&    2\\
 181& 3:32:34.11&-27:48:49.6&24.10&0.04&23.81&0.06&23.07&0.09&22.41&0.01&        2.3436&0.0022& 1&  10.9& 1.4&  2,4&    4\\
 249& 3:32:22.42&-27:48:33.6&25.97&0.23&25.17&0.16&23.47&0.13&22.23&0.01&        2.3342&0.0039& 0&   2.2& 2.0&1,2,4&    3\\
1711& 3:32:27.11&-27:44:44.1&25.72&0.18&25.09&0.16&23.32&0.08&22.69&0.02&        2.3235&0.0040& 1&   4.7& 1.8&  2,4&    3\\
2450& 3:32:43.64&-27:43:47.9&24.99&0.09&24.24&0.09&22.70&0.06&22.04&0.01&        2.3134&0.0017& 1&   7.9& 1.5&  2,4&    3\\
 796& 3:32:28.50&-27:46:58.2&26.16&0.36&25.61&0.32&22.49&0.04&20.72&0.01&        2.3091&0.0019& 1&   0.9& 2.4&  1,4&    4\\
2443& 3:32:24.20&-27:42:57.5&25.00&0.10&24.42&0.09&21.99&0.04&21.10&0.01&        2.2979&0.0037& 1&   5.0& 1.0&  1,2&  1,5\\
2099& 3:32:31.53&-27:43:50.9&24.95&0.08&24.38&0.08&23.06&0.08&22.79&0.03&        2.1934&0.0026& 1&   4.0& 1.5&    2&    2\\
 459& 3:32:26.59&-27:47:50.1&25.39&0.13&24.82&0.12&23.39&0.09&22.56&0.02&        2.1621&0.0054& 1&   4.8& 1.5&  1,2&  2,4\\
2572& 3:32:36.89&-27:42:25.9&25.03&0.10&24.29&0.09&22.83&0.09&22.02&0.01&        2.1375&0.0028& 1&   7.2& 1.6&1,2,4&    3\\
 881& 3:32:31.32&-27:46:46.9&25.63&0.16&25.22&0.19&23.50&0.10&23.09&0.03&        2.1336&0.0018& 1&   3.8& 1.3&  2,4&  4,6\\
1372& 3:32:21.72&-27:45:29.6&25.59&0.15&24.79&0.14&23.24&0.08&22.85&0.03&        2.0799&0.0051& 1&   5.5& 1.2&1,2,4&    3\\
 949& 3:32:23.69&-27:46:32.9&25.54&0.15&26.21&0.42&23.42&0.09&21.96&0.01&        2.0764&0.0052& 1&   3.5& 1.1&  2,4&  3,5\\
1663& 3:32:24.73&-27:44:50.3&24.83&0.11&24.23&0.08&23.16&0.08&22.43&0.02&        2.0249&0.0032& 1&   6.5& 1.9&  2,4&  3,6\\
 502& 3:32:34.14&-27:47:43.5&25.20&0.11&24.70&0.13&23.47&0.11&22.70&0.02&        2.0156&0.0032& 1&   2.6& 2.1&    2&    2\\
 149& 3:32:21.95&-27:48:55.6&24.53&0.06&23.85&0.06&22.64&0.05&21.93&0.01&        2.0069&0.0018& 1&   8.4& 1.9&  2,4&    3\\
1314& 3:32:26.73&-27:45:40.0&24.97&0.09&24.40&0.08&23.39&0.10&22.67&0.02&        2.0065&0.0023& 1&   7.3& 1.8&  2,4&3,4,5\\
 426& 3:32:40.06&-27:47:55.4&24.46&0.06&23.64&0.05&21.65&0.02&20.40&0.01&        1.9962&0.0014& 1&   6.1& 1.7&  1,2&    2\\
 271& 3:32:41.69&-27:48:29.6&24.73&0.07&24.29&0.09&23.07&0.08&22.52&0.01&        1.9957&0.0021& 1&   4.1& 1.8&    2&    2\\
2559& 3:32:42.34&-27:42:04.2&99.00&0.00&25.39&0.27&22.00&0.04&20.71&0.01&        1.9807&0.0014& 0&   1.1& 1.9&  1,3&  1,6\\
2219& 3:32:39.69&-27:43:06.6&23.66&0.03&23.42&0.04&22.90&0.09&21.92&0.01&        1.9646&0.0013& 1&  16.6& 1.7&  2,4&    3\\
2032& 3:32:45.19&-27:44:01.7&25.03&0.09&24.46&0.10&24.05&0.18&22.67&0.03&        1.9622&0.0046& 1&   2.8& 1.9&  2,4&    2\\
2018& 3:32:44.72&-27:44:01.4&24.36&0.05&23.93&0.06&22.89&0.07&21.89&0.01&        1.9621&0.0053& 1&   5.6& 1.7&    2&    2\\
8005& 3:32:21.35&-27:46:54.8&23.90&0.03&23.71&0.05&23.90&0.27&22.89&0.03&        1.9403&0.0030& 1&  15.5& 1.5&  2,4&    3\\
 472& 3:32:38.12&-27:47:49.6&99.00&0.00&25.99&0.37&21.80&0.02&21.04&0.01&        1.9213&0.0037& 1&   1.2& 1.1&  1,3&    6\\
1427& 3:32:33.15&-27:45:22.8&25.91&0.15&25.61&0.21&23.63&0.12&23.02&0.03&        1.9182&0.0014& 1&   1.7& 1.0&  1,2&  1,6\\
 870& 3:32:28.16&-27:46:48.4&25.32&0.10&24.65&0.10&23.20&0.08&22.34&0.01&        1.9092&0.0018& 1&   4.7& 1.7&  2,4&    3\\
 508& 3:32:33.74&-27:47:44.2&24.93&0.09&24.32&0.09&23.00&0.07&22.08&0.01&        1.9090&0.0024& 1&   4.6& 1.9&  2,4&    4\\
 656& 3:32:35.84&-27:47:18.7&24.58&0.07&23.73&0.05&22.35&0.04&21.77&0.01&        1.9057&0.0006& 1&  10.4& 1.5&  2,4&    3\\
2275& 3:32:17.28&-27:43:29.7&25.45&0.14&24.77&0.13&23.17&0.12&22.34&0.02&        1.9050&0.0033& 1&   1.1& 1.1&  1,2&    1\\
  90& 3:32:34.09&-27:49:11.4&24.75&0.08&24.29&0.10&22.91&0.07&21.53&0.01&        1.9029&0.0023& 1&   4.0& 1.8&  2,4&    4\\
 923& 3:32:27.26&-27:46:38.9&24.74&0.11&24.39&0.09&23.54&0.11&22.82&0.02&        1.8849&0.0014& 1&   8.1& 1.7&  2,4&    3\\
2107& 3:32:30.09&-27:42:42.9&24.76&0.07&24.26&0.08&23.49&0.16&22.16&0.02&        1.8843&0.0021& 1&   3.8& 1.5&  2,4&  3,5\\
1789& 3:32:31.84&-27:44:35.4&25.36&0.10&24.53&0.09&23.11&0.08&22.11&0.01&        1.8834&0.0014& 1&   3.3& 1.8&  2,4&    4\\
 250& 3:32:30.91&-27:48:32.1&24.19&0.06&23.96&0.08&22.97&0.07&22.33&0.01&        1.8828&0.0016& 1&   5.8& 1.7&    2&  2,6\\
 679& 3:32:35.52&-27:47:15.7&25.21&0.12&24.63&0.12&23.64&0.12&23.01&0.02&        1.8827&0.0019& 1&   4.4& 1.5&  2,4&  4,6\\
 183& 3:32:31.16&-27:48:48.2&24.29&0.07&23.74&0.06&23.03&0.08&22.42&0.01&        1.8820&0.0013& 1&   9.2& 1.6&  2,4&    4\\
 118& 3:32:31.53&-27:48:53.8&25.65&0.18&24.79&0.14&22.33&0.04&20.87&0.01&        1.8795&0.0041& 1&   1.8& 2.4&1,2,4&    4\\
1486& 3:32:18.73&-27:45:14.4&23.80&0.03&23.53&0.04&22.86&0.06&22.28&0.02&        1.8761&0.0013& 1&  19.0& 1.0&  2,4&  4,5\\
 355& 3:32:37.08&-27:48:14.4&26.94&0.46&25.38&0.23&23.42&0.10&22.51&0.01&        1.8686&0.0013& 0&   1.4& 1.4&  1,3&    6\\
1748& 3:32:20.20&-27:44:38.9&25.29&0.12&24.79&0.13&23.54&0.11&22.55&0.02&        1.8668&0.0023& 1&   2.8& 2.0&  2,4&    4\\
 894& 3:32:35.82&-27:46:43.7&25.08&0.08&24.48&0.09&23.13&0.07&22.25&0.01&        1.8502&0.0020& 1&   4.6& 1.8&  2,4&  3,6\\
 220& 3:32:37.18&-27:48:33.9&24.69&0.07&23.79&0.06&22.22&0.04&21.08&0.01&        1.8499&0.0013& 1&   4.7& 1.7&    2&    2\\
 875& 3:32:37.09&-27:46:47.1&24.78&0.08&24.40&0.10&23.90&0.15&\multicolumn{2}{c}{-}&        1.8491&0.0010& 1&   5.1& 2.0&     &    3\\
1498& 3:32:41.91&-27:45:12.1&26.56&0.35&25.63&0.26&21.71&0.02&20.72&0.01&        1.8478&0.0016& 0&   1.4& 0.9&    1&    1\\
 858& 3:32:37.36&-27:46:45.5&24.46&0.06&23.76&0.05&22.00&0.03&21.11&0.01&        1.8463&0.0017& 1&   8.4& 1.7&1,2,4&  3,4\\
1224& 3:32:17.58&-27:45:51.8&26.78&0.42&25.97&0.35&21.95&0.03&20.35&0.01&        1.8432&0.0027& 1&   2.0& 0.5&  1,3&    5\\
1822& 3:32:15.35&-27:44:31.9&24.69&0.07&24.32&0.08&22.94&0.07&22.45&0.02&        1.8419&0.0022& 1&   5.7& 1.9&  2,4&    4\\
 675& 3:32:38.81&-27:47:14.8&24.97&0.09&24.16&0.08&22.21&0.03&21.16&0.01&        1.8361&0.0076& 1&   3.3& 1.9&  1,2&    2\\
2526& 3:32:37.91&-27:42:15.4&26.60&0.36&25.20&0.19&22.11&0.05&20.33&0.01&        1.8139&0.0027& 1&   2.3& 0.9&  1,3&  1,5\\
 390& 3:32:22.09&-27:48:06.7&24.02&0.03&23.90&0.05&23.55&0.15&22.58&0.02&        1.7738&0.0022& 1&   6.8& 1.8&    2&    2\\
 487& 3:32:36.40&-27:47:47.0&23.85&0.03&23.65&0.06&22.53&0.04&21.86&0.01&        1.7672&0.0006& 1&   9.3& 1.8&  2,4&    4\\
 178& 3:32:38.20&-27:48:49.4&25.25&0.12&24.56&0.11&22.97&0.10&22.09&0.01&        1.7669&0.0051& 1&   2.3& 1.9&  1,2&    2\\
 484& 3:32:35.65&-27:47:48.8&23.83&0.03&23.73&0.05&23.23&0.08&22.70&0.02&        1.7651&0.0019& 1&   8.9& 1.3&    2&    2\\
2403& 3:32:30.95&-27:42:48.3&25.85&0.20&24.56&0.11&22.85&0.09&21.96&0.01&        1.7647&0.0020& 1&   1.8& 1.4&  1,2&    1\\
 335& 3:32:40.99&-27:48:16.8&24.07&0.04&23.89&0.06&23.32&0.09&22.67&0.02&        1.7626&0.0016& 1&   6.9& 1.7&    2&    2\\
1938& 3:32:23.71&-27:44:11.8&24.70&0.10&23.64&0.05&21.59&0.02&20.33&0.01&        1.7596&0.0029& 1&   5.8& 0.9&  1,2&    1\\
1464& 3:32:28.31&-27:45:18.8&25.79&0.18&24.85&0.12&23.40&0.10&21.95&0.01&        1.7552&0.0039& 1&   2.7& 2.6&1,2,4&  3,4\\
1454& 3:32:28.55&-27:45:19.4&27.38&0.63&25.64&0.22&23.47&0.10&22.17&0.01&        1.7552&0.0015& 1&   1.7& 0.8&  1,3&    5\\
 316& 3:32:20.82&-27:48:22.5&23.78&0.03&23.66&0.05&23.33&0.16&22.32&0.01&        1.7365&0.0015& 1&  10.6& 1.5&    2&    2\\
1155& 3:32:22.54&-27:46:03.8&24.97&0.09&24.22&0.08&22.23&0.04&20.40&0.01&        1.7269&0.0029& 1&   3.8& 1.5&  1,2&  1,5\\
1133& 3:32:22.87&-27:46:07.2&25.05&0.10&24.59&0.11&23.03&0.08&22.20&0.02&        1.7245&0.0018& 1&   3.6& 1.4&    2&    2\\
1624& 3:32:18.11&-27:44:55.1&26.08&0.23&25.17&0.18&23.52&0.11&22.32&0.02&        1.7166&0.0036& 0&   1.2& 1.3&    1&    1\\
1274& 3:32:31.33&-27:45:44.7&25.47&0.14&24.72&0.12&22.64&0.05&21.18&0.01&        1.6697&0.0029& 1&   2.1& 1.1&  1,2&    1\\
9112& 3:32:36.31&-27:47:22.4&\multicolumn{2}{c}{-}&\multicolumn{2}{c}{-}&\multicolumn{2}{c}{-}&\multicolumn{2}{c}{-}&        1.6390&0.0008& 0&   0.4& 0.0&     &    6\\
1399& 3:32:41.66&-27:45:25.6&25.54&0.15&25.26&0.19&23.79&0.14&22.71&0.02&        1.6146&0.0026& 1&   3.2& 1.4&  2,4&    3\\
2196& 3:32:36.67&-27:42:58.5&99.00&0.00&25.13&0.17&21.33&0.02&20.34&0.01&        1.6138&0.0038& 1&   1.2& 2.0&    3&    6\\
1667& 3:32:40.99&-27:44:50.2&26.25&0.27&25.46&0.23&23.76&0.14&22.51&0.02&        1.6134&0.0020& 1&   1.4& 1.2&    1&    5\\
2540& 3:32:30.33&-27:42:40.3&24.69&0.07&24.02&0.07&22.54&0.07&21.40&0.01&        1.6128&0.0015& 1&   4.1& 1.1&  2,4&  3,6\\
2113& 3:32:22.00&-27:42:43.5&26.18&0.26&24.77&0.13&22.15&0.05&20.37&0.01&        1.6128&0.0016& 1&   1.5& 1.5&    1&    1\\
2368& 3:32:17.10&-27:43:41.9&25.18&0.11&24.44&0.09&22.48&0.07&21.15&0.01&        1.6121&0.0014& 1&   3.5& 1.9&  1,2&  2,4\\
1380& 3:32:25.25&-27:45:29.0&24.91&0.09&24.60&0.09&22.99&0.07&22.30&0.02&        1.6121&0.0027& 1&   3.5& 1.7&    2&1,2,6\\
2603& 3:32:27.85&-27:43:05.7&24.39&0.06&23.80&0.05&22.76&0.08&21.74&0.01&        1.6120&0.0008& 1&  12.1& 1.6&  2,4&    3\\
2543& 3:32:35.92&-27:42:41.0&99.00&0.00&25.43&0.18&21.53&0.03&20.26&0.01&        1.6119&0.0030& 1&   1.0& 1.3&  1,3&    1\\
1691& 3:32:31.90&-27:44:45.0&24.97&0.08&24.12&0.07&22.11&0.03&21.11&0.01&        1.6119&0.0009& 1&   4.0& 1.9&1,2,4&    4\\
1495& 3:32:35.36&-27:45:12.6&25.17&0.10&24.64&0.10&22.50&0.04&21.20&0.01&        1.6113&0.0013& 1&   2.1& 2.6&1,2,4&    4\\
2055& 3:32:26.77&-27:43:58.1&26.58&0.36&25.49&0.23&23.02&0.07&22.21&0.02&        1.6112&0.0015& 1&   1.8& 1.0&  1,3&    6\\
1979& 3:32:24.64&-27:44:07.8&24.12&0.04&23.51&0.04&22.10&0.03&20.95&0.01&        1.6112&0.0011& 1&   2.7& 1.6&    2&    2\\
1254& 3:32:20.17&-27:45:49.3&23.91&0.03&23.71&0.05&23.49&0.10&22.21&0.02&        1.6105&0.0013& 1&   7.1& 1.7&    2&    2\\
2327& 3:32:26.12&-27:43:25.0&25.39&0.10&25.00&0.13&23.88&0.22&\multicolumn{2}{c}{-}&        1.6103&0.0014& 1&   1.6& 1.9&     &    6\\
2111& 3:32:27.94&-27:42:45.7&99.00&0.00&24.79&0.13&21.52&0.03&20.64&0.01&        1.6102&0.0025& 1&   3.0& 1.2&    1&  1,5\\
2142& 3:32:23.54&-27:42:49.3&24.78&0.08&24.34&0.09&22.83&0.09&22.22&0.02&        1.6098&0.0028& 1&   7.1& 1.0&  2,4&  4,5\\
2355& 3:32:14.32&-27:43:32.9&26.48&0.26&25.02&0.14&21.91&0.04&20.92&0.01&        1.6095&0.0015& 1&   1.2& 1.4&    1&    1\\
2251& 3:32:29.48&-27:43:22.0&25.11&0.10&24.28&0.08&21.68&0.03&20.64&0.01&        1.6094&0.0019& 1&   6.2& 1.6&1,2,4&  3,6\\
2361& 3:32:26.05&-27:42:36.6&99.00&0.00&24.99&0.15&21.25&0.02&20.40&0.01&        1.6086&0.0016& 1&   2.3& 1.1&    3&    5\\
2247& 3:32:27.86&-27:43:13.5&25.98&0.19&24.81&0.13&22.56&0.07&21.77&0.01&        1.6086&0.0007& 0&   1.9& 1.1&  1,2&    1\\
2148& 3:32:36.30&-27:42:49.5&27.15&0.54&24.47&0.10&20.90&0.02&19.80&0.01&        1.6086&0.0031& 1&   3.9& 1.0&    3&    5\\
 365& 3:32:27.80&-27:48:12.0&24.43&0.06&24.24&0.08&23.72&0.13&23.10&0.02&        1.6086&0.0015& 1&   4.9& 1.6&     &    2\\
1808& 3:32:26.15&-27:44:33.3&24.84&0.09&24.28&0.08&23.24&0.09&22.61&0.02&        1.6085&0.0012& 1&   4.8& 1.6&     &  4,6\\
2180& 3:32:29.56&-27:42:56.0&24.59&0.05&23.72&0.04&22.00&0.04&21.00&0.01&        1.6077&0.0016& 1&   9.1& 2.0&1,2,4&  3,6\\
2493& 3:32:38.51&-27:42:28.0&24.48&0.04&23.67&0.04&22.10&0.05&20.88&0.01&        1.6073&0.0016& 1&   8.4& 1.8&  2,4&    3\\
1708& 3:32:23.12&-27:44:42.2&99.00&0.00&25.27&0.19&22.22&0.03&20.81&0.01&        1.6069&0.0012& 1&   1.3& 1.7&  1,3&  1,5\\
 781& 3:32:17.71&-27:47:02.9&23.77&0.03&23.67&0.05&23.40&0.15&22.54&0.02&        1.6052&0.0016& 1&   9.1& 1.5&    2&    2\\
2352& 3:32:33.88&-27:42:04.1&25.10&0.10&24.15&0.08&21.31&0.02&20.09&0.01&        1.6042&0.0027& 1&   3.6& 1.0&  1,2&    1\\
2286& 3:32:29.99&-27:43:22.6&99.00&0.00&25.20&0.17&21.80&0.04&20.82&0.01&        1.6036&0.0025& 1&   1.9& 1.1&    3&    5\\
2454& 3:32:28.91&-27:43:03.6&25.31&0.11&24.41&0.09&22.24&0.05&20.80&0.01&        1.6019&0.0015& 1&   4.0& 1.2&1,2,4&4,5,6\\
2341& 3:32:17.52&-27:43:36.6&24.55&0.06&24.21&0.08&23.35&0.16&22.47&0.02&        1.6018&0.0037& 1&   7.7& 2.1&  2,4&    3\\
2550& 3:32:30.08&-27:42:12.2&24.78&0.07&24.14&0.08&23.02&0.11&22.03&0.01&        1.6012&0.0015& 1&   7.7& 1.6&  2,4&    3\\
2081& 3:32:29.86&-27:43:54.8&24.37&0.06&24.53&0.10&23.57&0.14&22.09&0.01&        1.6011&0.0010& 1&   4.5& 1.4&    2&    2\\
1084& 3:32:39.74&-27:46:11.5&26.42&0.31&25.16&0.19&21.64&0.02&19.73&0.01&        1.5518&0.0037& 1&   1.5& 1.8&  1,4&4,5,6\\
2573& 3:32:31.10&-27:42:05.3&26.55&0.33&24.74&0.13&21.97&0.04&20.06&0.01&        1.5496&0.0013& 1&   1.8& 1.3&    1&    1\\
1050& 3:32:17.88&-27:46:20.8&24.57&0.06&24.10&0.07&23.77&0.20&22.61&0.02&        1.5392&0.0022& 1&   5.9& 2.5&  2,4&    3\\
1146& 3:32:15.75&-27:46:04.6&24.92&0.09&24.05&0.07&22.08&0.04&20.86&0.01&        1.5368&0.0017& 1&   7.7& 1.8&1,2,4&    3\\
 408& 3:32:23.56&-27:48:02.6&24.94&0.08&24.46&0.10&23.79&0.13&22.89&0.02&        1.5076&0.0033& 1&   2.9& 2.0&    2&    2\\
 685& 3:32:25.33&-27:47:15.5&25.74&0.17&25.62&0.27&24.61&0.26&\multicolumn{2}{c}{-}&        1.4877&0.0007& 1&   1.0& 1.0&     &    1\\
 512& 3:32:22.84&-27:47:42.5&26.83&0.42&25.55&0.26&22.94&0.07&21.97&0.01&        1.4692&0.0001& 1&   1.9& 0.8&  1,3&    5\\
9110& 3:32:22.90&-27:47:42.3&\multicolumn{2}{c}{-}&\multicolumn{2}{c}{-}&\multicolumn{2}{c}{-}&\multicolumn{2}{c}{-}&        1.4685&0.0007& 1&   1.3& 0.0&     &    5\\
2484& 3:32:37.69&-27:42:19.5&24.98&0.09&24.09&0.07&22.58&0.07&21.59&0.01&        1.4356&0.0008& 1&   4.3& 1.7&  2,4&    4\\
2381& 3:32:38.77&-27:42:18.4&25.50&0.15&24.54&0.11&22.83&0.09&21.55&0.01&        1.4302&0.0009& 0&   3.2& 2.1&  2,4&    3\\
2470& 3:32:43.15&-27:42:42.1&99.00&0.00&24.15&0.07&21.15&0.02&20.16&0.01&        1.4156&0.0019& 1&   4.7& 1.1&  1,3&    5\\
2239& 3:32:31.32&-27:43:16.2&99.00&0.00&24.93&0.15&21.58&0.03&20.61&0.01&        1.4151&0.0018& 1&   2.6& 1.5&    3&  5,6\\
 996& 3:32:36.92&-27:46:28.5&99.00&0.00&25.56&0.26&22.14&0.03&21.36&0.01&        1.3844&0.0032& 0&   1.6& 1.3&    3&    5\\
1652& 3:32:41.58&-27:44:52.8&24.98&0.10&24.93&0.15&23.44&0.11&23.01&0.03&        1.3527&0.0015& 1&   4.0& 1.2&  2,4&    4\\
 793& 3:32:45.98&-27:46:57.7&24.61&0.07&23.86&0.06&22.31&0.05&21.73&0.01&        1.2951&0.0014& 1&   4.6& 1.7&     &    2\\
1682& 3:32:41.50&-27:44:40.2&26.92&0.45&24.63&0.11&21.12&0.01&19.84&0.01&        1.2950&0.0010& 1&   3.1& 1.1&    3&    6\\
2135& 3:32:38.60&-27:42:37.0&25.31&0.12&24.55&0.11&22.64&0.07&21.70&0.01&        1.2458&0.0002& 1&   3.2& 1.4&  2,4&    4\\
9107& 3:32:22.74&-27:46:02.6&\multicolumn{2}{c}{-}&\multicolumn{2}{c}{-}&\multicolumn{2}{c}{-}&\multicolumn{2}{c}{-}&        1.2273&0.0008& 1&   1.0& 0.0&     &    1\\
1217& 3:32:21.23&-27:45:54.8&24.28&0.05&24.03&0.07&23.45&0.10&\multicolumn{2}{c}{-}&        1.2251&0.0016& 1&   4.7& 1.8&     &    2\\
1952& 3:32:34.65&-27:44:08.1&25.94&0.21&24.19&0.07&20.98&0.01&20.04&0.01&        1.2239&0.0063& 1&   4.5& 1.2&    3&    5\\
1227& 3:32:21.30&-27:45:54.6&24.57&0.06&24.24&0.08&23.76&0.13&22.65&0.02&        1.2232&0.0014& 1&   4.3& 1.8&    2&    2\\
2235& 3:32:41.25&-27:43:09.7&99.00&0.00&25.54&0.26&22.11&0.05&20.80&0.01&        1.2216&0.0014& 1&   1.4& 1.3&    3&    5\\
1020& 3:32:24.60&-27:46:20.3&26.66&0.37&23.92&0.06&21.16&0.01&20.17&0.01&        1.2209&0.0017& 1&   4.5& 1.3&     &    4\\
2580& 3:32:42.97&-27:42:04.2&24.89&0.06&23.89&0.06&22.74&0.08&21.58&0.01&        1.2199&0.0008& 1&   4.2& 1.1&     &    1\\
 886& 3:32:47.14&-27:46:44.4&27.30&0.64&25.06&0.18&21.68&0.03&20.92&0.01&        1.2193&0.0003& 1&   2.0& 1.6&    3&    6\\
 774& 3:32:16.45&-27:47:02.3&23.92&0.04&23.44&0.04&22.64&0.08&22.35&0.02&        1.2171&0.0015& 1&   6.5& 2.1&     &    2\\
 773& 3:32:16.26&-27:47:03.2&24.33&0.05&24.04&0.07&23.32&0.14&22.66&0.02&        1.2169&0.0022& 1&   3.1& 2.8&     &    2\\
2092& 3:32:25.74&-27:43:47.1&99.00&0.00&24.24&0.08&20.93&0.01&19.89&0.01&        1.2168&0.0015& 1&   3.9& 1.7&    3&    6\\
2283& 3:32:12.27&-27:43:24.3&99.00&0.00&25.72&0.28&22.10&0.04&20.99&0.01&        1.2161&0.0007& 0&   0.5& 1.7&    1&    1\\
2231& 3:32:33.28&-27:42:36.1&99.00&0.00&25.46&0.22&22.66&0.08&21.73&0.01&        1.2131&0.0003& 1&   0.9& 2.2&  1,3&    6\\
 712& 3:32:38.64&-27:47:11.5&24.80&0.08&24.04&0.07&23.18&0.08&22.55&0.02&        1.1338&0.0015& 1&   4.6& 1.5&     &    6\\
 795& 3:32:42.77&-27:46:59.1&24.03&0.05&23.52&0.05&22.65&0.05&22.51&0.01&        1.1191&0.0016& 1&   6.9& 1.8&     &    2\\
1567& 3:32:24.01&-27:45:04.0&24.71&0.07&23.97&0.06&22.90&0.06&22.74&0.02&        1.1101&0.0023& 1&   5.2& 1.4&     &    2\\
2158& 3:32:26.38&-27:43:21.5&24.30&0.05&23.64&0.05&22.93&0.10&22.62&0.02&        1.1097&0.0006& 1&   5.1& 1.1&     &    1\\
9108& 3:32:28.31&-27:42:44.4&\multicolumn{2}{c}{-}&\multicolumn{2}{c}{-}&\multicolumn{2}{c}{-}&\multicolumn{2}{c}{-}&        1.1080&0.0002& 1&   0.9& 0.0&     &    1\\
 692& 3:32:47.44&-27:47:11.1&26.77&0.42&24.64&0.12&21.86&0.03&21.00&0.01&        1.0981&0.0019& 1&   2.6& 1.2&    3&    5\\
1315& 3:32:39.26&-27:45:32.3&25.58&0.16&23.44&0.04&20.65&0.01&19.74&0.01&        1.0952&0.0007& 1&   5.5& 1.4&     &    4\\
 428& 3:32:18.44&-27:47:57.0&24.05&0.04&23.36&0.04&22.38&0.06&22.02&0.01&        1.0793&0.0011& 1&   7.4& 2.9&     &    3\\
2191& 3:32:29.29&-27:42:44.8&25.82&0.16&23.94&0.05&21.80&0.04&21.23&0.01&        1.0404&0.0015& 1&   5.6& 0.9&     &    1\\
9105& 3:32:29.61&-27:43:20.3&\multicolumn{2}{c}{-}&\multicolumn{2}{c}{-}&\multicolumn{2}{c}{-}&\multicolumn{2}{c}{-}&        1.0356&0.0001& 0&   0.3& 0.0&     &    3\\
 839& 3:32:42.98&-27:46:50.0&99.00&0.00&24.35&0.09&21.10&0.01&20.28&0.01&        1.0356&0.0010& 1&   5.6& 0.9&    3&    5\\
 983& 3:32:15.79&-27:46:29.9&23.31&0.02&22.63&0.02&21.94&0.04&21.05&0.01&        1.0210&0.0007& 1&  22.6& 2.1&  2,4&    3\\
1704& 3:32:26.43&-27:44:43.7&99.00&0.00&24.87&0.14&21.54&0.02&20.61&0.01&        1.0143&0.0021& 1&   1.6& 2.0&    3&    6\\
1585& 3:32:38.59&-27:45:00.0&24.71&0.07&23.68&0.05&22.40&0.04&22.08&0.01&        0.9789&0.0024& 1&   5.6& 1.6&     &    4\\
2296& 3:32:41.68&-27:43:21.5&99.00&0.00&24.89&0.14&21.91&0.04&20.94&0.01&        0.9783&0.0015& 1&   1.9& 1.0&     &    1\\
1394& 3:32:28.90&-27:45:25.4&23.71&0.03&22.92&0.02&22.54&0.04&22.43&0.02&        0.9520&0.0003& 1&   1.7&10.2&     &    6\\
1309& 3:32:20.39&-27:45:42.1&25.63&0.16&24.49&0.10&24.43&0.23&\multicolumn{2}{c}{-}&        0.9143&0.0017& 1&   2.3& 1.3&     &    1\\
2461& 3:32:39.09&-27:42:44.2&23.53&0.03&22.77&0.02&22.36&0.06&\multicolumn{2}{c}{-}&        0.8939&0.0007& 1&   9.8& 1.7&     &    5\\
9113& 3:32:29.60&-27:42:54.6&\multicolumn{2}{c}{-}&\multicolumn{2}{c}{-}&\multicolumn{2}{c}{-}&\multicolumn{2}{c}{-}&        0.8570&0.0007& 1&   0.6& 0.0&     &    6\\
9114& 3:32:41.53&-27:44:36.8&\multicolumn{2}{c}{-}&\multicolumn{2}{c}{-}&\multicolumn{2}{c}{-}&\multicolumn{2}{c}{-}&        0.8373&0.0001& 1&   0.6& 0.0&     &    6\\
1579& 3:32:25.76&-27:44:59.3&24.12&0.05&23.03&0.03&21.91&0.03&\multicolumn{2}{c}{-}&        0.8335&0.0008& 1&  13.2& 1.8&     &    4\\
1592& 3:32:25.80&-27:45:00.0&24.78&0.08&23.81&0.05&22.45&0.04&21.43&0.01&        0.8319&0.0004& 1&   8.6& 1.8&  2,4&  4,6\\
 942& 3:32:36.49&-27:46:29.2&23.61&0.03&22.26&0.01&21.44&0.02&21.04&0.01&        0.7649&0.0003& 1&  24.0& 1.4&     &    5\\
1501& 3:32:35.45&-27:45:14.3&26.42&0.30&24.84&0.12&23.91&0.15&\multicolumn{2}{c}{-}&        0.7380&0.0006& 1&   1.2& 2.1&     &    4\\
9109& 3:32:28.41&-27:45:19.2&\multicolumn{2}{c}{-}&\multicolumn{2}{c}{-}&\multicolumn{2}{c}{-}&\multicolumn{2}{c}{-}&        0.7375&0.0003& 1&   0.8& 0.0&     &    5\\
 210& 3:32:39.43&-27:48:38.8&25.53&0.15&23.81&0.06&22.31&0.05&22.01&0.01&        0.7361&0.0032& 1&   4.1& 1.6&     &    2\\
9106& 3:32:21.23&-27:44:01.7&\multicolumn{2}{c}{-}&\multicolumn{2}{c}{-}&\multicolumn{2}{c}{-}&\multicolumn{2}{c}{-}&        0.7344&0.0002& 1&   0.8& 0.0&     &    1\\
 558& 3:32:46.54&-27:47:35.9&24.62&0.07&23.19&0.03&22.18&0.04&22.63&0.02&        0.7064&0.0012& 1&   6.2& 1.9&     &    2\\
2223& 3:32:30.89&-27:43:16.1&25.16&0.11&22.99&0.03&21.71&0.03&\multicolumn{2}{c}{-}&        0.6799&0.0001& 1&   1.7&10.0&     &    5\\
2109& 3:32:35.60&-27:42:43.3&25.74&0.13&24.25&0.07&23.85&0.21&\multicolumn{2}{c}{-}&        0.6766&0.0003& 1&   2.4& 1.4&     &    1\\
1700& 3:32:22.66&-27:44:45.2&24.90&0.09&23.69&0.05&23.27&0.09&\multicolumn{2}{c}{-}&        0.6677&0.0007& 1&   1.1& 5.6&     &    1\\
2357& 3:32:29.64&-27:42:42.6&22.48&0.01&20.97&0.00&19.82&0.01&19.47&0.01&        0.6676&0.0009& 1&  30.2& 2.1&     &    1\\
1388& 3:32:25.57&-27:45:28.9&26.04&0.32&24.47&0.08&24.00&0.17&\multicolumn{2}{c}{-}&        0.6669&0.0002& 1&   2.7& 1.6&     &    2\\
1920& 3:32:33.17&-27:44:15.2&26.65&0.36&23.54&0.04&21.56&0.02&21.71&0.01&        0.6657&0.0007& 1&   4.5& 1.9&     &    4\\
 643& 3:32:48.47&-27:47:19.7&24.84&0.08&22.53&0.02&21.12&0.02&21.45&0.01&        0.5328&0.0007& 1&  13.3& 1.8&     &    4\\
 192& 3:32:27.42&-27:48:45.8&25.73&0.17&24.63&0.12&24.11&0.21&\multicolumn{2}{c}{-}&        0.4758&0.0026& 1&   1.0& 0.0&     &    2\\
1232& 3:32:32.97&-27:45:45.6&23.18&0.02&20.41&0.00&19.07&0.00&\multicolumn{2}{c}{-}&        0.3659&0.0005& 1&  62.8& 1.7&     &    2\\
 453& 3:32:33.83&-27:47:48.0&23.66&0.03&22.14&0.01&21.38&0.02&21.73&0.01&        0.3444&0.0006& 1&   3.6&11.0&     &    4\\
1304& 3:32:26.89&-27:45:42.0&25.11&0.10&23.85&0.05&23.69&0.12&\multicolumn{2}{c}{-}&        0.3373&0.0001& 1&   0.7&10.2&     &    4\\
 815& 3:32:42.35&-27:46:57.2&26.02&0.24&24.93&0.16&23.53&0.11&22.30&0.01&        0.3331&0.0002& 1&   2.5& 1.7&  1,4&    4\\
9104& 3:32:29.70&-27:42:54.5&\multicolumn{2}{c}{-}&\multicolumn{2}{c}{-}&\multicolumn{2}{c}{-}&\multicolumn{2}{c}{-}&        0.2325&0.0001& 1&   1.5& 0.0&     &    3\\
1323& 3:32:41.52&-27:45:32.5&21.80&0.01&20.73&0.00&20.46&0.01&20.84&0.01&        0.1469&0.0001& 1&  57.5& 1.2&     &    1\\
1901& 3:32:44.80&-27:44:06.4&20.90&0.00&19.14&0.00&18.48&0.00&18.94&0.01&        0.1032&0.0002& 1&1002.2& 0.6&     &  5,6\\
 750& 3:32:31.74&-27:46:58.4&20.55&0.00&18.83&0.00&18.67&0.00&19.98&0.01&        0.0004&0.0003& 1& 280.0& 1.4&     &    4\\
1277& 3:32:19.95&-27:45:33.7&20.10&0.00&17.93&0.00&17.37&0.00&18.53&0.01&        0.0002&0.0005& 1& 603.1& 1.7&     &    3\\
2246& 3:32:25.90&-27:43:41.2&25.13&0.11&20.22&0.00&18.27&0.00&19.17&0.01&        0.0001&0.0004& 1&   4.0&22.7&     &    6\\
1581& 3:32:25.76&-27:45:01.7&23.02&0.02&22.20&0.01&22.58&0.05&\multicolumn{2}{c}{-}&        0.0000&0.0001& 1&  21.7& 1.7&     &    6\\
2494& 3:32:21.04&-27:43:10.2&24.01&0.04&20.79&0.00&20.92&0.01&21.37&0.01&        0.0000&0.0000& 0&     -& 0.0&     &    3\\
2210& 3:32:12.55&-27:43:06.0&99.00&0.00&25.58&0.25&22.85&0.08&21.08&0.01&        0.0000&0.0000&-1&     -& 0.0&    1&    1\\
2022& 3:32:14.79&-27:44:02.5&25.84&0.18&25.42&0.19&22.84&0.06&21.65&0.01&        0.0000&0.0000&-1&     -& 0.0&  1,4&    3\\
1805& 3:32:33.48&-27:44:30.5&23.60&0.03&22.65&0.02&22.53&0.04&22.85&0.03&        0.0000&0.0000& 0&     -& 0.0&     &    2\\
1794& 3:32:36.47&-27:44:31.8&24.80&0.08&23.62&0.04&23.13&0.08&22.64&0.02&        0.0000&0.0000& 0&     -& 0.0&     &    1\\
1619& 3:32:44.15&-27:44:53.6&25.08&0.10&23.13&0.03&22.07&0.03&22.43&0.02&        0.0000&0.0000& 0&     -& 0.0&     &    2\\
1588& 3:32:42.11&-27:44:58.4&24.30&0.05&23.37&0.04&22.15&0.03&21.76&0.01&        0.0000&0.0000& 0&     -& 0.0&     &    2\\
 809& 3:32:26.59&-27:46:48.9&22.88&0.01&20.78&0.00&19.34&0.00&19.71&0.01&        0.0000&0.0000& 0&     -& 0.0&     &    4\\
 676& 3:32:25.59&-27:47:14.4&27.59&0.72&25.49&0.24&21.98&0.03&21.18&0.01&\multicolumn{2}{c}{-}&-1&     -& 0.0&    1&    1\\
 603& 3:32:36.21&-27:47:26.2&27.13&0.55&99.00&0.00&22.44&0.04&21.23&0.01&\multicolumn{2}{c}{-}&-1&     -& 0.0&     &    6\\
2595& 3:32:26.21&-27:43:48.4&99.00&0.00&24.92&0.14&23.58&0.16&22.48&0.02&\multicolumn{2}{c}{-}&-1&     -& 0.0&  1,3&  1,5\\
2445& 3:32:42.29&-27:42:44.5&26.25&0.28&25.52&0.27&23.55&0.16&22.50&0.02&\multicolumn{2}{c}{-}&-1&     -& 0.0&  1,3&  1,5\\
2372& 3:32:23.01&-27:43:04.6&26.92&0.47&25.91&0.33&24.58&0.38&22.70&0.02&\multicolumn{2}{c}{-}&-1&     -& 0.0&  1,3&    5\\
2338& 3:32:38.24&-27:41:47.0&27.05&0.51&25.10&0.23&24.18&0.27&22.54&0.02&\multicolumn{2}{c}{-}&-1&     -& 0.0&  1,3&  1,6\\
2325& 3:32:26.10&-27:43:26.6&26.26&0.21&25.49&0.19&22.16&0.05&21.00&0.01&\multicolumn{2}{c}{-}&-1&     -& 0.0&     &    6\\
2253& 3:32:19.35&-27:43:14.8&27.45&0.67&25.30&0.20&23.32&0.12&21.89&0.01&\multicolumn{2}{c}{-}&-1&     -& 0.0&  1,3&    5\\
2171& 3:32:23.44&-27:42:55.0&25.94&0.22&24.75&0.12&21.95&0.04&20.99&0.01&\multicolumn{2}{c}{-}&-1&     -& 0.0&    1&    1\\
2087& 3:32:44.67&-27:43:51.8&25.19&0.11&22.83&0.02&21.86&0.03&22.50&0.02&\multicolumn{2}{c}{-}&-1&     -& 0.0&     &    1\\
2076& 3:32:32.12&-27:43:55.3&27.12&0.51&25.91&0.30&23.22&0.09&21.23&0.01&\multicolumn{2}{c}{-}&-1&     -& 0.0&    1&    1\\
2015& 3:32:20.96&-27:44:03.1&26.15&0.25&25.51&0.23&23.54&0.11&22.37&0.02&\multicolumn{2}{c}{-}&-1&     -& 0.0&    1&  1,3\\
1846& 3:32:15.81&-27:44:27.0&27.02&0.39&25.89&0.24&22.82&0.07&21.82&0.01&\multicolumn{2}{c}{-}&-1&     -& 0.0&  1,3&    5\\
1672& 3:32:25.02&-27:44:47.6&25.31&0.16&24.74&0.12&22.10&0.03&20.70&0.01&\multicolumn{2}{c}{-}&-1&     -& 0.0&1,2,4&    4\\
1528& 3:32:33.74&-27:45:07.6&27.51&0.69&25.85&0.27&22.67&0.05&21.23&0.01&\multicolumn{2}{c}{-}&-1&     -& 0.0&  1,3&    5\\
1485& 3:32:18.18&-27:45:15.9&26.60&0.31&25.81&0.28&23.92&0.15&22.76&0.03&\multicolumn{2}{c}{-}&-1&     -& 0.0&  1,3&    5\\
1298& 3:32:20.15&-27:45:43.1&26.88&0.43&25.89&0.32&24.02&0.16&22.63&0.02&\multicolumn{2}{c}{-}&-1&     -& 0.0&    1&    1\\
1070& 3:32:32.28&-27:46:15.3&25.82&0.19&24.89&0.14&22.31&0.04&21.44&0.01&\multicolumn{2}{c}{-}&-1&     -& 0.0&     &    2\\
1018& 3:32:44.01&-27:46:25.5&26.46&0.34&25.21&0.20&23.83&0.16&22.64&0.02&\multicolumn{2}{c}{-}&-1&     -& 0.0&  1,4&  3,5\\
 824& 3:32:35.78&-27:46:55.1&25.85&0.18&24.79&0.12&23.38&0.09&22.40&0.01&\multicolumn{2}{c}{-}&-1&     -& 0.0&1,2,4&  3,5\\
 739& 3:32:48.57&-27:47:07.6&99.00&0.00&25.57&0.27&21.91&0.03&21.15&0.01&\multicolumn{2}{c}{-}&-1&     -& 0.0&    3&    5\\
 463& 3:32:33.67&-27:47:51.1&99.00&0.00&25.32&0.21&22.36&0.04&21.53&0.01&\multicolumn{2}{c}{-}&-1&     -& 0.0&  1,3&    6\\
 441& 3:32:48.10&-27:47:56.1&26.42&0.32&25.69&0.30&23.58&0.15&22.78&0.02&\multicolumn{2}{c}{-}&-1&     -& 0.0&  1,3&    6\\
 410& 3:32:26.00&-27:47:51.4&25.78&0.18&24.77&0.12&22.78&0.05&21.12&0.01&\multicolumn{2}{c}{-}&-1&     -& 0.0&  1,2&    2\\
 396& 3:32:22.49&-27:48:04.7&25.66&0.12&25.44&0.18&22.98&0.07&21.49&0.01&\multicolumn{2}{c}{-}&-1&     -& 0.0&  1,2&    1\\
 190& 3:32:24.42&-27:48:44.2&26.05&0.22&25.49&0.24&23.21&0.09&21.98&0.01&\multicolumn{2}{c}{-}&-1&     -& 0.0&  1,4&    3\\
\end{longtable}
\tablefoot{
\tablefoottext{a}{Quality of the spectroscopic redshift determination: (1) good, (0) plausible, and (-1) guess or no redshift.\\}
\tablefoottext{b}{Average S/N per pixel (only averaged over pixels that have a noise value within 1\,$\sigma$ of the 3\,$\sigma$-clipped mean noise).\\}
\tablefoottext{c}{Multiplication factor needed to obtain spectral magnitudes consistent with imaging photometry (see Sec.\ \ref{sec:spec_cal}). Zero if imaging photometry is not available.\\}
\tablefoottext{d}{Part of sample: (1) red selection sample in P73, (2) blue selection sample in P73, (3) red selection sample in P74, (4) blue selection sample in P74 (see also Table \ref{table:samples}). If there is no number here, the galaxy was included as a filler or serendipitously.\\}
\tablefoottext{e}{Mask number where galaxy was observed. Multiple entries possible. Masks 1,5,6 were \emph{red} and masks 2,3,4 were \emph{blue}. Masks 1,2 were observed in P73 and masks 3,4,5,6 in P74 (or later).}
}
}

\section{GMASS spectra}\label{sec:app_spectra}

\begin{figure*}
\centering
  \includegraphics[width=\linewidth]{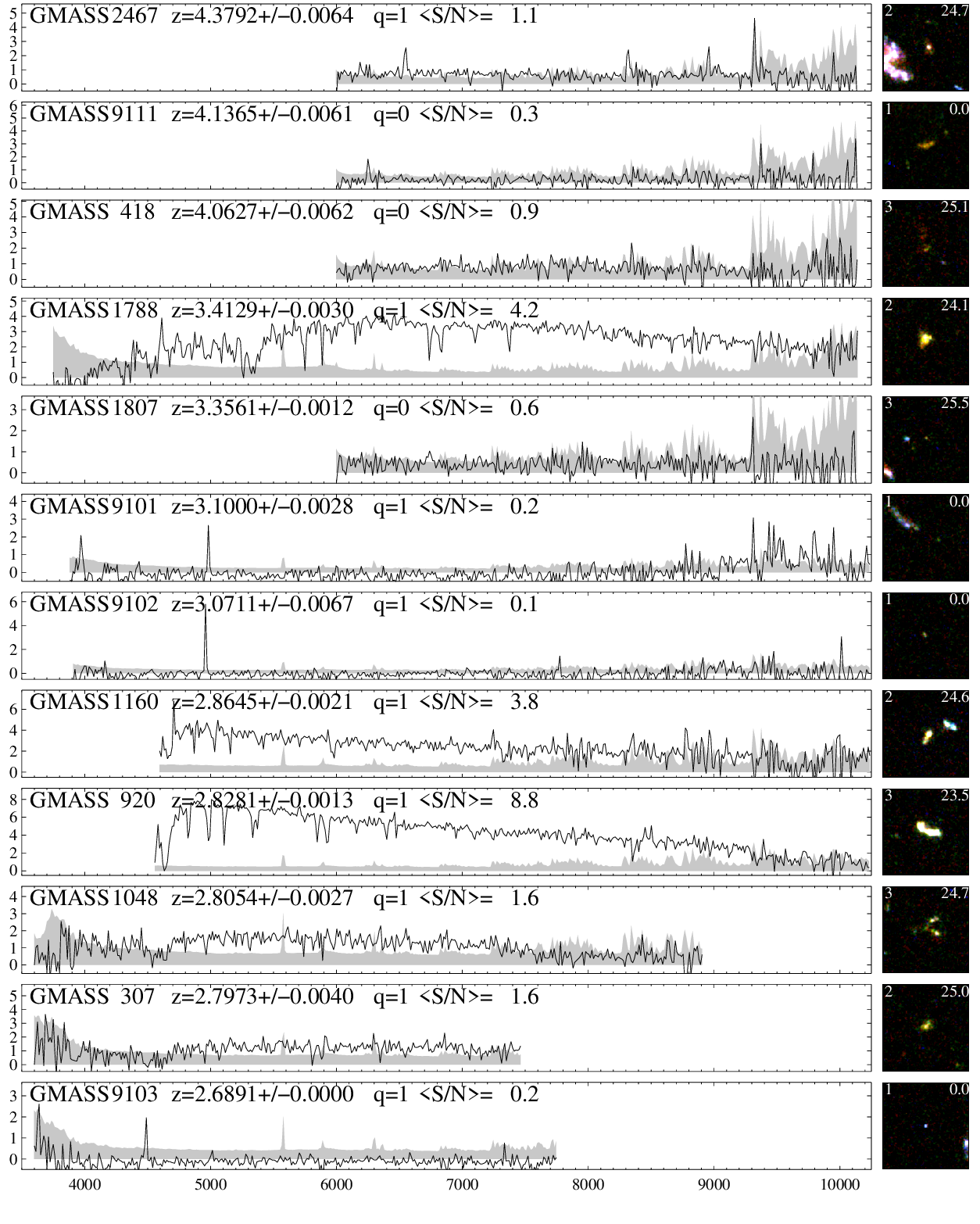}
  \caption{Spectra and postage stamp images of the 181 galaxies and
    stars with redshifts determined.  Wavelength in \AA\ on the
    horizontal axis and flux in $10^{-19}$
    erg\,s$^{-1}$\,\AA$^{-1}$\,cm$^{-2}$ on the vertical axis.
    Uncertainties caused by background noise are indicated by the
    underlying filled grey spectra.  Indicated in each spectrum are:
    GMASS identification number, redshift and its uncertainty,
    redshift quality (1: secure, 0: tentative), and mean S/N per
    pixel.  The postage stamps are constructed from HST/ACS
    observations in the $B$, $V$, and $I$ bands, convolved with a
    Gaussian kernel.  Indicated are morphological class (see text, on
    the left) and $z$ magnitude (on the right).  The spectra are
    sorted in descending order of redshift.}
  \label{fig:stamps_spectra01}
\end{figure*}

\begin{figure*}
\centering
  \includegraphics[width=\linewidth]{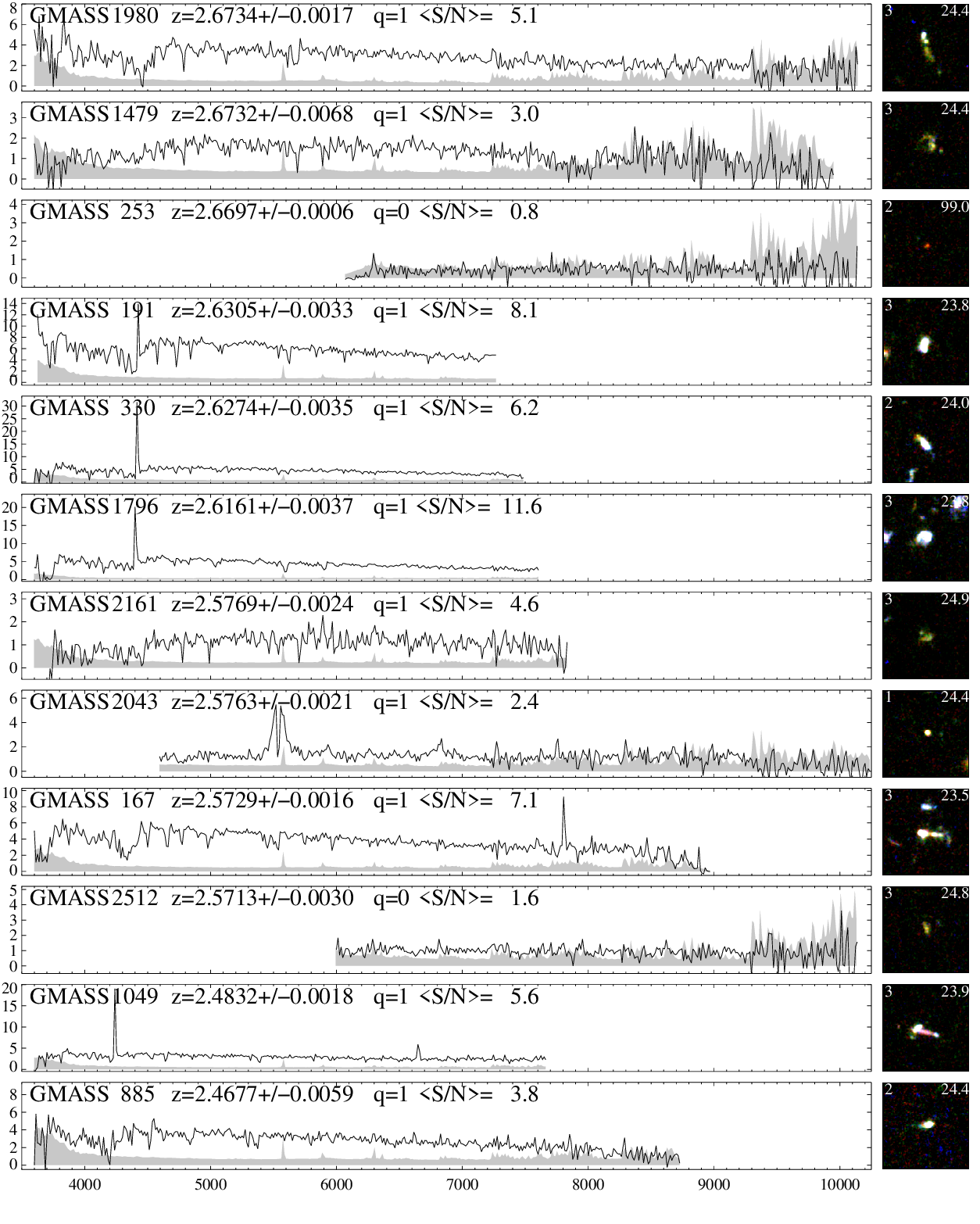}
  \caption{See Fig.~\ref{fig:stamps_spectra01} for description.}
  \label{fig:stamps_spectra02}
\end{figure*}

\begin{figure*}
\centering
  \includegraphics[width=\linewidth]{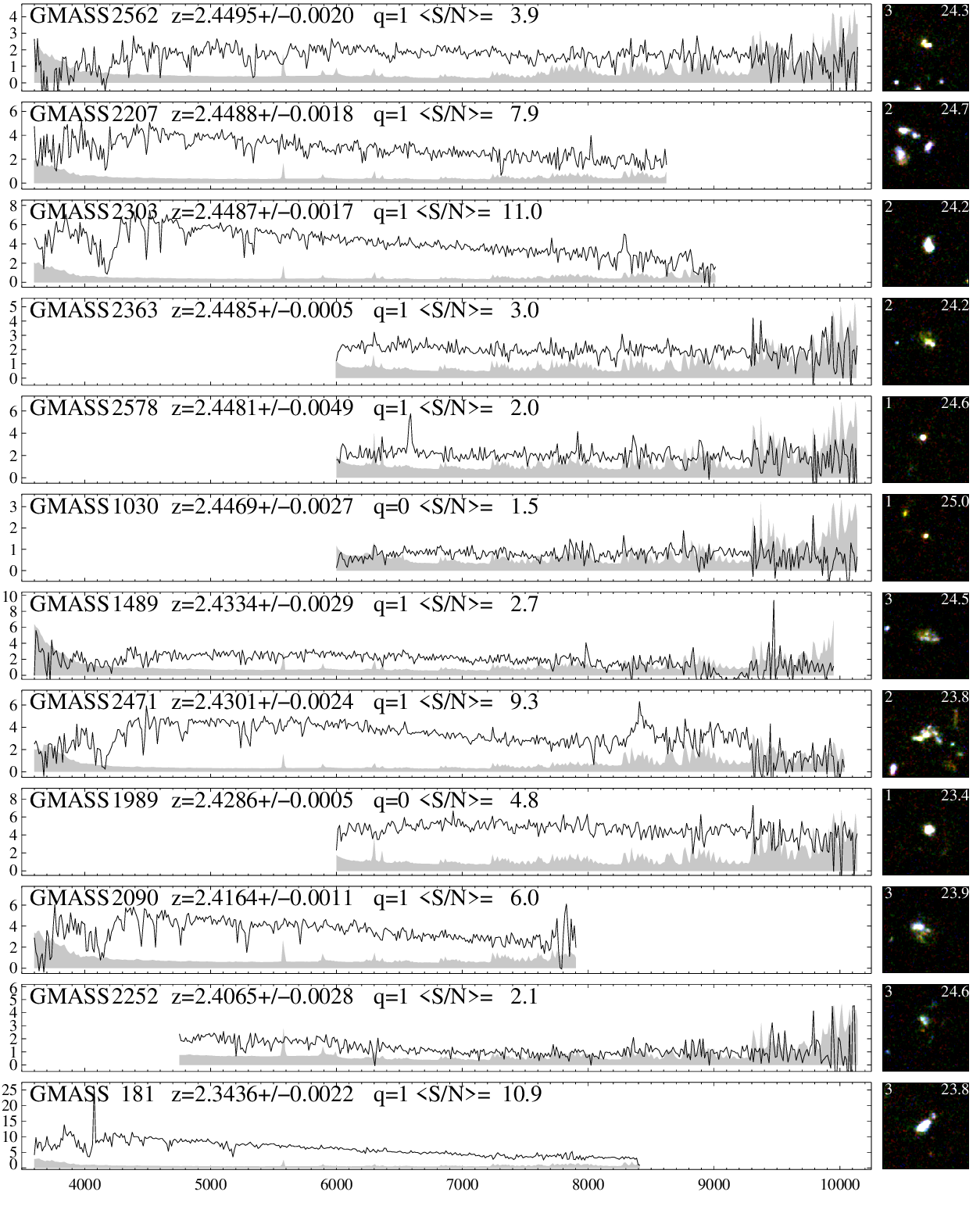}
  \caption{See Fig.~\ref{fig:stamps_spectra01} for description.}
  \label{fig:stamps_spectra03}
\end{figure*}

\begin{figure*}
\centering
  \includegraphics[width=\linewidth]{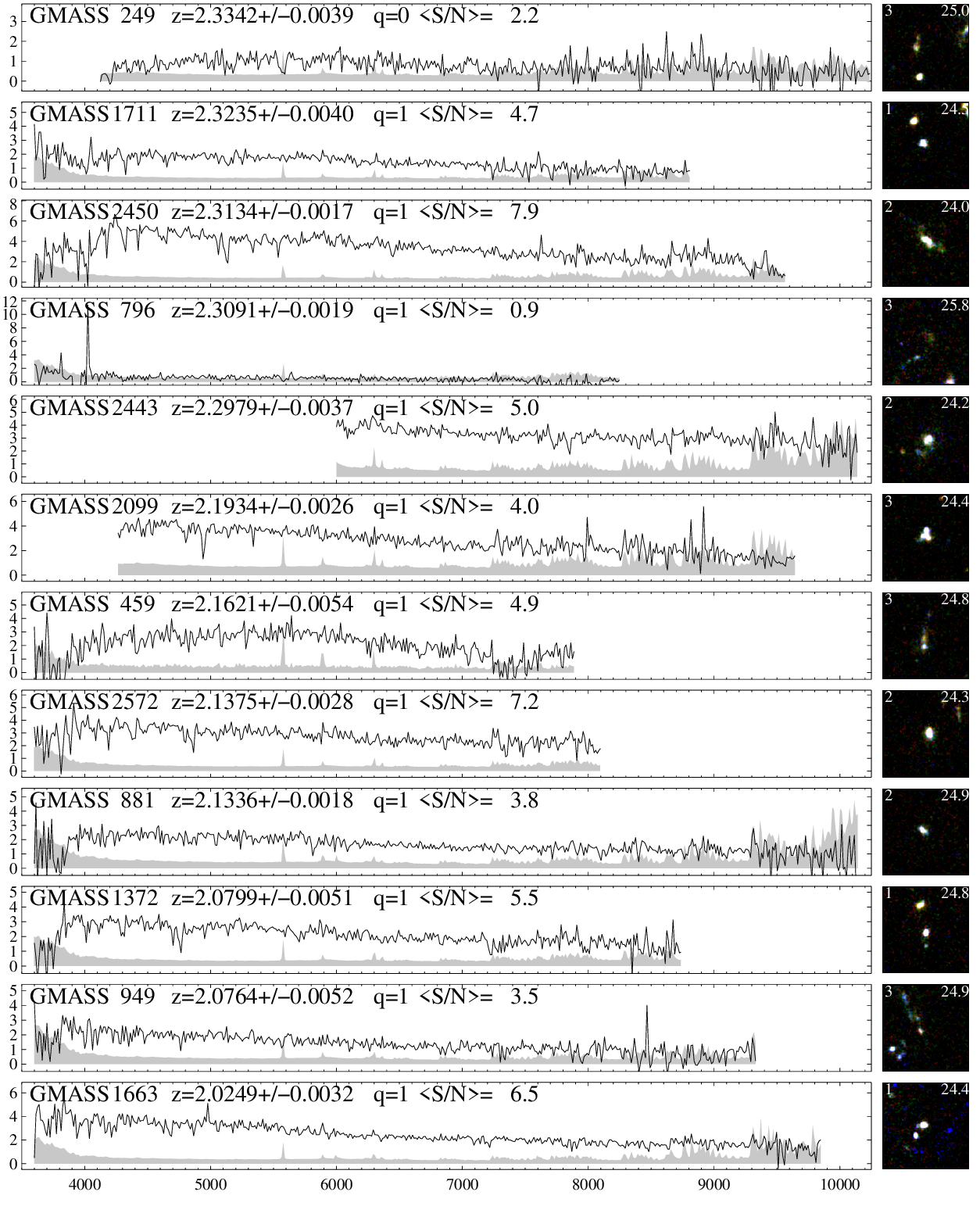}
  \caption{See Fig.~\ref{fig:stamps_spectra01} for description.}
  \label{fig:stamps_spectra04}
\end{figure*}

\begin{figure*}
\centering
  \includegraphics[width=\linewidth]{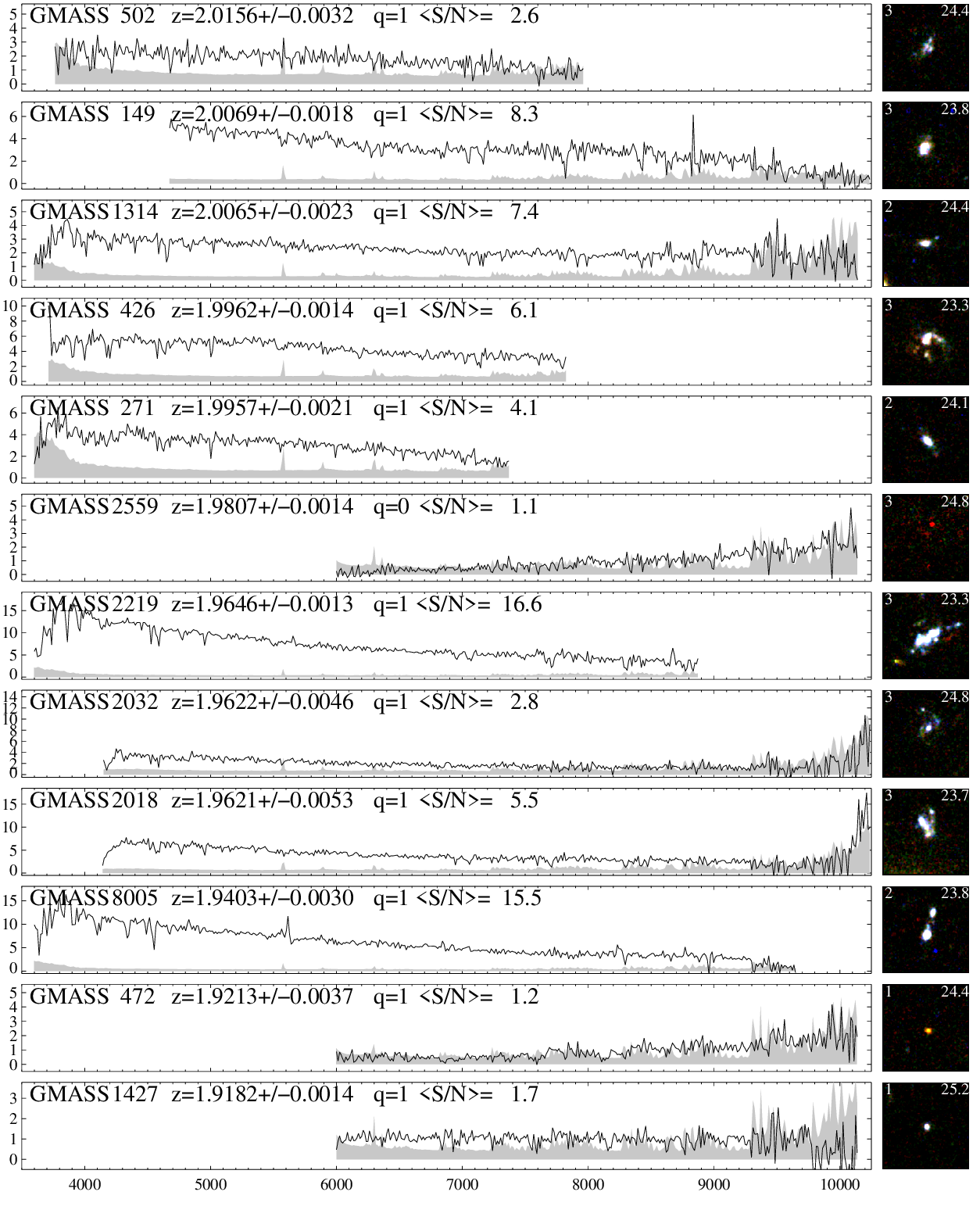}
  \caption{See Fig.~\ref{fig:stamps_spectra01} for description.}
  \label{fig:stamps_spectra05}
\end{figure*}

\begin{figure*}
\centering
  \includegraphics[width=\linewidth]{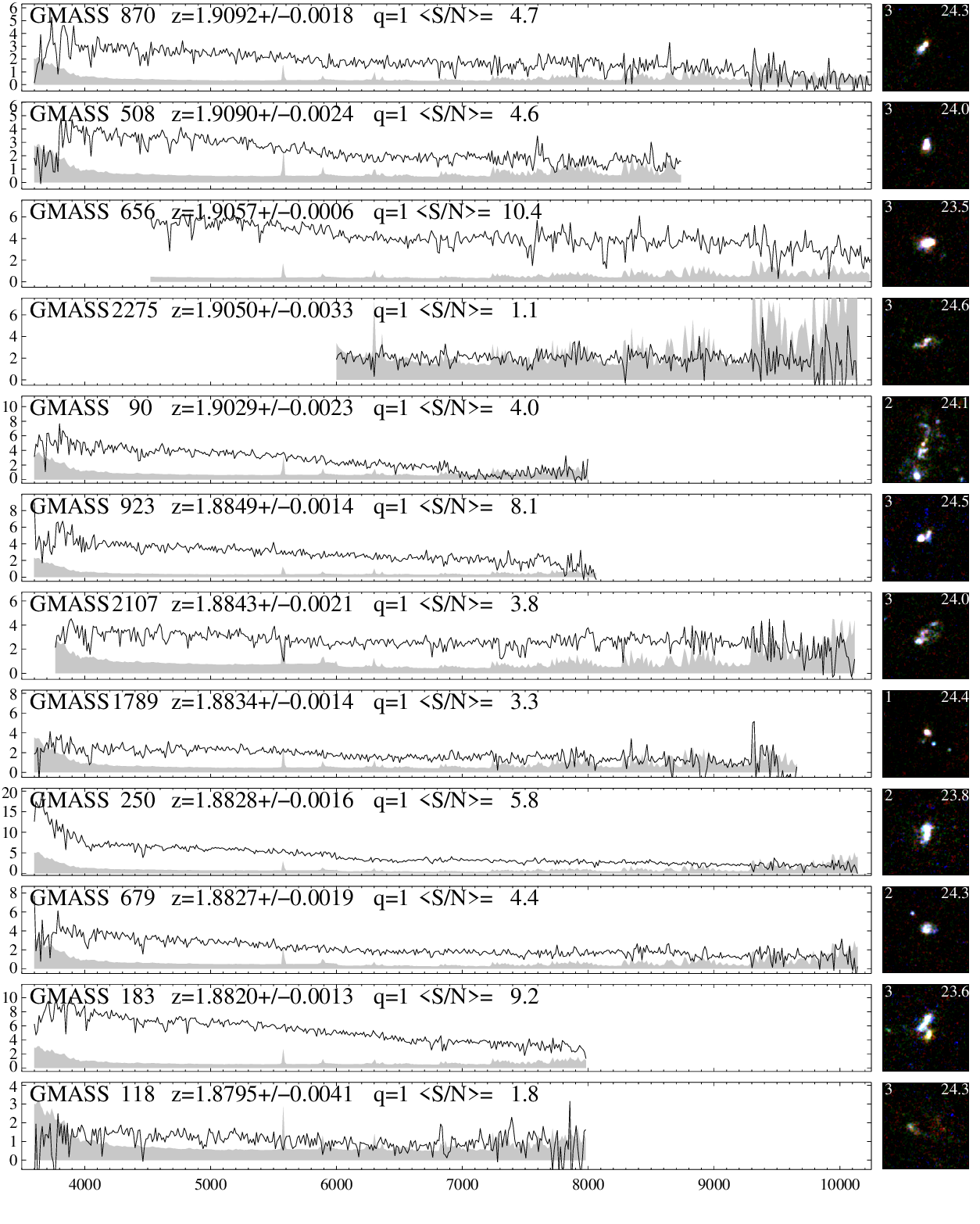}
  \caption{See Fig.~\ref{fig:stamps_spectra01} for description.}
  \label{fig:stamps_spectra06}
\end{figure*}

\begin{figure*}
\centering
  \includegraphics[width=\linewidth]{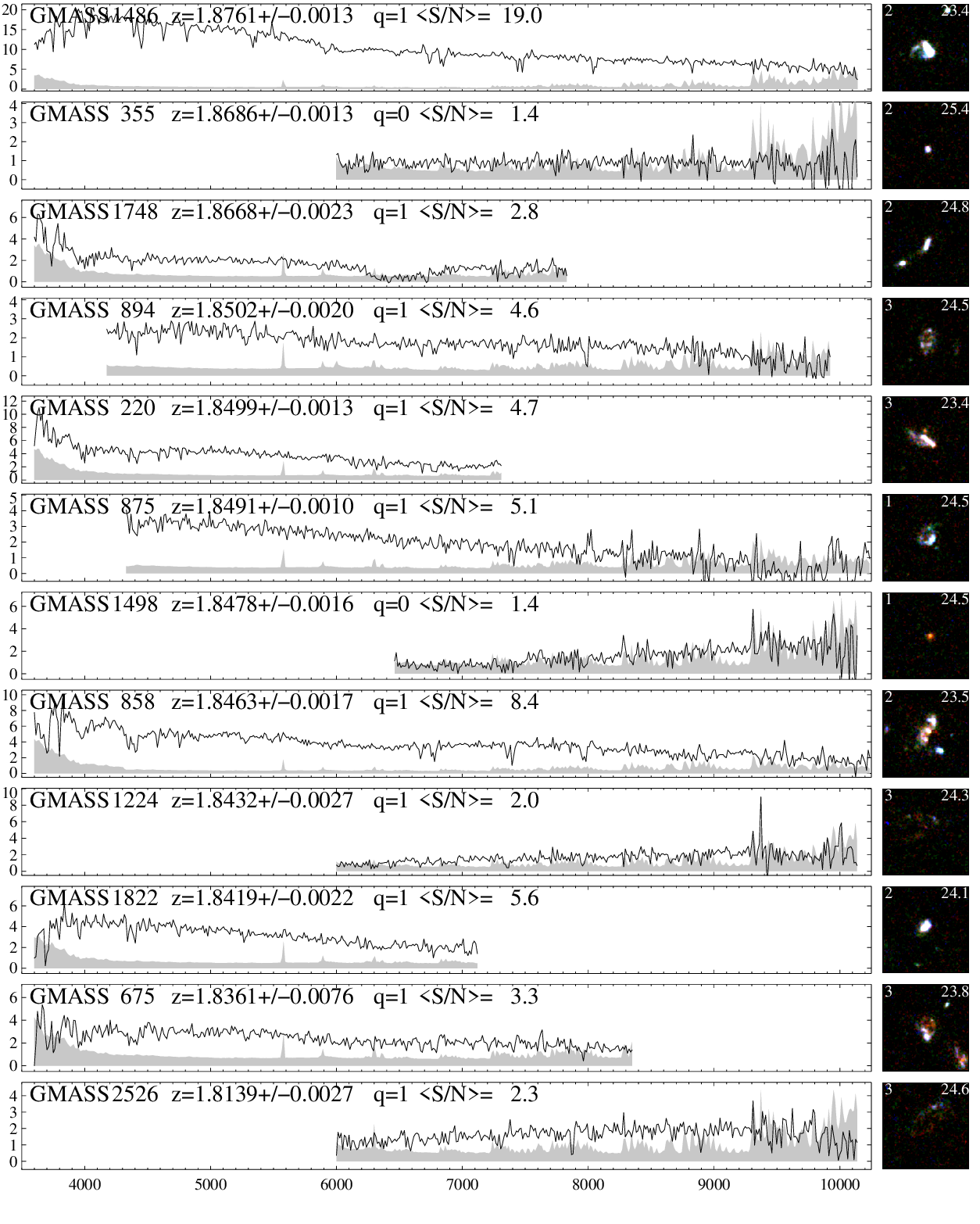}
  \caption{See Fig.~\ref{fig:stamps_spectra01} for description.}
  \label{fig:stamps_spectra07}
\end{figure*}

\begin{figure*}
\centering
  \includegraphics[width=\linewidth]{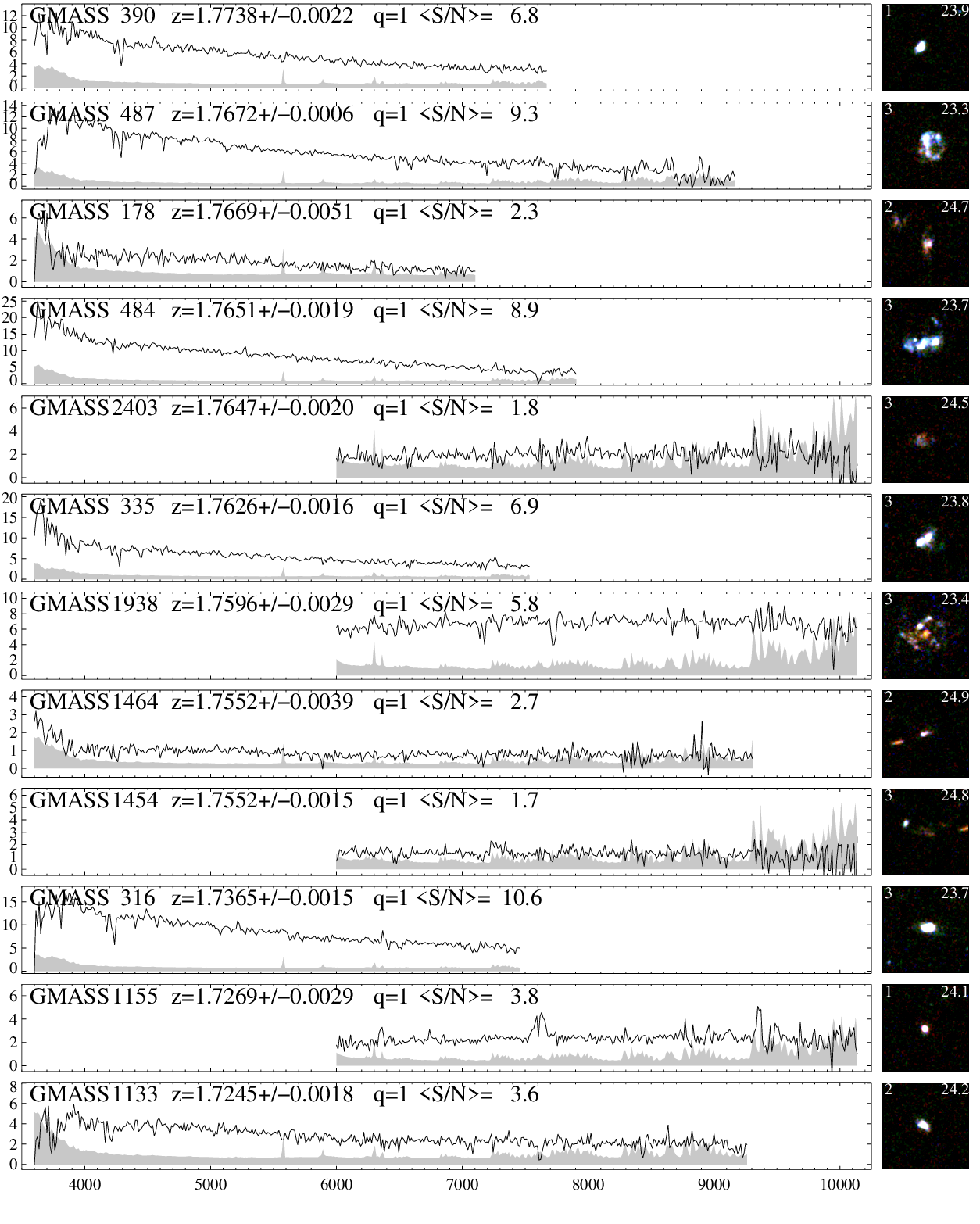}
  \caption{See Fig.~\ref{fig:stamps_spectra01} for description.}
  \label{fig:stamps_spectra08}
\end{figure*}

\begin{figure*}
\centering
  \includegraphics[width=\linewidth]{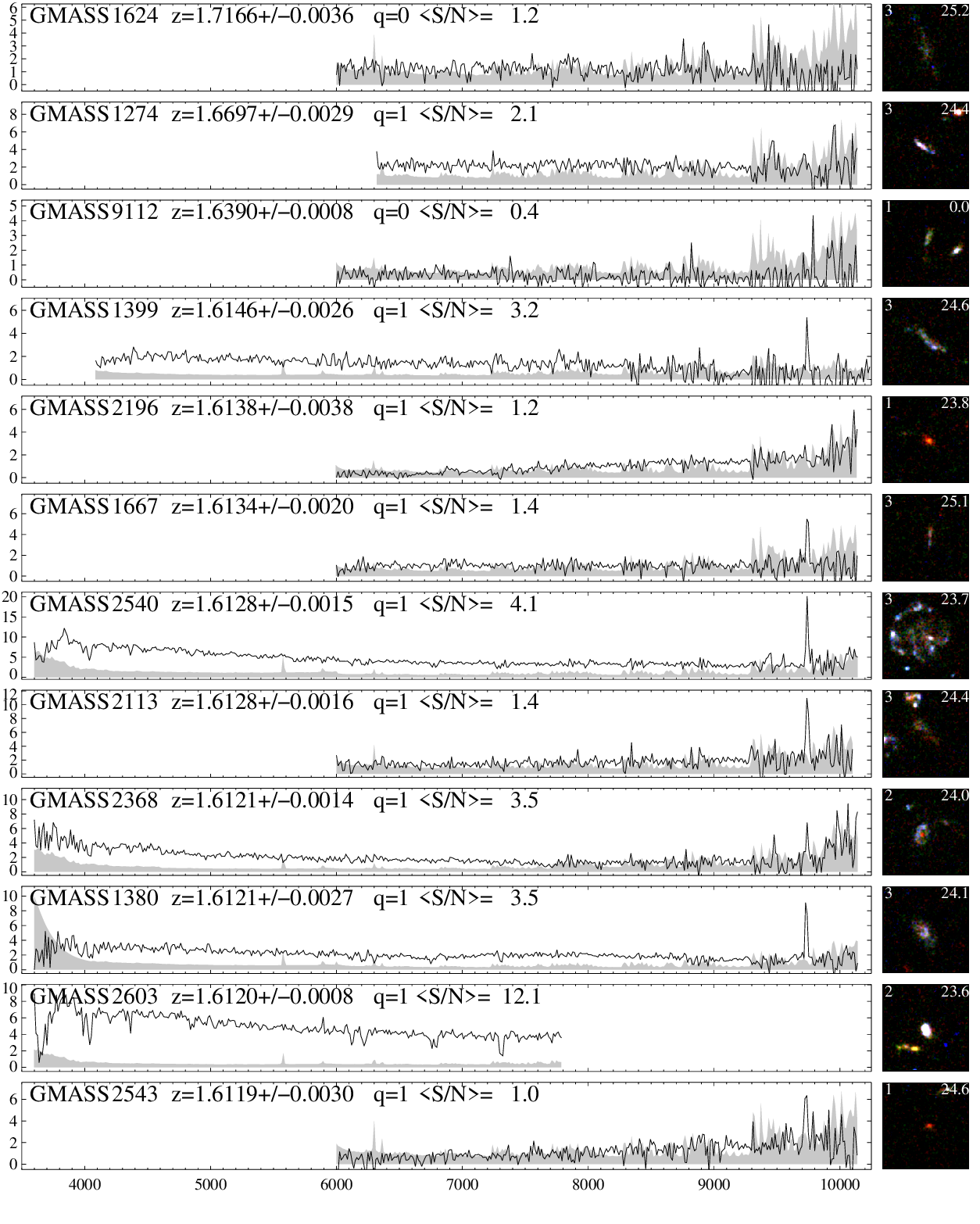}
  \caption{See Fig.~\ref{fig:stamps_spectra01} for description.}
  \label{fig:stamps_spectra09}
\end{figure*}

\begin{figure*}
\centering
  \includegraphics[width=\linewidth]{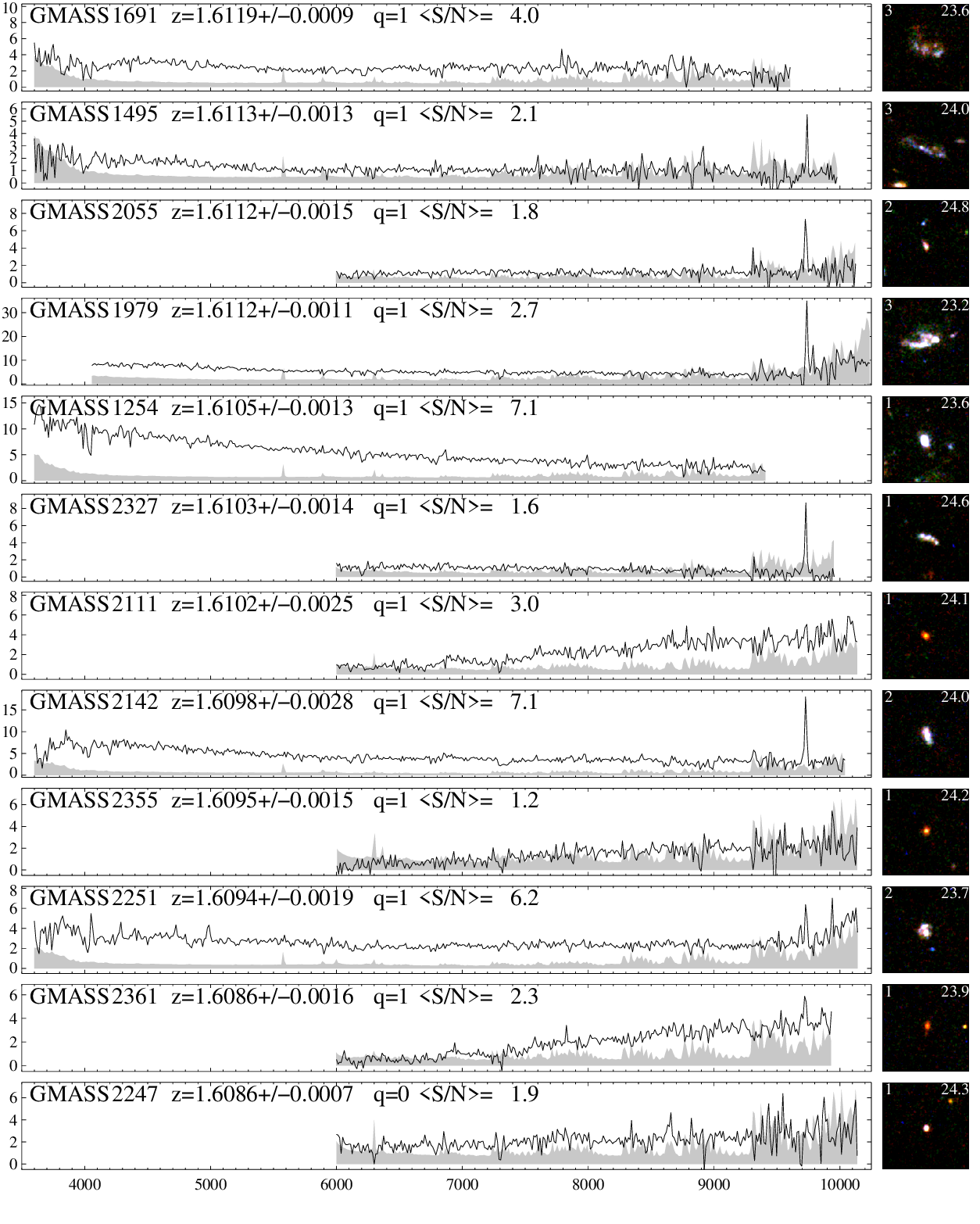}
  \caption{See Fig.~\ref{fig:stamps_spectra01} for description.}
  \label{fig:stamps_spectra10}
\end{figure*}

\begin{figure*}
\centering
  \includegraphics[width=\linewidth]{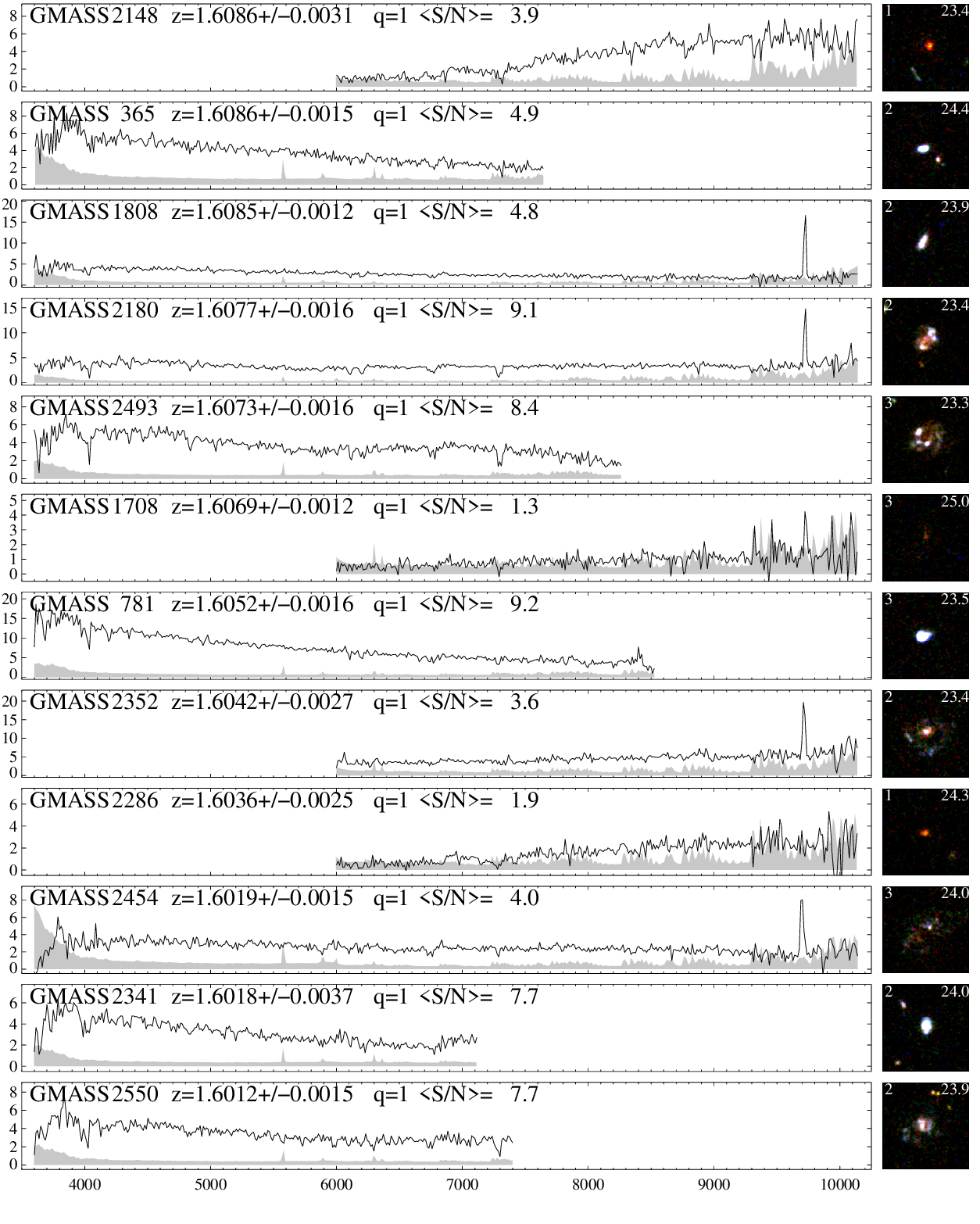}
  \caption{See Fig.~\ref{fig:stamps_spectra01} for description.}
  \label{fig:stamps_spectra11}
\end{figure*}

\begin{figure*}
\centering
  \includegraphics[width=\linewidth]{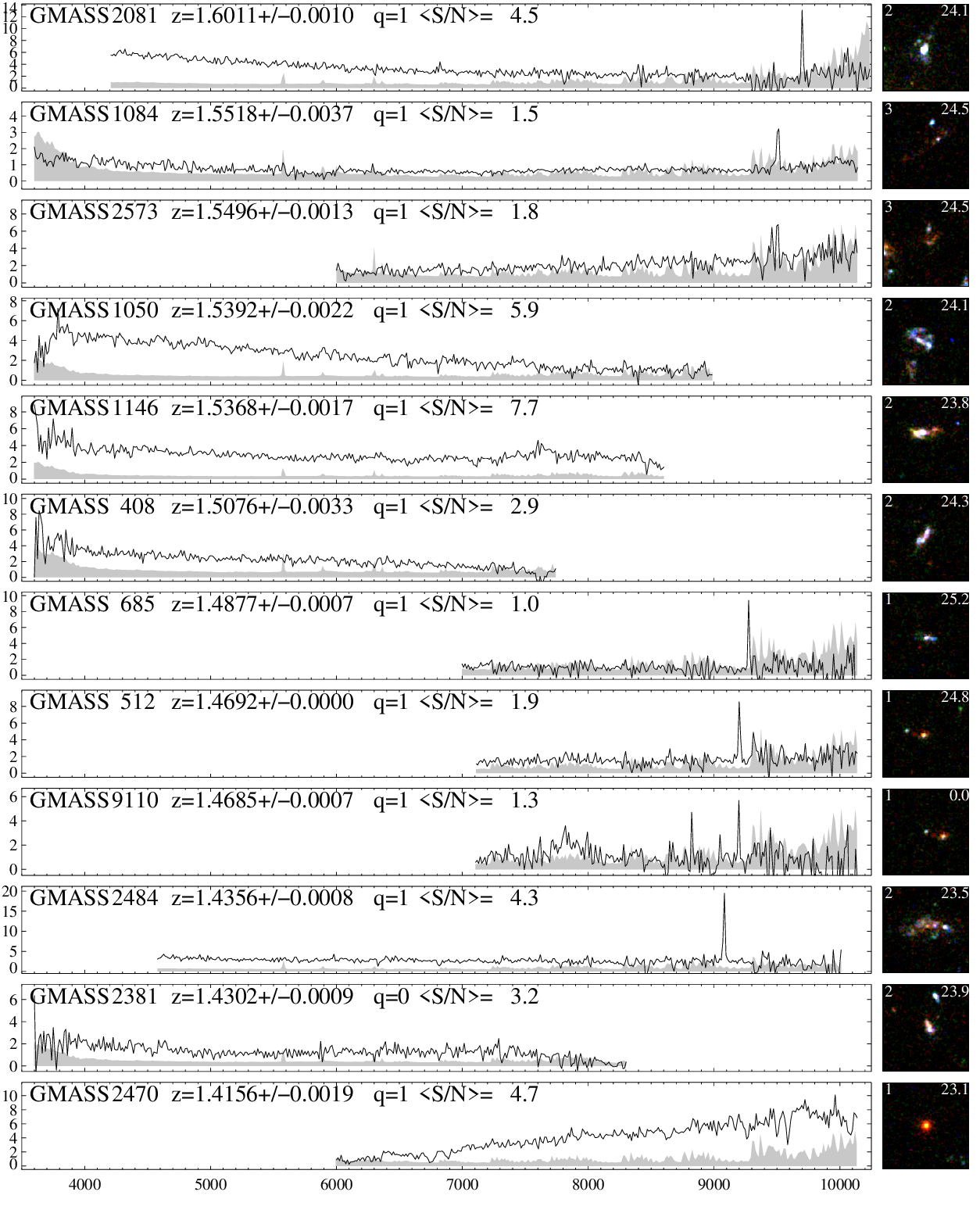}
  \caption{See Fig.~\ref{fig:stamps_spectra01} for description.}
  \label{fig:stamps_spectra12}
\end{figure*}

\begin{figure*}
\centering
  \includegraphics[width=\linewidth]{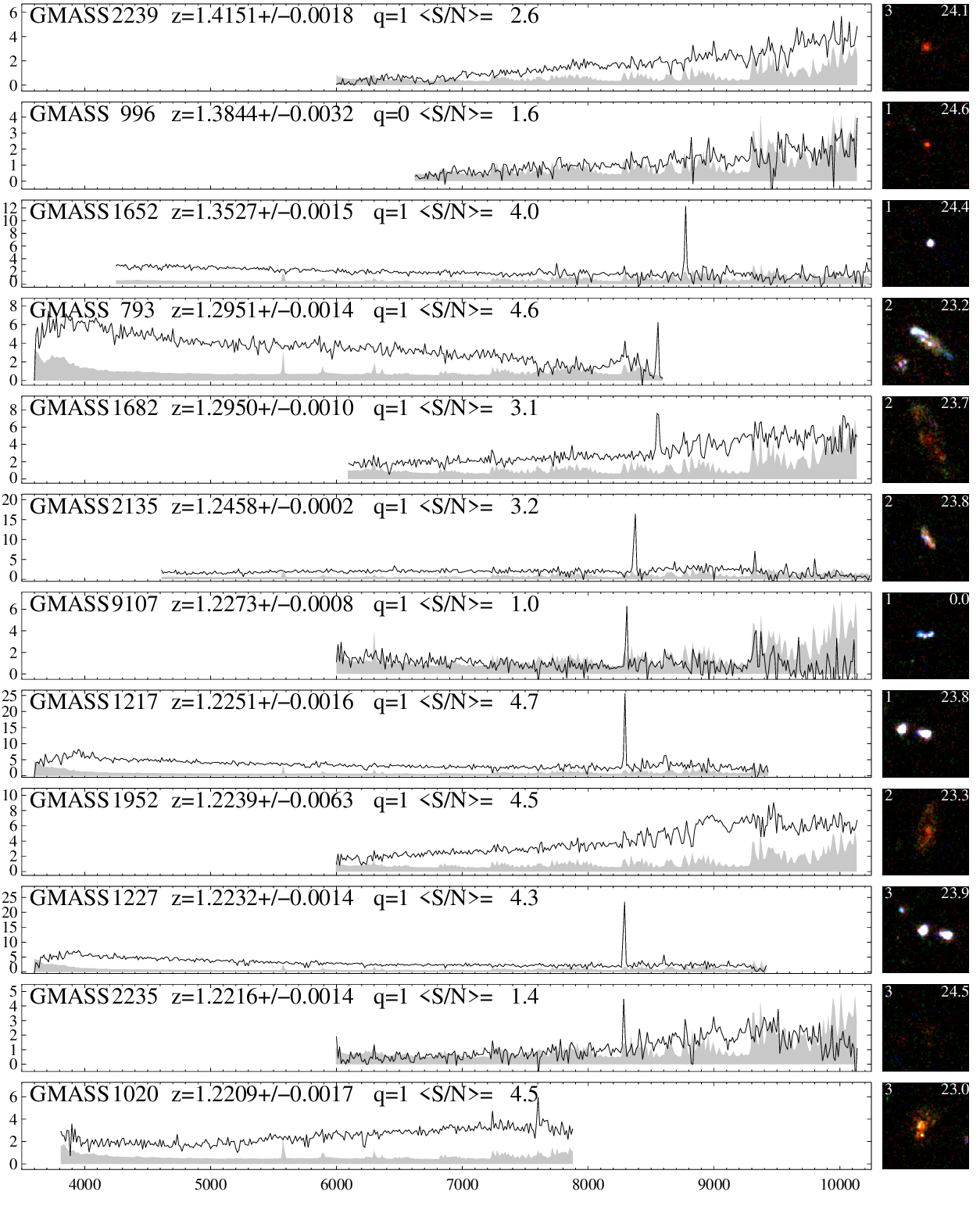}
  \caption{See Fig.~\ref{fig:stamps_spectra01} for description.}
  \label{fig:stamps_spectra13}
\end{figure*}

\begin{figure*}
\centering
  \includegraphics[width=\linewidth]{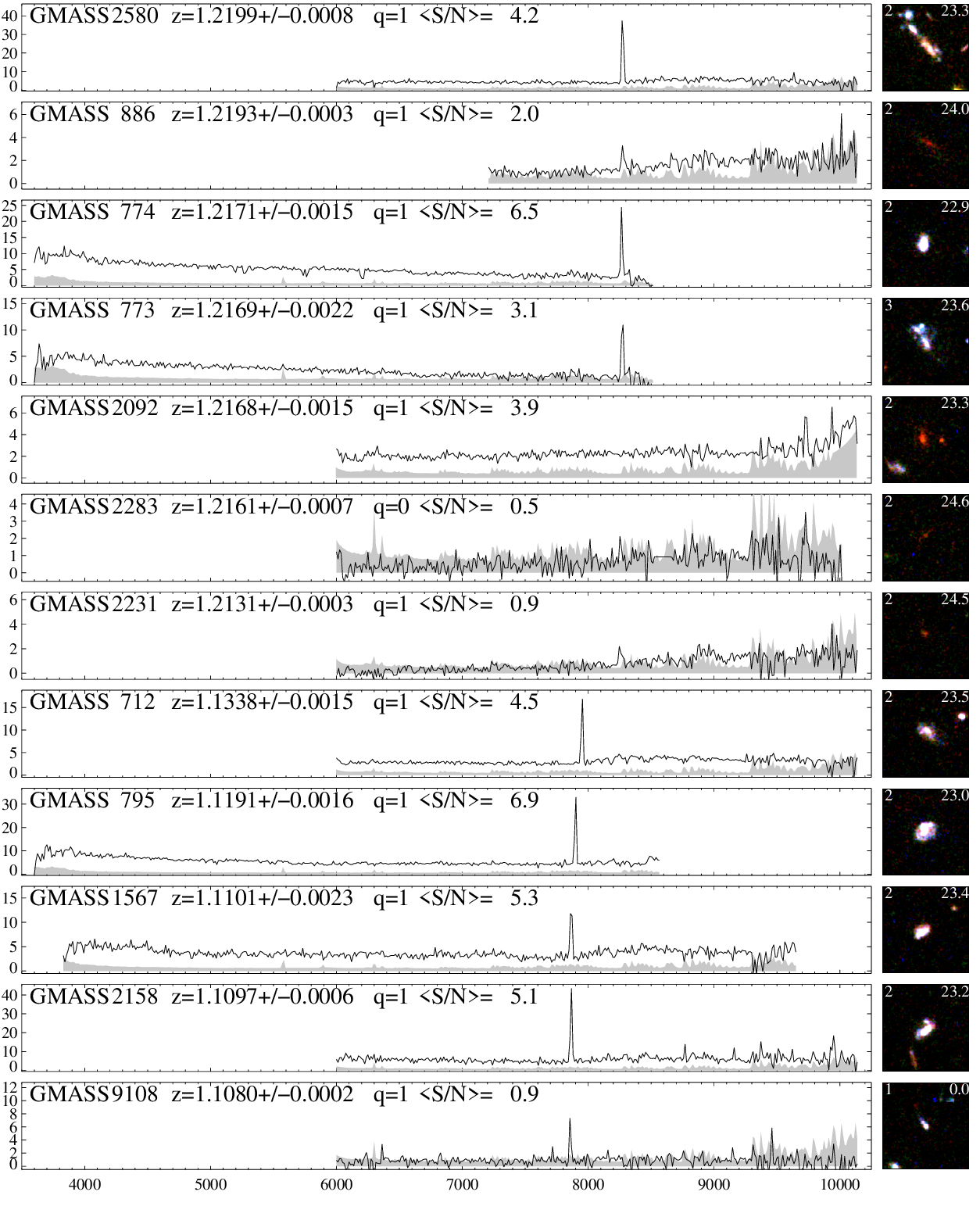}
  \caption{See Fig.~\ref{fig:stamps_spectra01} for description.}
  \label{fig:stamps_spectra14}
\end{figure*}

\begin{figure*}
\centering
  \includegraphics[width=\linewidth]{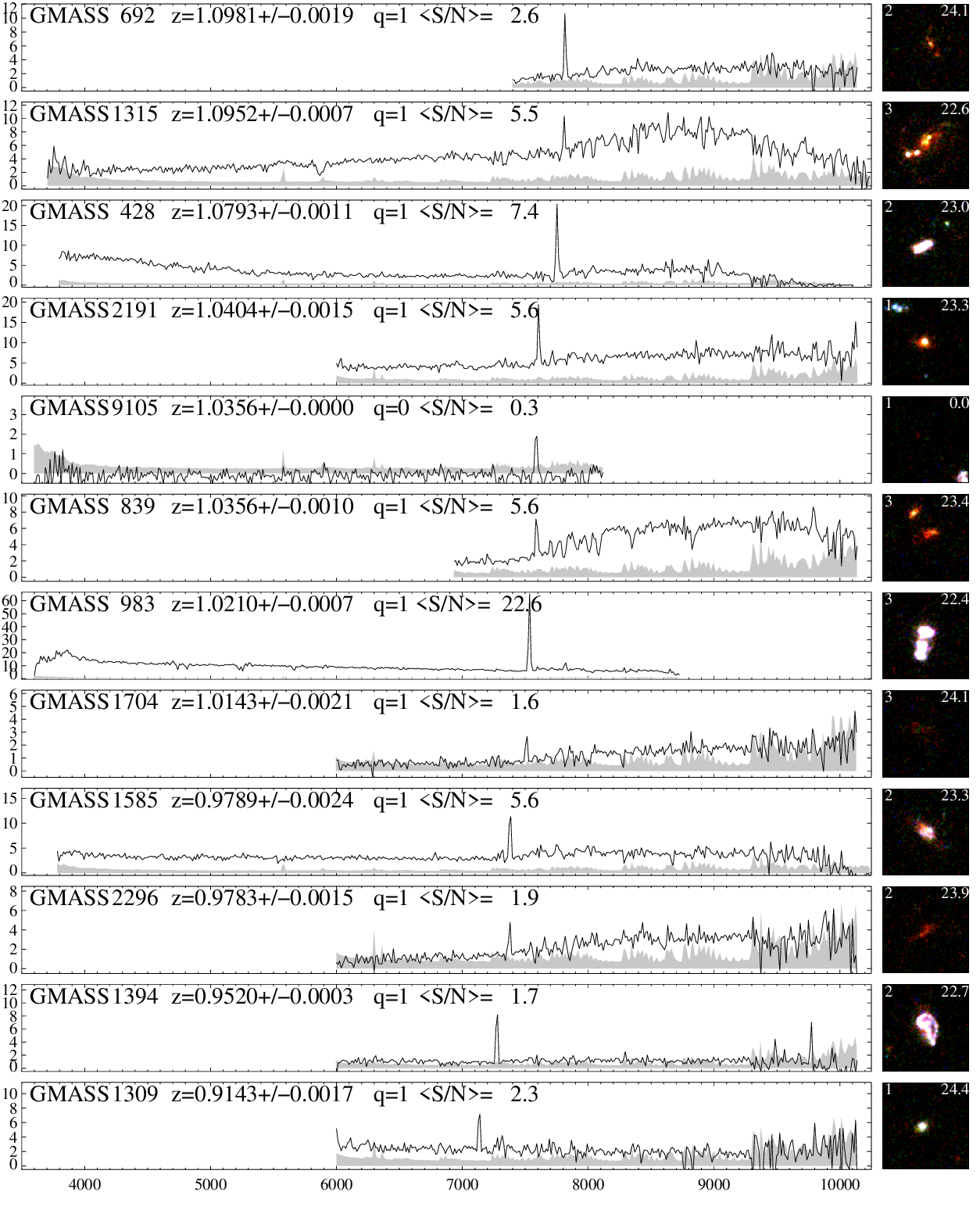}
  \caption{See Fig.~\ref{fig:stamps_spectra01} for description.}
  \label{fig:stamps_spectra15}
\end{figure*}

\begin{figure*}
\centering
  \includegraphics[width=\linewidth]{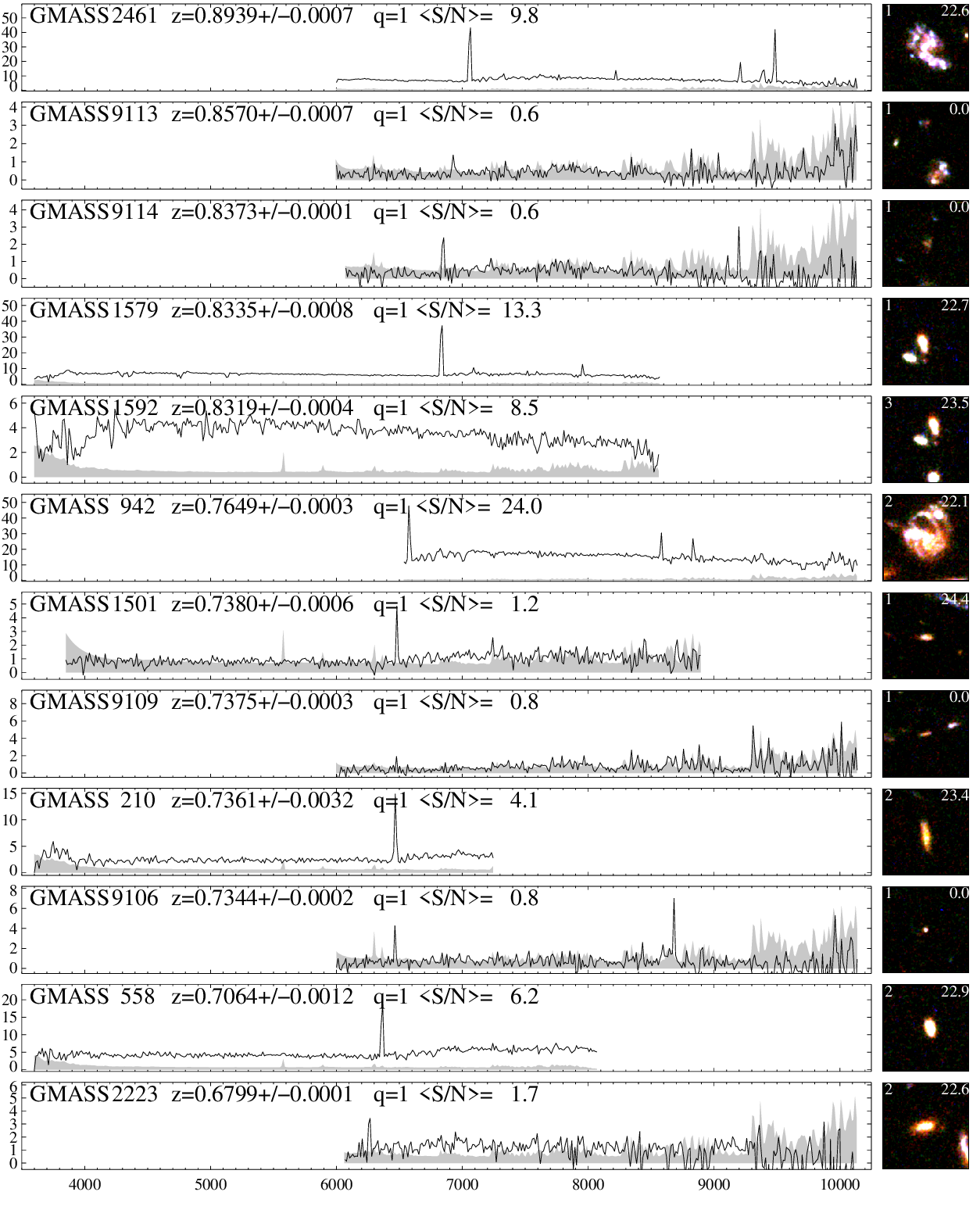}
  \caption{See Fig.~\ref{fig:stamps_spectra01} for description.}
  \label{fig:stamps_spectra16}
\end{figure*}

\begin{figure*}
\centering
  \includegraphics[width=\linewidth]{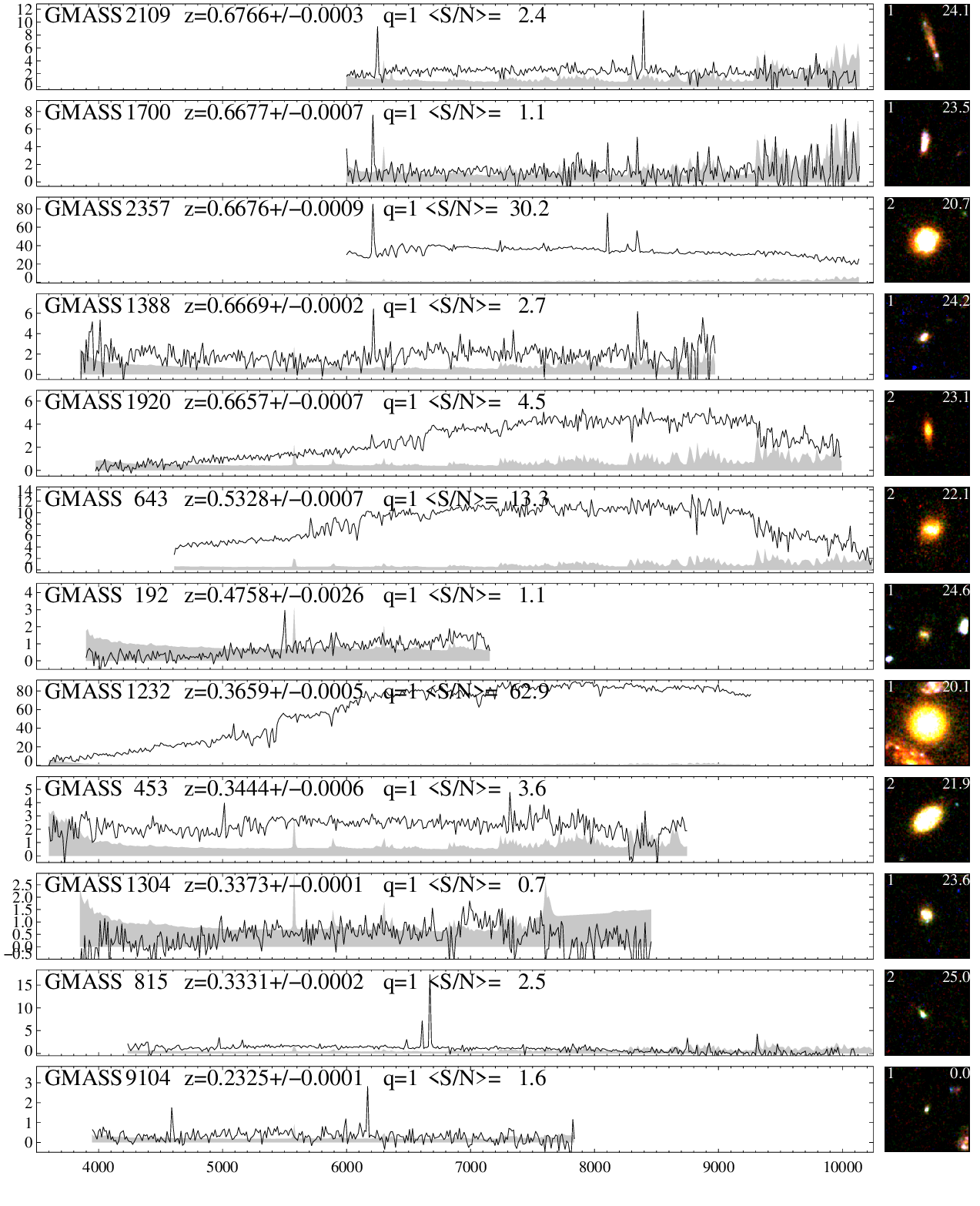}
  \caption{See Fig.~\ref{fig:stamps_spectra01} for description.}
  \label{fig:stamps_spectra17}
\end{figure*}

\begin{figure*}
\centering
  \includegraphics[width=\linewidth]{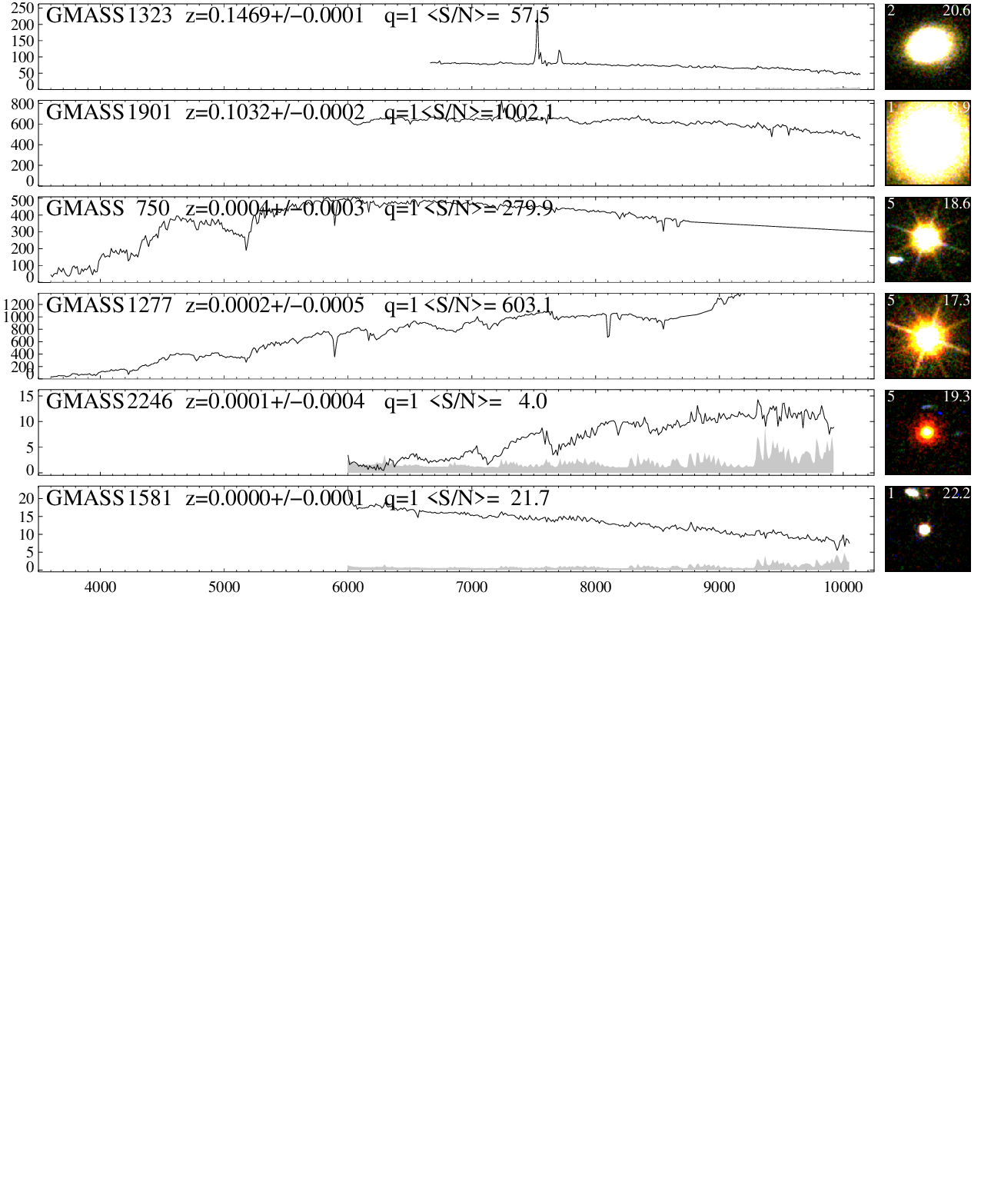}
  \caption{See Fig.~\ref{fig:stamps_spectra01} for description.}
  \label{fig:stamps_spectra18}
\end{figure*}

\end{appendix}

\end{document}